\documentclass[aps,prb,preprint,superscriptaddress]{revtex4-1}
\usepackage[utf8]{inputenc}
\usepackage{graphicx}
\usepackage{dcolumn}
\usepackage{bm,amsmath,verbatim}
\usepackage{amssymb}
\def\tr{{\text{tr}}\,}

\def\be{\begin{equation}}
\def\ee{\end{equation}}
\def\bea{\begin{eqnarray}}
\def\eea{\end{eqnarray}}
\def\bse{\begin{subequations}}
\def\ese{\end{subequations}}

\def\tr{{\text{tr}}\,}

\renewcommand{\v}[1]{{\bf #1}}

\newcommand{\s}{{\sigma}}

\def\be{\begin{eqnarray}}
\def\ee{\end{eqnarray}}

\newcommand{\Eq}[1]{Eq.~(\ref{#1})}
\newcommand{\e}{\epsilon}
\newcommand{\p}{\partial}

\newcommand{\ket}[1]{|#1\rangle}

\newcommand{\expect}[1]{\langle  #1\rangle}

\begin{document}

\title{From Majorana fermions to topological quantum computation in semiconductor/superconductor heterostructures}
\author{Jay D. Sau}
\affiliation{Department of Physics, Condensed Matter Theory Center and the Joint Quantum Institute, University of Maryland, College Park, Maryland 20742, USA}
\author{Sumanta Tewari}
\affiliation{Department of Physics and Astronomy, Clemson University, Clemson, SC 29634, USA}
\date{\today}

\maketitle

\section{Introduction}
Dirac is reported to have remarked in one of his talks that his equation was more intelligent than its author. But, as noted by Victor Weisskopf~\cite{Weisskopf1981}, it must be added that it was Dirac himself who found most of the additional insights. The concept of Majorana fermion, which also followed from the Dirac equation and was developed by Ettore Majorana \cite{Majorana1937}, was a notable exception. In relativistic quantum mechanics, physics of spin--1/2 fermions is described by the Dirac equation~\cite{Dirac1928, Peskin1995},
\begin{equation}
( i\gamma^{\mu}\partial_{\mu} -m )\psi = 0.
\label{eq:DE}
\end{equation}
Here, we have used the natural units ($\hbar=c=1$), and $\gamma^{\mu}$,  with $\mu = 0,1,2,3$, are a set of matrices satisfying the algebra, \begin{equation}\{\gamma^{\mu},\gamma^{\nu}\}\equiv \gamma^{\mu} \gamma^{\nu}+\gamma^{\nu} \gamma^{\mu}=2\eta^{\mu\nu},
\label{Eq:Gamma_1}
\end{equation}
 where  $\eta^{\mu\nu}$ is  the Minkowski metric ($\eta^{\mu \nu}=0, \mu \neq \nu; \eta^{00}=1; \eta^{ii}=-1, i=1,2,3$), and \begin{equation}\gamma_0\gamma_{\mu}\gamma_0 = \gamma_{\mu}^{\dagger}
 \label{Eq:Gamma_2}
 \end{equation}
 Eq.~{\ref{Eq:Gamma_1}}  
follows from the requirement that the energy eigenvalues satisfy the condition $E^2= p^2 + m^2$, while Eq.~{\ref{Eq:Gamma_2}} ensures hermiticity of Dirac Hamiltonian $H_D$ that follows from re-casting of Eq.~{\ref{eq:DE}} in terms of 
Schrodinger equation \begin{equation}i\frac{\partial \psi}{\partial t}=H_D \psi; \hspace*{.3 cm} H_{D}=-i\gamma^0 \vec{\gamma}\cdot \vec{\nabla}+\gamma^0 m. \label{Eq:Dirac_H}
\end{equation} 
  In $(3+1)$ dimension, the simplest representation of the Clifford algebra in Eqs.~{\ref{Eq:Gamma_1},\ref{Eq:Gamma_2}} can be found in terms of four $4\times 4$ matrices $\gamma^{\mu}$, known  as Dirac matrices. There are infinitely many unitarily equivalent representations of the Dirac matrices, each constituting a basis for solving the Dirac equation in Eq.~{\ref{eq:DE}}.

 In general the matrices have both real and imaginary elements, e.g., the so-called chiral or Weyl representation has, 
  \begin{align}
\begin{matrix}
\gamma^0\\
\end{matrix}
=
\begin{pmatrix}
0 & 1 \\
1 & 0 &
\end{pmatrix},
&&
\begin{matrix}
\gamma^i\\
\end{matrix}
=
\begin{pmatrix}
0 & \sigma^i \\
-\sigma^i & 0 &
\end{pmatrix}
\label{eq:Dirac_Matrices_Weyl}
\end{align}
where each element itself is a $2\times 2$ matrix, and the $\sigma^i$ are the Pauli matrices,
\begin{align}
\begin{matrix}
\sigma^1\\
\end{matrix}
=
\begin{pmatrix}
0 & 1 \\
1 & 0 &
\end{pmatrix},
&&
\begin{matrix}
\sigma^2\\
\end{matrix}
=
\begin{pmatrix}
0 & -i \\
i & 0 &
\end{pmatrix}
&&
\begin{matrix}
\sigma^3\\
\end{matrix}
=
\begin{pmatrix}
1 & 0 \\
0 & -1 &
\end{pmatrix}.
\end{align}
 Because of both real and imaginary elements in $\gamma^{\mu}$, Eq.~(\ref{eq:DE}) is a set of coupled differential equations with complex coefficients. Thus, the general solution $\psi(x)$ of Eq.~(\ref{eq:DE}) is a complex four-component spinor. This equation provides a relativistically covariant description of a spin$-1/2$ particle with charge, exactly as required by the electron. It also provides a natural 
 explanation for the gyromagnetic ratio of the electron being close to 2. 
 However, did not solve 
 the puzzle of negative energy solutions, which had plagued earlier versions of relativistic quantum mechanics and had
 confused the founders of quantum mechanics
 such as Pauli, Weiskopff, Wigner and Dirac himself. The presence of negative energy solutions leads to the possibility 
 that positive energy electron can scatter and reduce their energy without limit.  
 Dirac proposed a solution  to this problem by postulating  that all negative energy states were filled in the vacuum state.
 He also noticed 
 that corresponding to any negative energy solution $\psi(x,t)\propto e^{-i |E|t}$ of Eq.~\ref{eq:DE}, one could write a  positive energy charge conjugate wave-function~\cite{Sakurai} 
 \begin{equation}
 \psi^{(c)}(x,t)={\cal{C}}\psi^*(x,t)\propto e^{i|E|t},    \end{equation}
 where $\cal{C}$ is the charge conjugation matrix. The charge conjugation matrix is defined as a matrix that satisfies the constraint
 \begin{align}
    & {\cal{C}}^{-1}\gamma_\mu  {\cal{C}}=-\gamma_\mu^*.\label{eqC}
 \end{align}
  By considering the coupling to the vector potential, it is clear that the  wave-function $\psi^{(c)}(x,t)$
 represents a particle with the same mass and spin but opposite charge and magnetic moment to the particles described by \textbf{the pair} of
 positive energy solutions describing electrons. He suggested that these positive energy solutions would represent holes in the otherwise 
 filled vacuum. The hole excitations were termed positrons, which are the anti-particle of the electron, whose existence was 
 verified by Anderson~\cite{Anderson_Positron}. The observation of the positron was one of the first examples of a successful prediction of a new fundamental particle, and led to Pauli's memorable comment, ``Success seems to
have been on the side of Dirac rather than of logic''~\cite{Pais}. With the advent of semiconductor physics, 
 Dirac's argument became the standard description of hole doped semiconductors. Dirac's picture of a filled sea of negative energy electron states is an intrinsically many-particle picture, where, even in the absence of inter-particle interactions, a single particle can no longer be described by the solution of a single particle wave equation, e.g., the Dirac spinor $\psi$ ~\cite{Sakurai}. For example, Klein considered that the Dirac equation in a region of uniform electric field where the electrostatic 
 potential drops by $2m$, where $m$ is the mass parameter in Eq.~\ref{eq:DE}, would connect propagating negative kinetic energy states in one side to positive energy solutions 
 on the other side. This is called Klein tunneling of the negative energy states and has been seen in condensed matter systems \cite{Young2011}. In Dirac's picture, this process is viewed as the generation of electron-positron pairs by the electric field. While this process conserves charge, it does 
 not conserve number of particles because a pair of particles are created from vacuum. Thus, a more natural and indeed necessary~\cite{Sakurai} framework is to view the solution 
 $\psi(x,t)$ of Eq.~\ref{eq:DE} 
  as a second quantized field operator that unifies the description of particles and antiparticles through the introduction of creation and annihilation operators, 
 \begin{equation}
 \psi(x,t) = \int \frac{d^3p}{(2\pi)^3 \sqrt{2E_{\mathbf{p}}}} \sum_{s} [b_s(\mathbf{p})u_s(\mathbf{p})e^{-ipx+iEt}+ c_s^{\dagger}(\mathbf{p})v_s(\mathbf{p})e^{ipx-iEt}].
 \label{Eq:Dirac_Field}
 \end{equation} 
 Here, $s=\pm 1$ are the projection of the spin on the $z$ axis, $E_{\mathbf{p}}=p^0=\sqrt{\mathbf{p}^2+m^2}$, $u_s(\mathbf{p}), v_s(\mathbf{p})$ are positive and negative energy spinor solutions of Eq.~\ref{eq:DE} evaluated in the momentum space, and $b_s(\mathbf{p}) (c_s^{\dagger}(\mathbf{p}))$ is the annihilation (creation) operator for the particle (antiparticle). The Dirac field $\psi(x)$ annihilates a particle and creates its antiparticle and its hermitian conjugate $\psi^{\dagger}(x)$ creates a particle and annihilates its antiparticle. Imposing anti-commutation relations on the Dirac fields $\psi(x), \psi^{\dagger} (y)$ leads to similar anti-commutation relations among the particle and antiparticle creation and annihilation operators ($b_s^{\dagger}(\mathbf{p}), b_r(\mathbf{p}), c_s^{\dagger}(\mathbf{p}), c_r(\mathbf{p})$). One can also now write the total Hamiltonian as $H_D=\int \frac{d^3p}{(2\pi)^3}E_{\mathbf{p}}(b_s^{\dagger}(\mathbf{p})b_s(\mathbf{p}) + c_s^{\dagger}(\mathbf{p})c_s(\mathbf{p}))$ plus an unimportant constant that can be thrown away.  The Dirac equation (~\ref{eq:DE}) thus naturally predicts the existence of anti-fermion for each spin--$1/2$ fermion, e.g., positively charged positron for negatively charged electron as discovered by Anderson~\cite{Anderson_Positron}.
 
 In 1937, Ettore Majorana discovered a representation of the $\gamma$-matrices that satisfied the Clifford algebra (Eqs. (\ref{Eq:Gamma_1},\ref{Eq:Gamma_2})) but were purely imaginary.
  In this so-called Majorana basis ${\cal{C}}=\textbf{1}$, and the charge conjugation (CC) operation, which takes a particle to its antiparticle~\cite{Gellmann}, reduces to the complex conjugation operation. 
  Majorana then noticed that since the Dirac equation Eq.~\ref{eq:DE} becomes completely real, we can require that the solutions 
  $\psi(x,t)$ are also completely real, i.e. 
  \begin{equation}\psi=\psi^{c}=\cal{C}\psi^*=\psi^*.\label{Majcons}
  \end{equation} 
  With this constraint the Dirac equation would represent particles that are 
  identical to their own antiparticle.
  Majorana obsered that this constraint of particle being related to its own antiparticle can be ensured in any 
  representation of the Dirac matrices if we replace the Dirac mass term by the so-called Majorana mass $m$ in the Dirac 
  equation Eq.~\ref{eq:DE}  to construct the Majorana equation
  \begin{equation}
       i\gamma^{\mu}\partial_{\mu}\psi -m\psi^{(c)} = 0.\label{eq:ME}
  \end{equation}
  Because $\psi^{(c)}$ involves complex conjugation, the above equation is no longer invariant under the $U(1)$ gauge 
  transformation $\psi\rightarrow e^{i\Lambda}\psi$ needed to couple the particle to electromagnetic field in a gauge 
  invariant way. Therefore, $\psi(x,t)$ must represent a neutral particle, which makes sense for a particle which is 
  its own anti-particle.
     The physical equivalence of the Majorana particle and its antiparticle becomes clearer in the framework of quantum field theory.
     In analogy with Eq.~\ref{Eq:Dirac_Field}, a solution to the Majorana equation Eq.~\ref{eq:ME} can be expanded in terms of plane wave solutions~\cite{Pal2011},
     \begin{equation}
 \psi(x,t) = \int \frac{d^3p}{(2\pi)^3 \sqrt{2E_{\mathbf{p}}}} \sum_{s} [b_s(\mathbf{p})u_s(\mathbf{p})e^{-ipx+i E t}+ b_s^{\dagger}(\mathbf{p})C u_s^*(\mathbf{p})e^{ipx-iE t}].
 \label{Eq:Majorana_Field}
 \end{equation} 
 The field in Eq.~\ref{Eq:Majorana_Field} satisfies the Majorana self-conjugacy relation $\psi=\psi^{(c)}$ (charge neutral particle that is identical to its own antiparticle). More importantly, only one kind of creation operator $b_s^{\dagger}(\mathbf{p})$ operator 
 appears in the field equation as opposed to separate and electron and positron operators as in Eq.~\ref{Eq:Dirac_Field}. 
 An alternative way to see that the Majorana equation Eq.~\ref{eq:ME} involves half the degrees of freedom (i.e. does 
 not have a separate hole degree of freedom) is to use the Weyl representation of the $\gamma$ matrices in Eq.~\ref{eq:Dirac_Matrices_Weyl}. In this representation 
 the $\gamma$ matrices are block off-diagonal (see Eq.~\ref{eq:Dirac_Matrices_Weyl}) and the charge conjugation matrix $\cal{C}$ is purely block diagonal so that the bottom two components of  
 the Majorana equation Eq.~\ref{eq:ME} can be written in terms of a two-component spinor $\omega$: 
 \begin{equation}
       \bar{\sigma}^{\mu}\partial_{\mu}\omega + m\sigma_2 \omega^* = 0.\label{eq:MWE},
  \end{equation}
  where $\bar{\sigma}^{\mu}=[I_2,-\sigma^{1},-\sigma^{2},-\sigma^3]$. The full four component Majorana spinor satisfying Eq.~\ref{Majcons} and ~\ref{eq:MWE}
  is obtained as $\psi^T(x)=(\omega^T,(i\sigma^2\omega)^T)$.
 
 The twin properties of charge neutrality and the identification of the particle with its own antiparticle are not particularly uncommon among bosons, with photons and $\pi^0$-mesons being two examples. In fact, as is well known \cite{Peskin1995}, the Fourier expansion of the photon field $A_{\mu} (x)$ also contains creation and annihilation operators of only one kind, $a_{\lambda}(\mathbf{p})$ and $a_{\lambda}^{\dagger}(\mathbf{p})$, which is similar to the Fourier expansion in Eq.~\ref{Eq:Majorana_Field}, because photon is a boson that is identical to its own antiparticle. This is not counter-intuitive in bosons, which are considered as fields in the classical limits rather than 
 particles. But these properties are quite special among fermions, which are considered as particles in the classical limit. The neutron, for example, is a charge--neutral fermion, but it has an anti--particle (anti--neutron) that is different from the neutron by the sign of its magnetic moment. Also, neutrinos produced in beta-decay are thought to be charge--neutral, which follows from the conservation of electric charge. They also have a small but non--zero rest mass, so cannot be Weyl fermions~\cite{Weyl1929} which are massless. But whether neutrinos can be Majorana fermions, so that they are the same as their own antiparticle, is still an unsettled question in particle physics.

Although the jury is still out on whether any fermionic particle known in high energy physics can be a  Majorana fermion, in the last few years the concept of a fermionic particle being identical to its own antiparticle has entered the realm of condensed matter physics \cite{Nayak2008,Wilczek2009,Beenakker2011,Alicea2012,Leijnse2012,Sato2017,Aguado2017}. In condensed matter systems, however, the  ``elementary'' particles are necessarily electrons and protons, and so the term ``particle'' and ``antiparticle'' refer to the so-called \textit{quasiparticles} and \textit{quasiholes} corresponding to excitations of an underlying  many--body state.  As shown in a seminal paper by N. Read and D. Green~\cite{Read2000},  the Bogoliubov--de Gennes (BdG) equations satisfied by the superconducting  quasiparticles in the so--called weak--pairing phase of a two--dimensional (2D) spinless $p_x+ip_y$ superconductor are very similar to the Majorana form of the Dirac equation of relativistic quantum mechanics. These superconducting quasiparticles, then, represent the condensed matter analog of the Majorana fermions of high--energy physics. Similar to the Majorana fermions in high energy physics, in general, the Majorana-like BdG quasiparticles of 2D spinless $p_x+ip_y$ superconductors are characterized by a finite mass and finite energy. However, in the context of more recent developments with relevance to topological quantum computation (TQC), the term ``Majorana fermion'' has a slightly different meaning. The Bardeen-Cooper-Schrieffer (BCS) Hamiltonian of a superconductor has a so-called particle-hole (p-h) symmetry, according to which for every energy eigenvalue $+E$, there exists another energy eigenvalue $-E$. Additionally, if the hypothetical superconductor is made up of spinless electrons, the second-quantized operators corresponding to the excitations at $\pm E$ satisfy  $\gamma_E^{\dagger}=\gamma_{-E}$. Consequently, if the Hamiltonian admits a non-degenerate zero--energy eigenvalue, the corresponding quasiparticle is identical to its own anti-quasiparticle, with the second quantized operators satisfying $\gamma_0^{\dagger}=\gamma_{0}$. The zero-energy BdG quasiparticle states, should they exist, are bound states  localized by defects in the superconductor, e.g., vortices and sample edges~\cite{Read2000}, where the superconducting order parameter vanishes.  
In this chapter,  by the terms ``{\em Majorana zero mode (MZM)}'', or ``{\em Majorana bound state (MBS)}'' we will refer to such localized, charge--neutral, zero--energy bound states that may occur at defects and boundaries in appropriate superconductors.
The creation operator for such a zero energy state is a hermitian second quantized operator $\gamma^{\dagger}=\gamma$ that anti--commutes with other fermion operators and satisfy the relation $\gamma^2=1$. A collection of such Majorana bound states satisfy the algebra: 
\begin{align}
    &\{\gamma_i, \gamma_j\}=2\delta_{ij}, \gamma_i^{\dagger}=\gamma_i, \gamma_i^2=1.
\end{align} 
In two spatial dimensions, they obey a form of particle statistics known as non-Abelian statistics \cite{Moore1991,Nayak1996,Read2000,Ivanov2001,Stern2004}. In non-Abelian statistics, pairwise exchanges of particle coordinates represent  non--commutative operations, a fundamental property that can be used to implement fault-tolerant quantum gates. Recently, the interest in possible realization of zero energy Majorana bound states in condensed matter systems increased dramatically, motivated by the proposal~\cite{Kitaev2003} to use them as building blocks for fault tolerant topological quantum computation.

Charge-neutral fermionic quasiparticles are difficult to obtain even in condensed matter systems. The typical fermionic excitations in metals and semiconductors, e.g., electrons and holes, are charged quasiparticles. Bound states of electrons and holes, called excitons, can be charge neutral but they are bosonic quasiparticles. Superconductors may be a good system to look for charge neutral fermionic quasipartciles, because due to gauge symmetry breaking, charge is no longer a sharp observable. Indeed the Bogoliubov quasiparticles in a superconductor are linear superpositions of a particle and a hole, and hence are not charge eigenstates. However, satisfying the Majorana requirement $\gamma^{\dagger}=\gamma$ for a quasiparticle excitation is difficult even in a superconducting system. The reason can be traced back to the spin degeneracy of the constituent electrons and holes: The second quantized creation operator for a Bogoliubov excitation in a typical BCS superconductor can be written as, $d^{\dagger} = u c_{\uparrow}^{\dagger} + v c_{\downarrow}$. Therefore, even if the excitation energy vanishes (i.e., the Bogoliubov quasiparticle lies on the Fermi surface), forcing $u=v^{*}$, $d^{\dagger}$ is still not equal to $d$, because of the spins of electrons. It is for this reason a hypothetical spinless superconductor is introduced \cite{Read2000} as an ideal platform to realize MZMs, provided non-degenerate zero-energy localized states can be realized in such a system. 

In a superconductor, a non--degenerate localized zero--energy eigenstate, should it exist, enjoys a form of protection -- called \textit{topological protection} -- that makes it immune to weak local perturbations, as long as such perturbations do not close the superconducting gap. Such perturbations cannot move the state away from zero energy because of the particle--hole symmetry and the non--degeneracy condition. Under particle-hole symmetry, if the Hamiltonian of a superconductor has an energy eigenvalue $+E$, it must also have an energy eigenvalue $-E$. If the zero energy solution is non-degenerate, it is then its own particle-hole pair.  
 Since small perturbations to the BdG differential equation are not expected to change the total number of solutions,  it follows that weak local perturbations (i.e., perturbations that do not couple pairs of MFs) leave the non--degenerate zero energy eigenvalue unperturbed, because perturbing it to a non-zero value (say $+\epsilon$) will necessitate the introduction of its particle-hole counterpart ($-\epsilon$).
 This argument also implies that the zero--energy solutions are characterized by vanishing expectation values for \textit{any} local physical observable such as charge, mass, and spin. Otherwise, if there were a non-zero average of any of these observables in the zero energy wave function, a weak local field that couples to it (e.g., a weak magnetic field that couples to spin) would be able to shift the energy of the zero energy state. 
 As zero energy MFs in solid state systems are topologically protected, they can be removed from zero energy only by tuning the system through a topological quantum phase transition
 (TQPT)~\cite{Volovik1988} at which the bulk energy gap vanishes and the MFs become entangled with other gapless states at the topological quantum critical point.
 In the past decade, Majorana fermions have been discussed in a variety of low temperature systems 
\cite{Nayak1996,Read2000,Moore1991,Read1992,DSarma2005,DSarma2006,Kitaev2001,Bonderson2006,Zhang2008,Sato2009, Akhmerov2009,Ivanov2001,Stern2004,Chung2007,Fu2008,Fu2009,Fu2009a,Cook2011, Potter2012, Sato2009a,Ghosh2010, Martin2012, Sau2011,Klinovaja2011,Mao2011,Mao2012,Nadj-Perge602,Jack1255,Kimeaar5251,Zhang2019,Zhu189,Machida2019,dartiailh2021phase,ren2019topological,pientka2017topological}.
Perhaps the most important of these -- the semiconductor-superconductor (SM-SC) heterostructure -- has attracted intense attention as a result of an abundance of exciting experimental and theoretical results that have appeared steadily in the literature. In what follows, we will first discuss the theoretical background necessary to understand the emergence of MFs in defects and boundaries in condensed matter systems, followed by a detailed discussion of the theory and experiments looking for MFs in SM-SC heterostructures.     

\section{Theoretical background}

\subsection{Jackiw-Rebbi solution of a zero energy bound state in one dimension}\label{JackiwRebbi}

Before we discuss zero energy bound states localized in defects of the order parameter in superconductors, it will be instructive to discuss how such bound state solutions emerge in one-dimensional (1D) Dirac Hamiltonian. The zero energy bound state solutions for the various systems discussed in this chapter can be qualitatively understood by appropriate mapping of the corresponding Hamiltonians on the 1D Dirac problem.

One of the earliest examples of zero energy bound state solutions in a condensed matter system was investigated in domain wall states of polyacetylene\cite{Su1979,Su1980}.
In that case, starting with an ansatz for the dimerization order-parameter profile of polyacetylene, it was also possible to demonstrate the existence of a
zero energy bound state solution localized at a domain wall of the order parameter by explicitly solving the
mean-field equations \cite{Maki}. Remarkably, this domain
wall zero energy bound state was shown to be a condensed matter realization of the
zero mode associated with the mass solitons of a 1D Dirac problem
investigated by Jackiw and Rebbi\cite{Jackiw1,Jackiw2}.
The Jackiw and Rebbi soliton solution is a simple example of an
index theorem where fermionic zero modes can be used to count the
topological defects of a background
order parameter.

We begin with the Dirac Hamiltonian $H_D$ given in Eq.~\ref{Eq:Dirac_H}, $H_D=-i\gamma^0 \vec{\gamma}\cdot \vec{\nabla}+\gamma^0 m$. In one spatial dimension, we need only two Dirac matrices satisfying 
\begin{equation} 
\gamma^0\gamma^0=\textbf{1}, \gamma^0\gamma^1+\gamma^1\gamma^0=0, \gamma^1\gamma^1=-\textbf{1}.
\end{equation} 
To satisfy this algebra with only a pair of matrices, the following $2\times2$ matrices will do the trick:
\begin{align}
\begin{matrix}
\gamma^0\\
\end{matrix}
=\begin{matrix}
\sigma_z\\
\end{matrix},
&&
\begin{matrix}
\gamma^1\\
\end{matrix}
=
\begin{matrix}
i\sigma_x
\end{matrix}\label{gammaJR}
\end{align}
We can compute the charge conjugation matrix to satisfy Eq.~\ref{eqC} to be ${\cal{C}}=\sigma_x$.
The one dimensional Dirac Hamiltonian becomes, $H_D^1=-i\sigma_z(-i\sigma_x)\partial_x+\sigma_z m$. To discuss the Jakiew-Rebbi zero mode, we begin with the 1D  second quantized Dirac Hamiltonian,  
\be H_{D}^1=\int
dx\Big[-iv_F\psi^{\dagger}\sigma_y\p_x\psi
+m(x)\psi^{\dagger}\sigma_z\psi\Big],\label{h2}\ee where $
\psi^{\dagger}(x)=\begin{pmatrix}f^{\dagger}_1(x),&
f^{\dagger}_2(x)\end{pmatrix}$ with $f_{1,2}(x)$ being two independent fermion fields.
 In \Eq{h2}, we have used Fermi velocity $v_F$ to replace the velocity of light $c$ of the original Dirac equation ($c=1$ in natural units used in Eq.~\ref{eq:DE}) to indicate an effective Dirac equation valid in a lattice. The second term in \Eq{h2} has an effective mass and we assume 
$m(-x)=-m(x)$ is a spatially varying mass term that changes sign
at $x=0$ (the location of the domain wall).
 We now assume that the quasiparticle operator \begin{equation} q^{\dagger}=\int
dx~ [\phi_1(x)f^{\dagger}_1(x)+\phi_2(x) f^{\dagger}_2(x)]
\label{Eq:QP}
\end{equation}
satisfies the operator equation, \begin{equation}
[H_D^1,q^{\dagger}]=\epsilon q^{\dagger}. 
\label{Eq:Ladder}\end{equation} 
Computing the commutator in Eq.~\ref{Eq:Ladder} we find the
following Dirac equation for the two-component wave function
$\phi^{\rm{T}}(x)=(\phi_1(x),\phi_2(x))$: \be
-iv_F\sigma_y\partial_x\phi(x)+\sigma_z m(x)\phi(x)=\epsilon
\phi(x). \label{wa} \ee  First we note that because $\s_y$
anticommutes with $\s_x$ and $\s_z$, the first quantized Hamiltonian $H_{D}^1=-iv_F\sigma_y \partial_x+\sigma_z m(x)$ anticommutes with $\sigma_x$, $\{H_D^1,\sigma_x\}=0$. Then, if $\phi(x)$ is an
eigenfunction of $H_D^1$ with eigenvalue $\epsilon$, $H_D^1 \sigma_y \phi(x)=-\sigma_y H_D^1 \phi(x)=-\epsilon \sigma_y \phi(x)$, i.e., $\sigma_y\phi(x)$ is
also an eigenfunction of $H_D^1$ with eigenvalue $-\epsilon$. As a result,
the $\epsilon=0$ solutions of \Eq{wa} can be made a simultaneous
eigenstate of $H_D^1$ and $\s_x$. Let $\phi_0(x)$ denote such a solution and
\begin{equation}
    \s_x\phi_0(x)=\lambda\phi_0(x).\label{sxcond}
\end{equation} Setting $\e=0$ and left-multiplying
\Eq{wa} by $i\s_z$ we obtain \be \p_x\phi_0(x)={\lambda\over v_F}
m(x)\phi_0(x)\nonumber,\ee which implies \be
\phi_0(x)=e^{{\lambda\over v_F}\int_0^x
m(x')dx'}\phi_0(0).\label{zero}\ee For $m(x)=\pm {\rm
sign}(x)|m(x)|$, \Eq{zero} is a normalizable function for $\lambda=\mp 1$. In
this way we have now proven that for a sign change of $m(x)$ (i.e. a mass domain wall) there
is a single zero energy eigenvalue of the one-dimensional Dirac Hamiltonian with a wave function that is exponentially localized away from the mass domain wall. Note that, the zero energy solution is robust against variations in the mass distribution function $m(x)$, as long as there is a change of sign of the mass term at some value of $x$. The normalizable zero-energy solution exists  if \begin{eqnarray}
   m(x) &=& -m_1 \hspace*{1 cm} \rm{if} \hspace*{.5 cm} x < 0\nonumber\\
         &=& +m_2 \hspace*{1 cm}\rm{if} \hspace*{.5 cm} x>0
     \end{eqnarray}
     with $m_1, m_2 >0$. In particular, the solution exists even if the mass on the right hand side of the domain wall diverges $m_2 \rightarrow \infty$. In this case, the wave function corresponding to the zero energy eigenvalue vanishes for $x>0$. However, it remains non-zero and exponentially localized for $x\leq 0$. Now suppose the equation governing a topological system near a straight boundary (interface) with vacuum can be cast in the form a one dimensional Dirac equation (defined along the direction perpendicular to the boundary) with a negative mass term. Then modelling the vacuum just outside the boundary as a region with a positive infinite mass (so no particle can escape there) ensures the existence of a robust zero energy eigenfunction exponentially localized in the direction perpendicular to the boundary of the topological medium. This type of qualitative argument provides a visually appealing picture of topologically robust zero energy states localized at boundaries and order parameter defects in spinless $p_x+ip_y$ superconductors as discussed below. 
     
     It is important to mention here that despite the existence of a single topologically robust zero energy eigenstate localized at the mass domain wall in 1D Dirac theory, it does not represent a Majorana zero mode. In fact, the second quantized creation operator $q^{\dagger}$ in Eq.~{\ref{Eq:QP}} allows us to define the corresponding annihilation operator,
     \begin{equation} q=\int
dx~ [\phi_1^{*}(x)f_1(x)+\phi_2^{*}(x) f_2(x)].
\label{Eq:QP1}
\end{equation}
It should be clear that the operators $q,q^{\dagger}$ do not satisfy the Majorana condition $q^{\dagger}=q$, but, when properly normalized, satisfy the fermion anticommutation relation $\{q^{\dagger}, q\}=1$. The wave function $\phi_0(x)$ in 
Eq.~\ref{zero} and operators $q^{\dagger}, q$, thus, describe a zero-energy localized \textit{conventional} fermionic mode. Such a conventional fermion mode can in fact be viewed as the bound state of a \textit{pair} of Majorana zero modes with strongly overlapping wave functions. To see this, we define the operators, 
\begin{equation}
\gamma_{+}=q^{\dagger} + q \hspace*{1 cm} \gamma_{-} = i(q^{\dagger}-q)
\label{Eq:Overlapping}
\end{equation}
It should be easy to check that $\gamma_{+}$ and $\gamma_{-}$ both individually satisfy the Majorana condition, $\gamma_{+}^{\dagger}=\gamma_{+}, \gamma_{-}^{\dagger}=\gamma_{-}$, and further, $\gamma_{\pm}^2=q^{\dagger}q + q q^{\dagger} =1$, and they mutually anticommute with each other $\{\gamma_{+}, \gamma_{-}\}=0$. If we define a Fock state $|0\rangle$ with energy $E=0$ defined by the condition $q|0\rangle =0$ (i.e., the zero energy state at mass domain wall is unoccupied), then the occupied state $|1\rangle$ can be defined as $|1\rangle = q^{\dagger} |0\rangle$, which is degenerate ($E=0$) with $|0\rangle$. The conventional fermion occupation number operator $n=q^{\dagger}q$ can be expressed in terms of the Majorana operators $\gamma_{+}, \gamma_{-}$ as, $n=q^{\dagger}q= \frac{1}{2}(1 + i \gamma_{+}\gamma_{-})$. Conversely, the operator $i\gamma_{+}\gamma_{-}= 2q^{\dagger}q-1$, which takes the value $1$ in the state $|1\rangle$ and $-1$ in the state $|0\rangle$, is called the fermion parity operator. It is now clear that the conventional fermionic  mode described by $q^{\dagger}, q$ allows us to define a \textit{pair} of Majorana zero modes, but the two MZMs $\gamma_{+}, \gamma_{-}$ do not occur separately in space, describe the same localized wave function $\phi_0(x)$ in Eq.~\ref{zero}, and should really be viewed as a basis transformation from the creation and annihilation operators $q^{\dagger}, q$ to the Majorana operators $\gamma_{+}, \gamma_{-}$. The goal of the ongoing research on Majorana fermions in condensed matter systems is to create experimental conditions so that individual Majorana zero modes (e.g., $\gamma_{+}$ or $\gamma_{-}$) can occur spatially well-separated from each other. It is only in this limit they individually acquire topological protection.

\subsection{2D Spinless $(p_x+ip_y)$ superconductor}
The 2D spinless $(p_x+ip_y)$ superconductor (superfluid) is the canonical system that supports zero energy MZMs localized at the defects of the order parameter, such as vortices and sample edges~\cite{Read2000}. In 2D the mean field Hamiltonian describing the quasiparticle excitations for such a system is given by,
\begin{equation}
H_{2D}^{p} = \sum_{p}\xi_p c_p^{\dagger}c_p + \Delta_0 \sum_p\left[(p_x + ip_y)c_p^{\dagger}c_{-p}^{\dagger} + h.c.\right],
\label{eq:Hp}
\end{equation}
where $\xi_p = \epsilon_p - \mu$ with $\epsilon_p \rightarrow \frac{p^2}{2m*}$ for small $p$, $m^*$ is the effective mass, and $\mu$ is the chemical potential. Here, the spin indices of the electron operators are omitted because the system is considered to be spinless (or spin-polarized).  Read and Green~\cite{Read2000} showed that for the Hamiltonian in Eq.~\ref{eq:Hp}, the long distance behavior of Cooper pair wave function $g(\mathbf{r})$ undergoes a dramatic change as a function of $\mu$. For $\mu < 0$, $g(r) \sim e^{-r/r_0}$, indicating that the pairs are tightly bound and the superconductor (superfluid) is in a so-called strong pairing phase. On the other hand, for $\mu >0$, $g(r) \sim \frac{1}{r}$, and the long tail in the Cooper pair wave function indicates that the system is in a so-called weak pairing phase, which is continuously connected to the BCS weak coupling superconductor. The phase transition at $\mu=0$, at which the excitation gap vanishes at the momentum space point $\mathbf{k}=0$, is not associated with any change in symmetry of the superconducting state but is topological in nature. 

The weak and strong pairing phases of the Hamiltonian in Eq.~\ref{eq:Hp} are distinguished by distinct integer values of a topological invariant that can be defined as follows: The Hamiltonian in Eq.~\ref{eq:Hp} can be written in the Nambu basis as,
\begin{equation}
 H_{2D}^p = \sum_p \Psi^{\dagger}(p) {\cal{H}}_{2D}^p \Psi(p),
 \label{eq:Nambu}
\end{equation}
where $\Psi^{\dagger}(p)=(c_p^{\dagger}, c_{-p})$ and $\Psi(p)$ is its hermitian congugate. Here, the $2\times 2$ Hamiltonian matrix ${\cal{H}}_{2D}^p$ can be cast in the form,
\begin{equation}
{\cal{H}}_{2D}^p=\mathbf{d(\mathbf{p})}\cdot \boldsymbol{\sigma},
\label{eq:d}
\end{equation} 
where the three-component vector $\mathbf{d}(\mathbf{p})$ can be written as, $\mathbf{d}(\mathbf{p})=({\rm{Re}}(\Delta_p), -{\rm{Im}}(\Delta_p), \xi_p)$, $\Delta_p=\Delta_0(p_x+ip_y)$, and $\boldsymbol{\sigma}$ is the three-component vector of the Pauli matrices. The unit vector corresponding to $\mathbf{d}(\mathbf{p})$, namely, ${\hat{\mathbf{d}}(\mathbf{p})}=\frac{\mathbf{d}(\mathbf{p})}{|\mathbf{d}(\mathbf{p})|}$, provides a mapping of the two-dimensional momentum space on the surface of a unit sphere defined by $|{\hat{\mathbf{d}}(\mathbf{p})}|=1$ . As $\mathbf{p}$ moves over the 2D momentum space, ${\hat{\mathbf{d}}(\mathbf{p})}$ sweeps out area over its unit sphere.  Starting with $|\mathbf{p}|\rightarrow \infty$ as one covers the momentum space ending at $|\mathbf{p}|= 0$, the number of times the unit vector ${\hat{\mathbf{d}}(\mathbf{p})}$  wraps around the unit sphere  is a topological invariant called Chern number ($C$). Mathematically, the quantity $C$ is given by, 
\begin{equation}
C=\int \frac{d^2p}{4\pi} (\hat{\mathbf{d}}\cdot(\partial_{p_x} \hat{\mathbf{d}} \times \partial_{p_y} \hat{\mathbf{d}})) 
\label{eq:Chern}
\end{equation} 
For $|\mathbf{p}|\rightarrow \infty$ in any direction, $\xi_p \sim \frac{p^2}{2m^*}$ dominates and $\hat{\mathbf{d}}$ points along the north pole of the unit sphere. In the strong pairing phase of Eq.~\ref{eq:Hp}, $\mu <0$, and for $|\mathbf{p}|= 0$, $\hat{\mathbf{d}}_x, \hat{\mathbf{d}}_y =0, \hat{\mathbf{d}}_z=-\mu >0$. Hence in the strong pairing phase, for $|\mathbf{p}|= 0$, $\hat{\mathbf{d}}$ continues to point along the north pole. Conversely, in the weak pairing phase given by $\mu >0$, at the origin of the momentum space $\hat{\mathbf{d}}_z=-\mu <0$ and the unit vector $\hat{\mathbf{d}}$ points along the south pole of its unit sphere. Mathematically it can be shown that the Chern number in Eq.~\ref{eq:Chern} vanishes in the strong pairing phase while it acquires a value $C=+1$ in the weak pairing phase of Eq.~\ref{eq:Hp}. The value of the Chern number can only change at a topological quantum phase transition which, in the present case, occurs through a gap closing at the origin of the momentum space for $\mu=0$.

To elucidate the topological difference between the weak and the strong pairing phases, Read and Green showed that \cite{Read2000}, in the weak pairing phase ($\mu > 0$),  the BdG equations $H_{2D}^p\Psi(r)=E\Psi(r)$ near a vortex or the sample edges (where the order parameter $\Delta_0$ vanishes) admit zero energy ($E=0$) solutions, while the  corresponding second--quantized operator (the creation operator for the Bogoliubov state) is hermitian, $\gamma^{\dagger}=\gamma$. No such zero energy Majorana solutions exist near the vortices or sample edges in the strong pairing phase ($\mu >0$). To understand this, consider a long boundary or edge of the system parallel to the $y$-axis, separating the system in weak pairing phase situated at $x<0$ and vacuum or free space at $x>0$. The vacuum is characterized by the absence of particles, which can be implemented by having a potential $V$ large and positive for $x>0$. Since in the Hamiltonian \ref{eq:Hp}, $V$ modifies the chemical potential as $\mu \rightarrow (\mu -V)$, a large and positive potential $V$ in vacuum implies a large and negative $\mu$ for $x>0$. Since for $x<0$ we have the weak pairing phase with $\mu >0$, there must be a domain wall in $\mu$ where it changes sign near the boundary parallel to the $y$-axis. Using the small-$p$  approximation $\xi_p = \frac{p^2}{2m^*}-\mu \simeq -\mu$, that is, ignoring the term quadratic in $p$ in comparison to the linear $p$ term in the order parameter, it can be shown that \cite{Read2000}, for $E=0$, the BdG equations for the spinor $\Psi(r)=(u(r), v(r))^{T}$ near the boundary parallel to the $y$-axis can be written as,
\begin{eqnarray}
&&i\Delta_0 \frac{\partial v}{\partial x}=\mu(x) u\nonumber\\&&i\Delta_0 \frac{\partial u}{\partial x}=-\mu(x) v.
\label{eq:Edge}
\end{eqnarray}
It can be easily checked that this equation, when written in terms of $\Psi(r)$ and using Pauli matrices in the space of ($u, v$), is identical to the 1D Dirac equation in Eq.~\ref{wa} with $\epsilon=0$ and $\mu(x)$ playing the role of the spatially varying Dirac mass $m(x)$. From the Jackiw-Rebbi solution (Eq.~{\ref{zero}}) it then follows that, in the weak pairing phase ($\mu >0$) of Eq.~\ref{eq:Hp}, there is a topologically robust zero energy solution of the BdG equations near the boundary that acts as a domain wall in the chemical potential, and the corresponding wave function is exponentially localized in the direction perpendicular to the boundary. For the strong pairing phase ($\mu<0$) of Eq.~\ref{eq:Hp}, $\mu <0$ both for $x>0$ (vacuum) and $x<0$ (system), and in the absence of a domain wall in the chemical potential no such guaranteed zero energy solution exists at the boundary.

The simplest vortex excitation in a 2D superconductor can be modelled as a point with vanishing $\Delta_0$ (vortex core) and the superconducting phase $\phi$ having a $2\pi$ phase winding around that point. This, in turn, can be modelled as a puncture (hole) in the superconductor, where $\Delta_0$ is automatically zero, and a $2\pi$ phase winding of the order parameter around the hole. Thus, for a vortex excitation in the weak pairing phase ($\mu >0$) of 2D spinless $p_x + ip_y$ superconductor with flux quantum $\frac{hc}{2e}$, we may consider the edge of the vortex core as a circular ring of radius $r=r_0$, separating a region ($r>r_0$) with $\mu >0$ from the region inside the vortex core ($r<r_0$) modelled as vacuum ($\mu <0$). Assuming azimuthal symmetry in the presence of a single vortex situated at the origin, and writing the superconducting order parameter $\Delta(r,\theta)=|\Delta(r)| e^{i \theta}$ and $\mu(r)=\mu_0 h(r)$, the BdG equations near the vortex core can be written in polar coordinates. It can then be shown that \cite{Read2000}, in the weak pairing phase and for a vortex with flux quantum $\frac{hc}{2e}$, the BdG equations near the vortex core admit a zero energy solution analogous to the Jackiw-Rebbi solution near a mass domain wall as in the case of the boundary or the sample edge. 
The zero energy solution near the vortex core exists in the orbital angular momentum channel $l=0$. More generally, the low-energy BdG solutions describing chiral edge modes propagating along the circular edge at $r=r_0$ with energy $E$ can be written as, 
\begin{align}
 \chi_E(r, \theta) &= e^{il\theta}e^{-\int_{r_0}^{r} h(r')dr'}\begin{pmatrix} e^{-i\theta/2}\\e^{i\theta/2} \end{pmatrix}
 \label{eq:Vortex}
    \end{align}
    where $l$ is the orbital angular momentum and $E=\frac{\mu_0l}{r_0}$ \cite{Nayak2008}. If the vortex has an odd number of flux quantum $\frac{hc}{2e}$, the BdG wave function must be anti-periodic upon one full rotation around the vortex core, $\theta \rightarrow \theta + 2\pi$. However, since the spinor on the right hand side of Eq.~\ref{eq:Vortex} is also anti-periodic upon  $\theta \rightarrow \theta + 2\pi$, it follows that $l$ must be an integer (including zero). Thus, a $\frac{hc}{2e}$ vortex (or vortices with flux $\frac{nhc}{2e}$ with $n$ odd) admit a zero energy solution for $l=0$. Conversely, if the flux inside the vortex core is an even multiple of $\frac{hc}{2e}$, the BdG wave function must be periodic upon $\theta \rightarrow \theta + 2\pi$. This ensures that $l$ must be a half-odd-integer and there will be no zero energy mode inside the vortex core even in the weak pairing phase of 2D $p_x+ip_y$ superconductor.  
    Analogous to the zero mode solution near the boundary, the wave function of the zero energy solution near a vortex is exponentially localized away from the vortex core. A number of papers have demonstrated the vortex zero mode in spinless 2D $p_x+ip_y$ superconductor by explicitly solving the BdG equations~\cite{Stone2004,Stone2006,Tewari_Vortex} or mapping the problem on the Jackiw-Rebbi solution of 1D Dirac equation~\cite{Tewari_Index} 
    
    In the strong pairing phase ($\mu <0$), because $\mu$ is negative for both $r>r_0$ and $r<r_0$, no such zero energy solution is expected near the vortex core. The $E=0$ solution near the vortex core in the weak pairing phase of 2D $p_x+ip_y$ supercondctor can be contrasted with the case of the BdG equations near the vortex core of ordinary 2D $s$-wave superconductor which only admit $E\neq 0$ Caroli-de Gennes-Matricon \cite{Caroli} solutions. Furthermore, the spinlessness of the system under discussion guarantees that the second quantized operator corresponding to the non-degenerate $E=0$ solution near vortices and the sample boundary satisfies the Majorana condition $\gamma^{\dagger}=\gamma$.

\subsection{1D spinless p-wave superconductor: Kitaev lattice model}
In 1D spinless $p$-wave superconductor, in the absence of vortex excitations, the zero energy Majorana solutions occur near the boundaries or the ends of the wire. As argued by Kitaev \cite{Kitaev2001}, these zero modes should be observable in a fractional AC Josephson effect--type experiment. 
Although we could discuss the 1D spinless $p$-wave superconductor just as the 1D version of Eq.~\ref{eq:Hp}, it will be instructive to discuss this system using the real space lattice model introduced by Kitaev \cite{Kitaev2001}. The one-dimensional model of topological superconductivity proposed by Kitaev can be written as a tight binding Hamiltonian as follows,
\begin{equation}
H_K=-\sum\limits_{j=1}^{N}\mu c_j^\dagger c_j  -\sum\limits_{j=1}^{N-1}\left(tc_{j}^\dagger c_{j+1} + \Delta e^{i\phi} c_j c_{j+1}+ h.c.\right)\label{eq:KM}
\end{equation}
where $t >0$, $\mu >0$, and $\Delta e^{i\phi}$ with $\Delta >0$ are the nearest neighbor hopping amplitude, chemical potential, and superconducting order parameter, respectively, and $c_j^{\dagger}$ and $c_j$ are second quantized creation and annihilation operators on a 1D lattice with number of sites $N$.
\begin{figure}
\centering
\includegraphics[width=\linewidth]{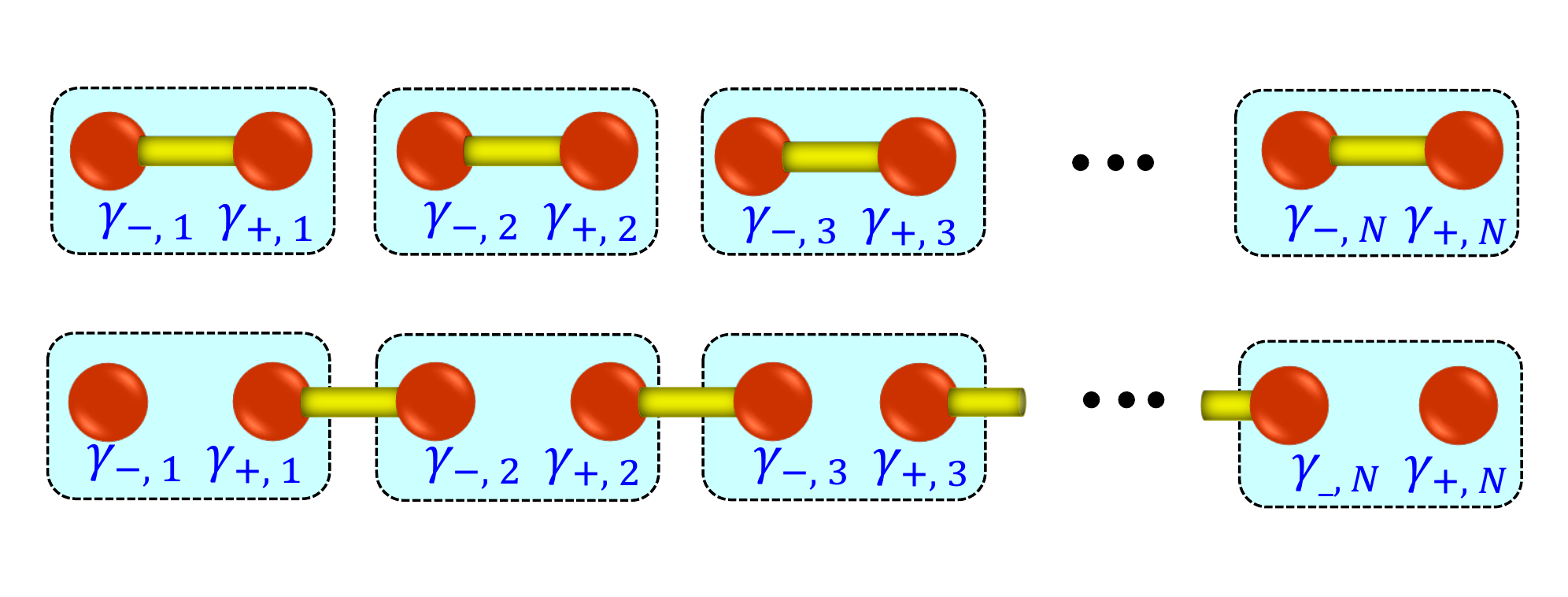}
\caption{Top: Topologically trivial phase of the Kitaev model. The Majorana fermions on the same site are paired and there are no unpaired Majorana zero mode at the ends. Bottom: Topologically non-trivial phase of the Kitaev model. Majorana fermions from nearest neighbor sites are paired, with a pair of unpaired, dangling, Majorana zero modes at the two ends. The system is gapped in the bulk in both topologically trivial and non-trivial phases, but because of the zero energy Majorana fermions at the ends, the ground state of the topologically non-trivial phase is two-fold degenerate.    
 }\label{fig:Kitaev_Model}
\end{figure}

To analyze the Hamiltonian in Eq.~\ref{eq:KM}, let's first consider the special case $\mu=0, t=\Delta$. In this limit, the Hamiltonian can be written as,
\begin{eqnarray}
H_K &=& -t\sum\limits_{j=1}^{N-1}\left(c_{j}^\dagger c_{j+1} +c_{j+1}^\dagger c_{j}+ e^{i\phi} c_j c_{j+1}+e^{-i\phi} c^{\dagger}_{j+1} c^{\dagger}_{j} \right)\nonumber\\
&=& -t\sum\limits_{j=1}^{N-1}\left(e^{i\phi} c_j c_{j+1}-c_{j}c^{\dagger}_{j+1}+c_{j}^\dagger c_{j+1}  -e^{-i\phi} c^{\dagger}_{j} c^{\dagger}_{j+1} \right)
\end{eqnarray}
In analogy with Eq.~\ref{Eq:Overlapping}, let's now introduce a pair of Majorana modes for each conventional fermionic mode described by $c_j, c^{\dagger}_j$,
\begin{eqnarray}
&\gamma_{+,j}&=e^{-i\phi/2}c^{\dagger}_j+ e^{i\phi/2}c_j\nonumber\\
&\gamma_{-,j}&=i(e^{-i\phi/2}c^{\dagger}_j - e^{i\phi/2}c_j)
\label{eq:KM1}
\end{eqnarray}
Despite the phase factors in Eq.~\ref{eq:KM1}, introduced to take into account the superconducting phase in Eq.~\ref{eq:KM}, it should be easy to check that $\gamma_{+,j}$ and $\gamma_{-,j}$ both satisfy the Majorana condition $\gamma_{\pm,j}^{\dagger}=\gamma_{\pm,j}$. In terms of these operators, the Hamiltonian in Eq.~\ref{eq:KM1} can be written as,
\begin{equation}
H_K=-it\sum\limits_{j=1}^{N-1}\gamma_{+,j}\gamma_{-,j+1}
\label{eq:KM2}\end{equation}
Note that the two Majorana operators $\gamma_{-,1}$ and $\gamma_{+,N}$ do not even appear in Eq.~\ref{eq:KM2}, and the rest of the Majorana operators are paired between neighboring lattices sites by the hopping parameter $t$. This situation is pictorially represented in the lower panel of Fig.~\ref{fig:Kitaev_Model}. As a result of nearest neighbor pairing, a pair of MZMs at the two ends remain unpaired. Interestingly, because in this limit ($\mu=0, t=\Delta$), the dangling Majorana operators $\gamma_{-,1}, \gamma_{+,N}$ do not appear in the Hamiltonian, they trivially commute with $H_K$ in Eq.~\ref{eq:KM2}: $[H_K, \gamma_{-,1}]=0=[H_K, \gamma_{+,N}]$. With these Majorana operators we can construct a conventional fermion operator, $q^{\dagger}=(\gamma_{-,1}-i\gamma_{+,N})$, and by virtue of the commutation relations of $H_K$ with $\gamma_{-,1}, \gamma_{+,N}$, we find $[H_K, q^{\dagger}]=0$. It follows that if $|G\rangle$ is the ground state of $H_K$ with energy $E_G$, $q^{\dagger} |G\rangle$ is also an eigenstate of $H_K$ with the same energy $E_G$. Thus, the ground state energy is two-fold degenerate and the pair of degenerate ground states differ in the total fermion number by one, i.e., they have opposite fermion parity. Introducing new conventional fermion operators $d_j=(\gamma_{-,j+1}+i\gamma_{+,j})$ and corresponding $d^{\dagger}_j$, we can rewrite the Hamiltonian $H_K$ as,
\begin{equation}
    H_K=\frac{t}{2}\sum\limits_{j=1}^{N-1} d^{\dagger}_j d_j - (N-1)t
\end{equation}
Therefore, the system has a bulk gap to conventional fermion excitations and a pair of degenerate ground states which differ by the total fermion parity (i.e., they differ by one in the total fermion number). Note that one of the degenerate ground states, $q^{\dagger}|G\rangle= (\gamma_{-,1}-i\gamma_{+,N})|G\rangle$, is associated with fermion occupation of a non-local state composed out of the pair of dangling MZMs at the two ends.

Now let's consider the Hamiltonian in Eq.~\ref{eq:KM} in another simple limit, $\mu<0, t=\Delta=0$. In this case, using Eq.~\ref{eq:KM1}, the Hamiltonian reduces to
\begin{equation}
H_K=-\frac{\mu}{2} \sum\limits_{j=1}^{N}(1+i\gamma_{+,j}\gamma_{-,j})
\label{eq:KM3}
\end{equation}
As shown in Fig.~\ref{fig:Kitaev_Model} top panel, in this case, the Majorana modes $\gamma_{+,j},\gamma_{-,j}$ on the same site $j$ are paired. As a result, there are no longer any dangling MZMs at the two ends.
The system is gapped in this limit as well  since introducing a new spinless
fermion excitation costs a non-zero energy $-\mu >0$. This phase is topologically trivial and there is no degeneracy of the ground state associated with different fermion parity. It is important to emphasize that although we have discussed the properties of the topologically non-trivial and trivial phases of Kitaev model only in simple limits, both these phases are gapped. Because of the spectral gap, the properties of these phases are valid more generally even away from the simple limits as long as the gap in the spectrum remains non-zero. The topological properties of phases can change only at a quantum phase transition where the bulk gap collapses.  

 More generally, applying periodic boundary conditions, and Fourier transforming the Hamiltonian in Eq.~\ref{eq:KM} into momentum space, the Bogoliubov-de Gennes Hamiltonian can be written as (for simplicity we take $\phi=0$ and the lattice constant $a=1$),
\begin{align}
\begin{split}
&H_K=\int dk\Psi^{\dagger}\left(k\right){\cal{H}}_{K}\Psi\left(k\right)\hspace{10mm}\Psi^\dagger (k) =\left(c_k^\dagger,c_{-k}\right)
\\
&{\cal{H}}_{K}=(-2t\cos k-\mu)\tau_z+2\Delta\tau_y \sin k
\end{split}
\label{eq:HamKit}
\end{align}
where $k$ is the momentum and $\tau_z,\tau_y$ are the Pauli matrices operating in the particle-hole space. The bulk band structure for the wire, found by diagonalizing Eq.~\ref{eq:HamKit}, is given by, 
\begin{equation}
  E(k)=\pm \sqrt{(2t\cos k+\mu)^2+4\left|\Delta\right|^2\sin^2 k}, 
  \label{eq:Kitaev_Band}
\end{equation} 
which shows a bulk band gap closure at $k=\{0,\pi\}$ for $\mu=\pm 2t$, representing the topological quantum phase transitions as described in the Kitaev model. The topological superconducting phase with a pair of MZMs at the end occurs for $|\mu|<2t$ while the topologically trivial phase emerges for $|\mu|>2t$. 

We will now describe how to formally characterize the topological phase of superconductors such as the Kitaev model 
using a topological invariant. A topological invariant is a quantity that can change only at TQPT's where the bulk gap closes. For a superconductor, these are points in the parameter space where the energy eigenvalues of the BdG Hamiltonian go through zero. One might think that the determinant of the BdG Hamiltonian may be used as a topological invariant because it is given by the product of the energy eigenvalues. Unfortunately, however, the sign of the determinant of the BdG Hamiltonian does not change when one of the energy eigenvalues go through zero because of the particle-hole symmetry that ensures that the eigenvalues of the BdG Hamiltonian come in $(E,-E)$ pairs. In order to find a suitable quantity that changes sign at each gap closing we define the Pfaffian of the BdG Hamiltonian as explained below.

Any BdG Hamiltonian anti-commutes with a particle-hole symmetry operator, which, for ${\cal{H}}_K$, is given by, $\Lambda=\tau_x K$, 
so that 
\begin{align}
&\Lambda H_{BdG}\Lambda=-H_{BdG}\\\nonumber
&\textrm{or }\tau_x H_{BdG}\tau_x=-H_{BdG}^*=-H_{BdG}^T.\label{eqPHS}
\end{align}
This relation is analogous to the charge-conjugation constraint for Majorana fermions and guarantees that energy eigenvalues of 
the BdG Hamiltonian come in $(\epsilon,-\epsilon)$ pairs.
Thus, the low energy projected BdG Hamiltonian near such a zero-energy level crossing where the energy eigenvalue is  $\epsilon$ 
can be written as
\begin{align}
&H_{BdG}\approx \left(\begin{array}{cc}\epsilon & 0\\0&-\epsilon\end{array}\right),
\end{align}
which is particle-hole symmetric with the operator $\tau_x K$.
Such a crossing of a pair of levels can be characterized using a Pfaffian of the BdG Hamiltonian~\cite{Kitaev2001}, 
which is written as 
\begin{align}
&Pf[\tau_x H_{BdG}]=Pf[\left(\begin{array}{cc}0&\epsilon\\-\epsilon&0\end{array}\right)]=\epsilon,\label{PfHBdG}
\end{align}
and can be seen to change sign when $\epsilon$ crosses zero. For a two-level system, writing the energy eigenvalue $\epsilon$ 
as a Pfaffian appears as a coincidence. To see that this way of writing $\epsilon$ is more than a coincidence, 
let us recall that the Pfaffian of a $2n\times 2n$ anti-symmetric matrix $A$ is defined as
\begin{align}
&Pf[A]=(2^n n\!)^{-1}\sum_\pi \epsilon_{i_1 j_1\dots i_n j_n}\prod_k A_{i_k j_k},
\end{align}
where permutations $\pi$ are constrained so that $i_k<j_k$ and $i_k<i_{k+1}$. The determinant of $H_{BdG}$ will not 
serve this purpose because $Det[H_{BdG}]=-\epsilon^2$ does not change sign as $\epsilon$ crosses zero.
To compute the pfaffian associated with $H_{BdG}$ i.e. Eq.~\ref{PfHBdG} we need to convert $H_{BdG}$ 
into an anti-symmetric Hamiltonian. This can be done using the particle-hole symmetry of $H_{BdG}$ (i.e. Eq.~\ref{eqPHS}),
which  can be re-written as 
\begin{align}
&\tau_x H_{BdG}=-[\tau_x H_{BdG}]^T.
\end{align}
Thus, the matrix $A=\tau_x H_{BdG}$ associated with the BdG Hamiltonian $H_{BdG}$ is anti-symmetric and can be used to compute the Pfaffian.
Similar to the Pfaffian of the projected Hamiltonian Eq.~\ref{PfHBdG}, the Pfaffian of the full Hamiltonian $Pf[H_{BdG}\tau_x]$ 
changes sign at a zero-energy level crossing. 
This suggests that the $\textrm{sgn}[Pf[H_{BdG}\tau_x]]$ as a topological invariant.

For translationally invariant systems, the BdG Hamiltonian is characterized by crystal momentum $k$. Out of these 
only $H_{BdG}(k)$ is particle-hole symmetric only at $k=0,\pi$, which are the only values of $k$ that satisfy $k\equiv -k$ modulo $2\pi$.
While we might be tempted to use $\textrm{sgn}[Pf[H_{BdG}(k=0,\pi)\tau_x]]$ as a two component $Z_2$ topological invariant, one can see that a 
disconnected chain of systems with $\textrm{sgn}[Pf[H_{BdG}\tau_x]]=\pm 1$, which would be trivial.
Such trivial systems would be characterized by $\textrm{sgn}[Pf[H_{BdG}(k=0,\pi)\tau_x]]=\textrm{sgn}[Pf[H_{BdG}\tau_x]]$.
This suggests the topological invariant for superconductors constructed by Kitaev~\cite{Kitaev2001}, 
\begin{align}
&Q=\textrm{sgn}\left[Pf[\tau_x {\cal{H}}_{K}\left(k=0\right)]Pf[\tau_x {\cal{H}}_{K}\left(k=\pi\right)]\right].\label{topinvKitaev}
\end{align}
This topological invariant also provides information about the two dimensional topological superconductors, which are usually 
described by a Chern number topological invariant. Specifically, the topological invariant Eq.~\ref{topinvKitaev} determines 
the parity of the Chern number in two dimensional topological superconductors, which is exactly the condition for obtaining 
odd number of Majorana modes in vortex cores~\cite{Ghosh2010}.

The change in the sign of the Pfaffian between $k=0$ and $k=\pi$ i.e. the topological invariant Eq.~\ref{topinvKitaev} can be 
associated with the existence of end Majorana modes  for open boundary conditions, thus proving a bulk-boundary correspondence.
To see this, let us start by considering a long ring of the system with periodic boundary conditions with an odd number of unit 
cells $L$. The allowed momentum in that case is $k=2\pi n/L$, includes $k=0$ and a set of $(k,-k)$ pairs for $n<(L+1)/2$. 
The pairs $(k,-k)$ can be combined into a particle-hole symmetric Hamiltonian, which in the absence of gap closure at $k\neq 0$ 
is adiabatically connected to $k\simeq 0$. Thus, the Pfaffian for the pair $(k,-k)$ would be positive so that the $\textrm{sgn}[Pf[H_{BdG}(\textrm{periodic})\tau_x]]=\textrm{sgn}[Pf[H_{BdG}(k=0)\tau_x]]$. Similarly for anti-periodic boundary 
conditions $\textrm{sgn}[Pf[H_{BdG}(\textrm{anti-periodic})\tau_x]]=\textrm{sgn}[Pf[H_{BdG}(k=\pi)\tau_x]]$. Thus, 
the topological invariant Eq.~\ref{topinvKitaev} amounts to a change in the Pfaffian of the BdG Hamiltonian in going from 
periodic to anti-periodic boundary conditions. 
Changing the boundary conditions 
from periodic to anti-periodic is equivalent to changing the hopping across the ring from $t$ to $-t$. 
BdG Hamiltonians are however $Z_2-$gauge invariant in the sense 
that one can change the fermion operators $\psi^\dagger\rightarrow -\psi^\dagger$ by a unitary transformation $U=exp(i\pi\psi^\dagger\psi)$, which does not affect 
the spectrum of the Hamiltonian. Applying such a $Z_2-$gauge transformation to a segment of the SC loop, flips the sign of the hopping $t$ at the ends of the segment. 
If the segment is longer than the coherence length of the superconductor, the two ends cannot affect each other. This shows that the spectrum of the junction cannot be changed by the transformation of boundary conditions from $t\rightarrow-t$. Let us now imagine that we change the hopping adiabatically as $\tilde{t}(\lambda)=\lambda t$
where the boundary conditions go from periodic to anti-periodic as the parameter $\lambda$ goes from $1$ to $-1$. Because of the $\tilde{t}\rightarrow-\tilde{t}$ symmetry of the spectrum, any zero-level crossing that occurs at $\lambda$ also occurs at $-\lambda$. Thus, the only way for the Pfaffian of the BdG Hamiltonian to change sign 
is for there to be a level crossing at $\lambda=0$. This means that in the topological phase i.e. satisfying Eq.~\ref{topinvKitaev}, the spectrum has a pair of $E=0$ zero-energy Majorana modes for open boundary conditions (i.e. $\lambda=0$) at each end of the system. 

To compute this topological invariant for the Kitaev Hamiltonian ${\cal{H}}_{K}\left(k\right)$ at $k=0,\pi$ as 
\begin{align}
&\tau_x{\cal{H}}_{K}\left(k=0,\pi\right)=(\pm 2 t-\mu)\tau_x\tau_z=i(\pm 2 t-\mu)\tau_y.
\end{align}
Substituting this into the topological invariant Eq.~\ref{topinvKitaev} becomes,
\begin{align}
&Q_{Kit}=\textrm{sgn}\left[\mu^2-(2t)^2\right],
\end{align}
which produces exactly the criterion $|\mu|>2t$ to obtain topological superconductivity in the Kitaev model.

\subsection{Majorana fermions and Majorana zero modes in one dimensional $p-$wave superconductors}
The Kitaev Hamiltonian $H_{K}$ in the continuum limit can be viewed as a $p-$wave superconductor. 
For small $k$ in Eq.~\ref{eq:Kitaev_Band}, using the approximations $\sin k \sim k$ and $\cos k \sim 1-\frac{k^2}{2}$, the Hamiltonian $H_{K}$ in Eq.~\ref{eq:HamKit} can be approximated as, 
\begin{align}
\begin{split}
&H_K=\int dk\Psi^{\dagger}\left(k\right){\cal{H}}_{K}\Psi\left(k\right)\hspace{10mm}\Psi^\dagger (k) =\left(c_k^\dagger,c_{-k}\right)
\\
&{\cal{H}}_{K}= (tk^2- {\tilde{\mu}})\tau_z - i\tilde{\Delta}\partial_x \tau_y 
\end{split}
\label{eq:small_k}
\end{align}
where $\tilde{\mu}=\mu + 2t$ and $\tilde{\Delta}=2\Delta$.
As shown by the energy eigenvalues in Eq.~\ref{eq:Kitaev_Band}, for small $k$, the topological quantum phase transition is at $\mu = -2t$ or ${\tilde{\mu}}=0$.  The Hamiltonian in Eq.~\ref{eq:small_k} can be mapped on the transverse-field Ising model by Jordan-Wigner transformations~\cite{Pfeuty}. Modifications of Eq.~\ref{eq:small_k} by longer range hopping and pairing terms \cite{Chakravarty,Sen} reveal the existence of multiple topological phases with more than one MZMs at each end protected by chirality symmetry \cite{tewari2012topological}. Exact analytical solutions of Eq.~\ref{eq:small_k} for a finite length wire $L$ reveal exponentially localized MZMs and splitting oscillations of MZM wave functions~\cite{Chuanchang2019}.  In the following, we will see how the bulk excitations of such a superconductor are Majorana fermions, while the 
end excitations can be viewed as Majorana bound states.

Let us  see how the excitations of  $p-$wave superconductor with the Hamiltonian Eq.~\ref{eq:small_k} are Majorana 
fermions~\cite{Read2000} in the sense of being described by the one dimensional version of the Majorana equation Eq.~\ref{eq:ME}.
The Hamiltonian of the $p-$ wave superconductor Eq.~\ref{eq:small_k}, in the small $k$ limit where we drop the $tk^2$ term, can be written in real space as: 
\begin{align}
    &H=\int dx \left[m\psi^\dagger(x)\psi(x)+i \Delta (\psi^\dagger(x)\partial_1 \psi^\dagger(x)-\psi(x)\partial_1\psi(x))\right],\label{Hpx}
\end{align}
where $\psi^\dagger(x)=\sum_k c_k^\dagger e^{i k x}$.
We note that $m$ plays the role of chemical potential ${\tilde{\mu}}$. 
The equation of motion for the fermion operator is written as 
\begin{align}
&\partial_0\psi = i[H,\psi]=-\partial_1\psi^\dagger-m\psi.\label{BdGpwave}
\end{align}
Let us now compare this to the Majorana equation in ($1+1$) dimension, 
which is the generalization of Eq.~\ref{eq:ME} to one spatial dimension and is written as,
\begin{align}
&i [\gamma_0\partial_0+\gamma_1\partial_1]\Psi+m{\cal{C}}\Psi^*=0,\label{ME1D}
\end{align}
Here, the $\gamma$ matrices are chosen to be the ones shown in Eq.~\ref{gammaJR}.
The solutions of this equation satisfy the Majorana constraint $\Psi={\cal{C}}\Psi^*=\sigma_x\Psi^*$, which  is solved automatically by 
writing $\Psi^T=(\psi,\psi^*)$.
Writing the equation by components, similar to Eq.~\ref{eq:MWE}, the above Majorana equation of motion becomes  
an equation for a single fermion component
\begin{align}
&i [\partial_0 \psi+\partial_1 \psi^*]+m\psi=0,\label{MajSC}
\end{align}
which is identical to the operator form of the time-dependent BdG equation for a $p-$wave superconductor in Eq.~\ref{BdGpwave}.
The two component Majorana equation Eq.~\ref{ME1D}, with the Majorana constraint is equivalent to the Dirac equation. Multiplying 
by $\gamma_0$ we can rewrite Eq.~\ref{ME1D} in a form that is identical to the familiar time-dependent Bogoliubov-de Gennes equation corresponding
to the Hamiltonian Eq.~\ref{Hpx}, which is written as
\begin{align}
    &i \partial_0\Psi=-i\sigma^2\partial_1\Psi+m\sigma^3\Psi,\label{tBdG}
\end{align}
where the spinor $\Psi^T=(\psi,\psi^*)$ would be called a Nambu spinor. For stationary states, we can expand $\Psi\propto \Psi_\epsilon e^{-i \epsilon t}$, this reduces to the conventional BdG equation 
\begin{align}
    &-i\sigma^2\partial_1\Psi+m(x)\sigma^3\Psi=\epsilon\Psi,\label{pxBdG}
\end{align}
where the charge conjugation constraint for the Majorana mode Eq.~\ref{Majcons} is now equivalent to the particle-hole symmetry of BdG equations and  maps energy eigenvalues of the above equation from $\epsilon$ to $-\epsilon$.
Thus, the bulk quasiparticles in the $p-$wave superconductor realize a solid state analog of Majorana fermions~\cite{Read2000}, exactly in 
the same way that semiconductors with a massive Dirac dispersion realize Dirac fermions. Despite the BCS Hamiltonian being 
discovered more than fifty years ago, it was only recently pointed out that the Majorana nature of such superconducting quasiparticleas 
can be tested by observing the pair annihilation of pairs of Bogoliubov quasiparticles~\cite{beenakker2014annihilation}.

Let us now discuss zero energy bound states localized in defects of the Majorana mass in Eq.~\ref{ME1D} following the observation that 
Eq.~\ref{pxBdG} is formally identical to the Jackiw-Rebbi Eq.~\ref{wa}. 
Because of the correspondence between the Majorana equation Eq.~\ref{ME1D} and 
the $p-$wave superconductors discussed above, such defects correspond to defects in the $p-$wave superconductor Eq.~\ref{eq:small_k} 
where one goes from ${\tilde{\mu}}>0$ to ${\tilde{\mu}}<0$ as $x$ crosses $0$. Interestingly, zero-energy modes were predicted 
in high-energy physics~\cite{jackiw1981zero}
as bound states in vortices of scalar fields that are coupled to Dirac fermions.
 Since we are looking for bound states without momentum conservation, 
the appropriate Majorana field ansatz generalizing Eq.~\ref{Eq:Majorana_Field} is written as
\begin{align}
    &\Psi(x)=\Phi(x)q_\epsilon e^{i\epsilon t}+q_\epsilon^\dagger e^{-i\epsilon t}{\cal{C}}\Phi^*(x),\label{Psipwave}
\end{align}
where $\Psi(x)$ satisfies the Majorana equation Eq.~\ref{ME1D} or equivalently the time-dependent BdG equation Eq.~\ref{tBdG} if 
$\Phi(x)$ satisfies Eq.~\ref{wa}.
As discussed in subsection~\ref{JackiwRebbi}, Eq.~\ref{wa} has  real  solution at $\epsilon=0$,  $\Phi(x)=\phi_0(x)$, which satisfies Eq.~\ref{sxcond}, which is equivalent to the Majorana condition Eq.~\ref{Majcons} so that $\Phi(x)={\cal{C}}\Phi^*(x)$.
This ensures that $\Psi(x)=[q_{\epsilon=0}+q_{\epsilon=0}^\dagger]\Phi(x)=\gamma\Phi(x)$.
Thus, the operator $\gamma$ associated with the zero-energy state
\begin{equation}
    \gamma=[q_{\epsilon=0}+q_{\epsilon=0}^\dagger]=\gamma^\dagger,
\end{equation}
satisfy the Majorana condition $\gamma^{\dagger}=\gamma$ and $\gamma^2=1$ with proper normalization, which are the conditions that we found 
Majorana bound states in the Kitaev model to satisfy. 
Actually, for a finite length Kitaev wire, the vacuum outside can be modelled as a region of large and negative chemical potential $\tilde{\mu} \ll 0$. When the wire is in the topological phase, $\tilde{\mu}>0$, an edge in the Kitaev wire then acts as a mass domain wall considered above.
Thus, the zero-energy modes associated with the 
one dimensional Majorana equation Eq.~\ref{ME1D}, which is the equation obeyed by the excitations of a $p$-wave superconductor,
is a Majorana bound state that is adiabatically connected to the end Majorana modes of the Kitaev model.

\subsection{Non-Abelian statistics: quantum information processing using Majorana modes}\label{braiding}
As we discussed in the last section a pair of Majorana zero modes, say $\gamma_{1}$ and $\gamma_2$ is associated with a zero-energy fermion 
$q=\gamma_1+i\gamma_2$. The zero-energy of the fermion means that we can describe the quantum state of the pair of Majorana modes 
by two states of the number operator $n=q^\dagger q=0,1$. Alternatively, we will find it convenient to use the conserved fermion parity $F=1-2n=i\gamma_1\gamma_2$ 
instead of the number operator. The fermion parity of freedom 
can be used to store and manipulate quantum information. The basic operation for such quantum information processing is to measure 
the fermion parity of each pair of Majoranas. We will postpone the details of how such a measurement can be accomplished till the 
sub-section~\ref{teleportation}. Below we will discuss, how all topologically protected operations on Majorana systems, such as braiding can be 
accomplished by a sequence of fermion parity measurements through a scheme called measurement only quantum computation~\cite{bonderson2008measurement}.

The fundamental resource for manipulating quantum information stored in Majorana modes is 
non-Abelian statistics. Non-Abelian statistics is a generalization of quantum statistics of fermions and bosons in particle physics, which 
is defined by the transformation of the many-body quantum wave-function under interchange of a pair of particles.
Such an interchange is based on 
transport of the Majorana modes from one position to another~\cite{Alicea2011}. To understand how this transport can be accomplished by
only fermion parity measurement, without physically moving the particles through the system~\cite{vijay2016teleportation}, 
consider three Majorana modes $\gamma_1$, $\gamma_2$ and $\gamma_3$. Let us start in an eigenstate  $\ket{\Psi}$ of the fermion parity of 
$\gamma_2$ and $\gamma_3$ so that $i\gamma_2\gamma_3\ket{\Psi}=\zeta_{23}\ket{\Psi}$,
where $\zeta_{23}=\pm 1$. Such a fermion parity eigenstate may be prepared by measuring the fermion parity 
$i\gamma_2\gamma_3$. Following this 
we measure the fermion parity $i\gamma_1\gamma_2$ and obtain a result $\zeta_{12}$. 
The Majorana operators, in the Heisenberg picture, change following this measurement are now  labelled as 
$\gamma^{'}_1$, 
$\gamma^{'}_2$, 
$\gamma^{'}_3$. 
The measurement projects the wavefunction $\ket{\Psi}$ to $\ket{\Psi'}=2^{1/2} \Pi\ket{\Psi}$, where $\Pi=(1+i\zeta_{12}\gamma_1\gamma_2)/2$ 
is a projection operator into the eigenstate with eigenvalue $\zeta_{12}$ for the fermion parity operator $i\gamma_1\gamma_2$. 
We can check that the state $\ket{\Psi'}$ is normalized by noting that 
\begin{align}
    &\expect{\Psi|(i\gamma_1\gamma_2)|\Psi}\zeta_{23}=\expect{\Psi|(i\gamma_1\gamma_2)(i\gamma_2\gamma_3)|\Psi}=\expect{\Psi|(i\gamma_2\gamma_3)(i\gamma_1\gamma_2)|\Psi},
\end{align}
where comparing the latter two forms
\begin{align}
    &\expect{\Psi|(i\gamma_1\gamma_2)|\Psi}\zeta_{23}=-\expect{\Psi|(\gamma_1\gamma_3)|\Psi}=-\expect{\Psi|(\gamma_3\gamma_1)|\Psi}=\expect{\Psi|(\gamma_1\gamma_3)|\Psi}=0.\label{eqPsigamma12}
\end{align}
The key observation to understand the transformation of Majorana modes is that the total fermion parity  operator $i\gamma_1\gamma_2\gamma_3$ 
for the three Majorana operators involved is left invariant by any fermion parity measurement involving the Majorana bound states $\gamma_1$, 
$\gamma_2$ or $\gamma_3$. To see how this is true let us compute the expectation of the fermion parity $i\gamma_1\gamma_2\gamma_3$, 
together with an operator $\cal{O}$ that is independent of $\gamma_1$, 
$\gamma_2$ or $\gamma_3$, with respect to $\ket{\Psi'}$ as 
\begin{align}
&2^{-1}\expect{\Psi'|i\gamma_1\gamma_2\gamma_3{\cal{O}}|\Psi'}=\expect{\Psi|\Pi(i\gamma_1\gamma_2)\gamma_3{\cal{O}}\Pi|\Psi}=\expect{\Psi|\Pi(i\gamma_1\gamma_2)\gamma_3{\cal{O}}|\Psi},\label{eq59}
\end{align}
where we have used the fact that $[\Pi,(i\gamma_1\gamma_2)\gamma_3{\cal{O}}]=0$ and that $\Pi^2=\Pi$ as expected for a projection operator.
Continuing the above chain of equations 
\begin{align}
    &\expect{\Psi|\Pi(i\gamma_1\gamma_2)\gamma_3{\cal{O}}|\Psi}=2^{-1}[\expect{\Psi|(i\gamma_1\gamma_2)\gamma_3{\cal{O}}|\Psi}-\expect{\Psi|\gamma_3{\cal{O}}|\Psi}]=2^{-1}\expect{\Psi|(i\gamma_1\gamma_2)\gamma_3{\cal{O}}|\Psi}.\label{eq60}
\end{align}
Comparing the first term of Eq.~\ref{eq59} and the last term of Eq.~\ref{eq60}, we see that the operator $i\gamma_1\gamma_2\gamma_3$ is 
indeed preserved as promised. The next step is to observe that $\ket{\Psi}$ is an eigenstate of the fermion parity $i\gamma_2\gamma_3$, 
while $\ket{\Psi'}$ is an eigenstate of $i\gamma_1\gamma_2$. Applying this to the invariance relation we proved we see that
\begin{align}
    &\zeta_{12}\expect{\Psi'|\gamma_3{\cal{O}}|\Psi'}=\expect{\Psi'|i\gamma_1\gamma_2\gamma_3{\cal{O}}|\Psi'}=\expect{\Psi|(i\gamma_1\gamma_2)\gamma_3{\cal{O}}|\Psi}=\zeta_{23}\expect{\Psi|\gamma_1{\cal{O}}|\Psi}.
\end{align}
This shows that  $\gamma_1$ is transferred into to $\zeta_{23}\zeta_{12}\gamma_3$ by the fermion parity measurement of $\gamma_1$ and $\gamma_2$.

As an example of how to perform an exchange using such a transport process consider the exchange of Majorana modes $\gamma_{1,2}$, which will require 
two auxilliary pairs $\gamma_{3,4}$ and $\gamma_{5,6}$ in an eiegnstate with eigenvalues $\zeta_{34}$ and $\zeta_{56}$ respectively. For the first 
step we measure $i\gamma_2\gamma_3$ to obtain an eigenvalue $\zeta_{23}$. This transfers $\gamma_2\rightarrow \zeta_{34}\zeta_{23}\gamma_4$.
For the next step we measure $i\gamma_1\gamma_2$ to obtain an eigenvalue $\zeta_{12}$. This transfers $\gamma_{1}\rightarrow \zeta_{23}\zeta_{12}\gamma_3$.
The result of this pair of measurements is to transfer the pair $(1,2)$ to the pair $(3,4)$. The next steps repeat this by sending $(3,4)$ to $(5,6)$ followed by sending 
the pair $(5,6)$ back to $(2,1)$. The final result of this is the braiding relation
\begin{align}
&\gamma_1\rightarrow \zeta\gamma_2\nonumber\\
&\gamma_2\rightarrow -\zeta\gamma_1,
\end{align}
where $\zeta=\zeta_{34}\zeta_{23}\zeta_{56}\zeta_{45}\zeta_{21}\zeta_{62}$. Noting that the parity operator $(i\gamma_1\gamma_2)^2=1$, we can write the 
above exchange matrix as a unitary operator 
\begin{align}
&U_{12}(\zeta)=e^{-\pi\zeta\gamma_1\gamma_2/4}=\frac{1}{\sqrt{2}}(1-\zeta\gamma_1\gamma_2).
\end{align}
Using the relations $\{\gamma_i,\gamma_j\}=2\delta_{i,j}$, it should be easy to check that 
\begin{align}
&U_{12}(\zeta)\gamma_1 U_{12}^\dagger(\zeta)=\frac{1}{2}(1-\zeta\gamma_1\gamma_2)\gamma_1 (1+\zeta\gamma_1\gamma_2)=\frac{\gamma_1}{2}(1+\zeta\gamma_1\gamma_2)^2=\gamma_1(\zeta\gamma_1\gamma_2)=\zeta\gamma_2.
\end{align}
and similarly $U_{12}(\zeta)\gamma_2 U_{12}^\dagger(\zeta)=-\zeta \gamma_1$.
In the measurement based scheme the value of $\zeta$ for the sign of the exchange would be random but computable from the results of the measurements 
in the various steps. If one is unhappy with the result one can always continue to measure until one has the correct sign of $\zeta$. Ultimately, this is an 
overhead in the computation, which is not greatly significant, though it can be avoided in a deterministic Hamiltonian based approach~\cite{sau2011controlling} as opposed to the measurement-based approach discussed here.

We can convince ourselves of the non-Abelian nature of these operations by computing a pair of exchanges of $1$ and $2$ and then $2$ and $3$ follows by 
the reverse exchanges in the same order
\begin{align}
&U_{12}(\zeta_{12})U_{23}(\zeta_{23})U_{12}(\zeta_{12})^\dagger U_{23}(\zeta_{23})^\dagger = e^{(\pi/4)\zeta_{23}(1-\zeta_{12})\gamma_{2}\gamma_3}.\label{eqreverse}
\end{align}
Note that even though we did each operation in forward and reverse, the combined effect of the exchanges is non-trivial (i.e. not the identity 
operation) if $\zeta_{12}\neq 1$. By multiplying both sides of Eq.~\ref{eqreverse} by 
the $U_{23}(\zeta_{23})U_{12}(\zeta_{12})$ on the right, we see the two exchange operations if produced in different orders produces different outcomes
\begin{align}
&U_{12}(\zeta_{12})U_{23}(\zeta_{23}) = e^{(\pi/4)\zeta_{23}(1-\zeta_{12})\gamma_{2}\gamma_3}U_{23}(\zeta_{23})U_{12}(\zeta_{12})\neq U_{23}(\zeta_{23})U_{12}(\zeta_{12}).
\end{align}
Thus, Majorana modes obey non-Abelian exchange statistics.

The fermion parity degree of freedom associated with a pair of Majorana modes cannot by itself be used as a qubit (i.e. two level system) for quantum information 
processing. There are several ways of constructing two level systems from Majorana modes, which is a process known as encoding~\cite{bravyi2002fermionic}. To understand how qubits can be constructed from Majorana modes, we first note that 
the fermion parity of a superconductor is conserved. Therefore  
 the fermion parity of a system of Majorana modes cannot be changed by any of the allowed operations. One efficient, though not the 
simplest way to avoid this constraint is to use six Majorana modes $\gamma_{j=1,\dots,6}$ to construct two two-level systems described by spin$-1/2$ operators 
$S^{(j=1,2)}_{a=1,2,3}$~\cite{karzig2017scalable}. Even though we start from an eight-dimensional Hilbert space associated with six Majorana modes, the restriction of a total fermion parity 
constraint on exchange operators limited to these qubits limits us to  the four dimensional Hilbert space associated with the two spin$-1/2$.
One choice for the effective spin components we can use is written compactly as 
\begin{align}
&S^{(j)}_a=i\gamma_{3(j-1)+a}\gamma_{3(j-1)+a+1 \textrm{mod}(3)}.
\end{align}
More explicitly, for $j=1$, $S^{1}_{a=1,2,3}=i\gamma_1\gamma_2, i\gamma_2\gamma_3$ and $i\gamma_{3}\gamma_1$ respectively. It is easy to check that 
the two Pauli spin operations defined commute $[S^{(1)}_a,S^{(2)}_b]=0$ as well as obey the usual algebra for spin$-1/2$ i.e. $S^{(j)}_a S^{(j)}_b=i\epsilon_{abc}S^{(j)}_c$.  Measurements of pairs of Majorana modes on the island, that generate the non-Abelian statistics, also are equivalent to measurement of the components 
of the spin matrices. By using $j=2$ as an auxiliary spin, we can generate the Hadamard gap on the spin $j=1$ by measurements or by non-Abelian exchanges.
Together with joint measurements of sets of four or more Majoranas, that can be done via teleportation (more details in the subsection on teleportation), we can 
generate entanglement as well as all quantum operations in the so-called Clifford group~\cite{karzig2017scalable}. These operations by themselves 
are almost complete, in the sense of spanning the entire Hilbert space, and can be made complete   by adding a phase gate~\cite{nayak2008nonabelian}. 
The details of these procedures are beyond the scope of this chapter.

\section{Topological superconductivity in spin-orbit coupled semiconductors}

2D or 1D spinless $p_x+ip_y$ superconductors do not exist in nature. There is some evidence  that strontium ruthenate\cite{Mackenzie2003}  may be a layered quasi-2D $p_x+ip_y$ superconductor, but it is spinful. In this system, a certain type of vortex excitations, called half-quantum vortices (HQV)~\cite{Leggett,Kee2000}, which carry magnetic flux  $\frac{hc}{4e}$ as opposed to the usual superconducting flux quantum $\frac{hc}{2e}$, were proposed to support Majorana zero modes~\cite{Tewari_HQV}. Qualitatively, a HQV can be thought of as an ordinary single quantum vortex (i.e., a vortex carrying flux $\frac{hc}{2e}$), but only in one of the spin sector. Thus, the HQV in a spinful $p_x+ip_y$ superconductor can inherit some of the properties of $\frac{hc}{2e}$ vortices in spinless superconductors, specifically the occurrence of MZM at the vortex core.    
Despite the possibility of MZMs in HQV's in superconducting strontium ruthenate~\cite{Tewari_HQV,Chung,Jang}  and cold fermion systems in the presence of a p--wave Feshbach resonance~\cite{Regal,Ticknor,Schunck}, actually realizing MZMs in these systems are quite challenging. In strontium ruthenate, it is not known for sure if the symmetry of the superconducting order parameter is indeed spin-triplet $p_x+ip_y$. Also, even if the order parameter is of the appropriate symmetry and  the required HQV's can be realized experimentally, the mini--gap $\sim \Delta^2/\epsilon_F \sim 0.1$mK (with $\epsilon_F$ the Fermi energy and $\Delta$ the magnitude of the p--wave order parameter) that separates the MZM from the higher energy regular BdG excitations localized at the vortex cores should be very small. On the other hand, in cold fermion systems with p--wave Feshbach resonance in the unitary limit, even if the mini--gap $\sim \Delta^2/\epsilon_F \sim \epsilon_F$ may be relatively large, the p--wave pairs could be unstable and the short lifetimes of these pairs and molecules represents an important experimental challenge. Finding different and more realistic schemes of realizing MZMs in condensed matter systems is therefore of utmost experimental importance.
\begin{figure}
\centering
\includegraphics[width=\linewidth]{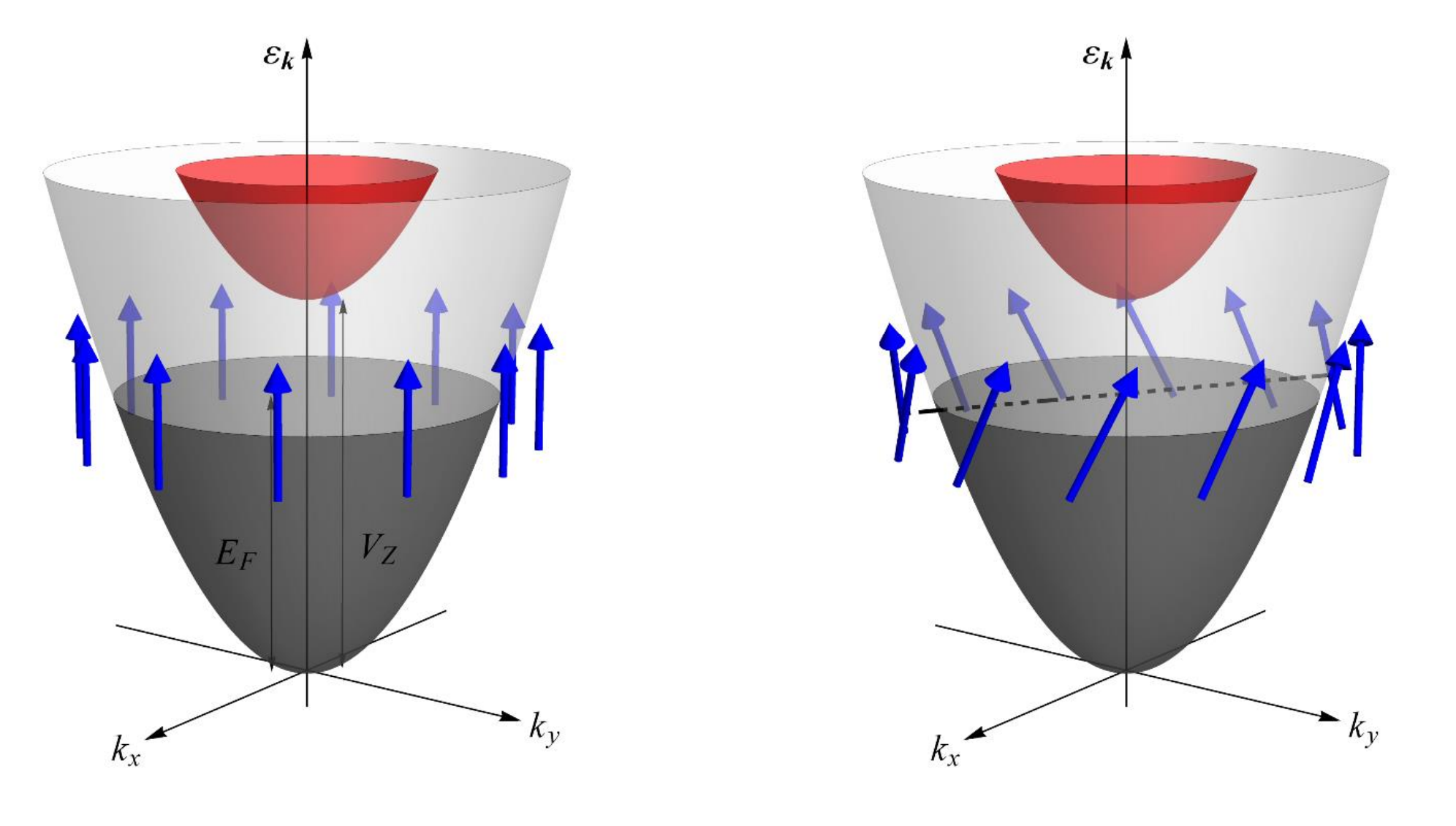}
\caption{Left: Fermi surface in a Zeeman-shifted spin-polarized band. Inducing $s$-wave superconductivity on the Fermi surface is impossible in this case because the spins of the electrons at opposite momenta ($\mathbf{k}$ and $-\mathbf{k}$) are aligned. Right: Fermi surface in a Zeeman-shifted spin-polarized band in the presence of Rashba spin-orbit coupling. Because of the combined effect of Rashba and Zeeman coupling the spins of the electrons make a constant angle with the momentum on the Fermi surface. Spins are no longer exactly aligned at momenta  $\mathbf{k}$ and $-\mathbf{k}$ and inducing $s$-wave superconductivity is now possible. Figure reproduced from Sau et al. arXiv:1012.3170~\cite{sau2011chiral}
 }\label{fig:Spin_Orbit}
\end{figure}

To devise realistic schemes for realizing MZMs, we recall that first and foremost we need an appropriate superconductor. To make the scheme experimentally feasible, an ordinary $s$-wave superconductor is most preferable. The second requirement is the existence of spinless fermions, which obviously do not occur naturally in solid state systems. One possible way out is the application of a Zeeman field to the electrons, $V_z \sigma_z$, where $V_z$ is the strength of the Zeeman field and $\sigma_z$ is a Pauli spin matrix. Since the energy of the electrons in the parallel and anti-parallel direction to the Zeeman field is different, this would lead to a shift of the parallel and anti-parallel energy bands. If now the Fermi energy could be tuned to fall in the lower band (Fig.~\ref{fig:Spin_Orbit}), we would have spin-polarized fermions on the Fermi surface. For the purposes of realizing MZMs, spin-polarized fermions are just as good as spinless fermions because in both cases the definition of the Bogoliubov operator does not involve mixing of different spins. One could now think of inducing $s$-wave superconductivity on the Fermi surface, but this is impossible because the spins of the electrons at opposite momenta $\mathbf{k}$ and $-\mathbf{k}$ on the Fermi surface are exactly aligned. To work around this problem one could think of electrons having a strong Rashba spin-orbit coupling $\alpha (\vec{\sigma} \times \vec{p})\cdot \hat{z}$, where $\alpha$ is the strength of the Rashba coupling. Due to the combined effects of  $\alpha$ and $V_z$, the spins of the electrons make a constant angle with the momenta on the Fermi surface. Since the spins of the electrons at momenta $\mathbf{k}$ and $-\mathbf{k}$ on the Fermi surface are no longer exactly aligned (Fig.~\ref{fig:Spin_Orbit}), $s$-wave superconductivity is now possible. One could now think of inducing $s$-wave superconductivity on the electrons with Rashba spin-orbit coupling and Zeeman field by proximity effect from a bulk ordinary $s$-wave superconductor. Ideas along these lines but without full analytical solutions of the resulting BdG equations and Majorana zero modes were proposed earlier in Refs.~[\onlinecite{Fujimoto2008,Zhang2008,Sato}].

To derive the BdG Hamiltonian for the spin-orbit coupled semiconductor with proximity induced $s$-wave superconductivity, let us start by writing down the non-superconducting 
part of the Hamiltonian i.e. that of the spin-orbit coupled semiconductor in a Zeeman field as 
\begin{equation}
H_N=\int d\mathbf{r}\, \sum_{\sigma,\sigma'}c^{\dagger}_{\sigma}(\mathbf{r})H_{0,\sigma\sigma'} c_{\sigma'}(\mathbf{r}),
\end{equation}
where the normal part of the Hamiltonian density is written as
\begin{equation}
    H_0=[\frac{p^2}{2 m^*}\!-\!\mu\!+\!V_z \sigma_z\!+\!\alpha (\vec \sigma \!\times \! \vec p)\!\cdot\! \hat{z}],
\end{equation}
and $\hat{c}_{\sigma}^{\dagger}(\mathbf{r})$ are the creation operators for electrons with spin $\sigma$. 
The proximity-induced superconducting pairing potential is written as
\begin{equation}
H_{p}=\int d\mathbf{r}\,\{\Delta(\mathbf{r})c^{\dagger}_{\uparrow}(\mathbf{r})c^{\dagger}_{\downarrow}(\mathbf{r})+\rm H.c\},
\end{equation}
where $\Delta(\mathbf{r})$ is the proximity-induced superconducting pair potential. 
The BdG equations describe the Bogoliubov excitation operators of this Hamiltonian which are of the form 
\begin{equation}
\gamma^\dagger=\int d\mathbf{r}\,\sum_{\sigma}u_{\sigma}(\mathbf{r})c_{\sigma}^{\dagger}(\mathbf{r})+v_{\sigma}(\mathbf{r})c_{\sigma}(\mathbf{r})
\label{eq:BCSqp}
\end{equation}
and are defined by the equation 
\begin{equation}\label{eq:qpeqn}
[H_{BdG},\gamma^\dagger]=E\hat{\gamma}^\dagger
\end{equation}
Here, $H_{BdG}$ is defined as $H_{BdG}=H_{N}+ H_{p}$.
By encoding the particle and the hole components of the wave-function $u_{\sigma}(\mathbf{r})$ and $v_{\sigma}(\mathbf{r})$ into a Nambu spinor 
$\Psi(\mathbf{r})=(u_{\uparrow}(\mathbf{r}), u_{\downarrow}(\mathbf{r}),-v_{\downarrow}(\mathbf{r}),v_{\uparrow}(\bm r))$, the above 
operator equation can be written as a BdG equation for the wave-function $\Psi(\mathbf{r})$ as
\begin{equation}
H_{BdG}\Psi(\mathbf{r})=\left(\begin{array}{cc}H_0&\Delta(\mathbf{r})\\\Delta^*(\mathbf{r})&-\sigma_y H_0^* \sigma_y\end{array}\right)\Psi(\mathbf{r})=E\Psi(\mathbf{r}).
\label{eq:H5}
\end{equation}
The spins for the hole components in $\Psi(\mathbf{r})$ are inverted, which leads to the $\sigma_y$ in the lower right corner of the 
Hamiltonian to ensure the manifest spin-rotation symmetry of the singlet superconductivity.
The matrix structure in the particle-hole space of $H_{BdG}$ can be captured by Pauli matrices, which allow one to 
write the BdG Hamiltonian for the system as 
\begin{equation}
H_{BdG}=[\frac{p^2}{2 m^*}\!-\!\mu\!+\!\alpha (\vec \sigma \!\times \! \vec p)\!\cdot\! \hat{z}]\tau_z\!+\!V_z \sigma_z+[\Delta(\bm r)\tau_++h.c],\label{HBdG}
\end{equation}
where $\tau_+=\tau_-^\dagger=\frac{\tau_x+\imath\tau_y}{2}$. 
By applying a Hermitian conjugation to Eq.~\ref{eq:qpeqn} we can see that any solution $\gamma^\dagger$ at energy $E$ of Eq.~\ref{eq:qpeqn} is accompanied by another solution $\gamma$ with energy $-E$. In the spinor language, this corresponds to a new spinor wave-function 
that is related by the particle-hole transformation
\begin{equation}
    \Xi\Psi(\mathbf{r})=\sigma_y\tau_y \Psi^*(\mathbf{r}),
\end{equation}
where $\Xi=\sigma_y\tau_y K$ is defined as the particle-hole symmetry operator with $K$ being complex conjugation.
The existence of spinor solutions that come in $(E,-E)$ pairs is guaranteed by the particle-hole symmetry 
of the BdG Hamiltonian
\begin{equation}
    \Xi H_{BdG}\Xi=\sigma_y\tau_y H_{BdG}^*\sigma_y\tau_y=-H_{BdG}.\label{PHS}
\end{equation}
As we show in the next section, a detailed analytical solution of the BdG equations following from Eq.~\ref{eq:H5} near defects of the pair potential $\Delta$, e.g., vortices and sample edges, reveal the existence of MZMs above a critical value of the Zeeman coupling $V_z$.

\subsection{Vortex bound states in semiconductor-superconductor structures}
In this section, we show how one can derive explicitly the existence of the Majorana bound states in 
a vortex in a spin-orbit coupled semiconductor superconductor heterostructure~\cite{sau2010nonabelian}. Let us start by 
introducing a vortex into $\Delta(\bm r)$ by assuming it to be 
of the form $\Delta(r,\theta)=|\Delta(r)|e^{i n\theta}$, where $n=0,1,\dots$ is the multiplicity of the vortex.
In the absence of a vortex (i.e. $n=0$), the BdG Hamiltonian has a rotation symmetry 
generated by the total angular momentum $J_z=L_z+\sigma_z/2$. The spin-orbit coupling
term proportional to $\alpha$ couples spin and orbital angular momentum so that only the total angular momentum $J_z$ is conserved. Adding a vortex changes this angular momentum operator, because application of rotation $\theta\rightarrow\theta+\varphi$ shifts the superconductor phase $\Delta(r,\theta)\rightarrow \Delta(r,\theta+\varphi)=\Delta(r,\theta)e^{i\varphi}$. This phase is generated by application of the unitary operator $U=e^{i n\tau_z\varphi/2}$ using the relation $U^\dagger \tau_+ U=e^{in\varphi}\tau_+$. We can include this unitary operator in the rotation operator
by modifying the total angular momentum operator to
\begin{equation}
J_{z}=L_z+\frac{1}{2}(\sigma_z-n\tau_z).
\label{eq:Jz}
\end{equation}
With this choice, it is a straightforward calculation to check that $J_z$ is conserved
i.e. $[J_z,H_{BdG}]=0$.
This allows us to assume the vortex solutions to be eigenstates of total angular 
momentum $J_z=m_J$. The angular (i.e. $\theta$) dependence of such eigenstates 
is constrained entirely by $m_J$, so that the wave function can be written entirely 
in terms of a radial spinor:
\begin{equation}
\Psi_{m_J}(r,\theta)=e^{\imath L_z \theta}\Psi_{m_J}(r)=e^{\imath (m_J-\sigma_z/2+n\tau_z/2) \theta}\Psi_{m_J}(r)\label{eq:theta},
\end{equation}
where the radial spinor $\Psi_{m_J}(r)=\left(u_{\uparrow,m_J}(r),u_{\downarrow,m_J}(r)e,v_{\downarrow,m_J}(r),-v_{\uparrow,m_J}(r)\right)^T$.

We are now at a point where we can write the full BdG Hamiltonian for an $n$-fold vortex in polar coordinates as
\begin{equation}
H_{BdG}=(-\eta\nabla^2-\mu)\tau_z + V_z\sigma_z+\imath\frac{\alpha}{2} (\sigma_+p_--\sigma_-p_+)\tau_z+\Delta(r)[\cos{(n\theta)}\tau_x+\sin{(n\theta)}\tau_y].
\end{equation}
where $\eta=\frac{\hbar^2}{2 m^*}$, $\sigma_+=\sigma_-^\dagger=\sigma_x+\imath \sigma_y$
and $p_+=p_x+\imath p_y=e^{\imath\theta}(-\imath\partial_r+\frac{1}{r}\partial_\theta)$ and $p_-=p_x-\imath p_y=e^{-\imath\theta}(-\imath\partial_r-\frac{1}{r}\partial_\theta)$.
With this Hamiltonian the BdG equation can be written as $H_{BdG}\Psi_{m_J}(r,\theta)=E\Psi_{m_J}(r,\theta)$. 
We can now use the angular dependence in Eq.~\ref{eq:theta} to write a purely 
radial (i.e. one dimensional) BdG equation in terms of a radial BdG Hamiltonian
\begin{align}
&\tilde{H}_{BdG,m_J}=e^{-\imath (m_J-\sigma_z/2+n\tau_z/2)\theta}H_{BdG}e^{\imath (m_J-\sigma_z/2+n\tau_z/2)\theta}.
\end{align} 
By substituting the full Hamiltonian $H_{BdG}$ into this equation, we get the radial 
Hamiltonian as,
\begin{align}
&\tilde{H}_{BdG,m_J}=-\{\eta(\partial_r^2+\frac{1}{r}\partial_r+\frac{(2 m_J-\sigma_z+n\tau_z)^2}{4 r^2})+\mu\}\tau_z + V_z\sigma_z\nonumber\\
&-\frac{\imath\alpha}{2} \{\sigma_+-\sigma_-\}\tau_z\partial_r-\imath\frac{\alpha}{2 r} \{\sigma_+\frac{2 m_J+n\tau_z+1}{2}+\sigma_-\frac{2 m_J+n\tau_z-1}{2}\}\tau_z+\Delta(r)\tau_x.
\end{align}
The resulting BdG equations are now much more tractable because they are one dimensional i.e. radial though complex. We can make this Hamiltonian real by performing a $\pi/4$ spin rotation via the unitary transformation $U=e^{i\pi \sigma_z/4}$ so that 
$U\sigma_+ U^\dagger=i\sigma_+$. Following this transformation, the solutions 
of the BdG equation can be assumed to be real without loss of generality.
 However, the BdG equations are still quite challenging because they are 
four component coupled second order differential equations.

We can reduce the complexity of this problem by half by taking advantage of 
the particle-hole symmetry of the zero-energy Majorana mode in the vortex. 
The particle-hole symmetry operator, $\Xi$,
 transforms the $J_z=m_J$ spinor eigenstate with energy $E$
  into a $-m_J$ eigenstate with energy $-E$ because
 $$\Xi e^{\imath (m_J-\sigma_z/2+n\tau_z/2) \theta}\Psi_{m_J}(r)= e^{\imath (-m_J-\sigma_z/2+n\tau_z/2) \theta}\Xi\Psi_{m_J}(r).$$
Since a Majorana mode is particle-hole symmetric, it must be associated with 
the quantum numbers $m_J=0$.
Focusing on this channel of the BdG Hamiltonian $\tilde{H}_{BdG, m_J =0}=\tilde{H}_{BdG}$ and the fact that particle-hole symmetry changes the sign 
of the energy $E$, such a particle-hole symmetry for real Hamiltonians is tantamount to a 
chiral symmetry
\begin{align}
&\sigma_y\tau_y \tilde{H}_{BdG}\sigma_y\tau_y=-\tilde{H}_{BdG}.\label{eqchiralsymmetry}
\end{align} 
It is easy to check that zero-energy solutions $\Psi_{m_J=0}(r)=\Psi(r)$ of such a chiral symmetric 
Hamiltonian would be an eigenstate of the chiral symmetry operator $S=\sigma_y\tau_y$ 
so that 
\begin{align}
&S\Psi(r)=\lambda \Psi(r),
\end{align}
where $\lambda=\pm 1$.
Since each eigenvalue $\lambda$ of $S$ is two-fold degenerate, by choosing 
an eigenvalue $\lambda$ we can write $\Psi(r)$ in terms of two functions $u_\sigma(r)$ 
as 
$\Psi(r)=\sum_\sigma u_\sigma(r)\eta_\sigma$ where $S\eta_\sigma=\lambda\eta_\sigma$ are the two eigenspinors of $S$. Thus, we can use the particle-hole constraint 
to reduce the 4 component BdG equations to two components.

We can see the reduction of the BdG Hamiltonian to two components by replacing 
$\tau_x$ in $\tilde{H}_{BdG}$ by 
$\tau_x=\imath\lambda\sigma_y\tau_z$, which follows from $\sigma_y\tau_y=\lambda$.
Making this substitution, which applies only to the $E=0$ states, 
 the BdG Hamiltonian for a given value of $\lambda$ becomes 
 \begin{align}
&\tilde{H}_{BdG}=-\{\eta(\partial_r^2+\frac{1}{r}\partial_r+\frac{(-\sigma_z+\tau_z)^2}{4 r^2})+\mu\}\tau_z + V_z\sigma_z\nonumber\\
&-\frac{\alpha}{2} \{\sigma_++\sigma_-\}\tau_z\partial_r-\frac{\alpha}{2 r} \{\sigma_+\frac{\tau_z+1}{2}+\sigma_-\frac{\tau_z-1}{2}\}\tau_z+\imath \lambda\sigma_y\tau_z\Delta(r)\label{eq:decBdg}.
\end{align}
While this still appears to be a $4\times 4$ matrix, the transformed Hamiltonian 
commutes with $\tau_z$, so that we can consider each of the 
$\tau_z=\pm 1$ sectors (electron and hole) separately. 
This allows one to write the BdG differential equation in the 
$(\tau_z=+1)$ 
in terms  of the spinor
$\Psi_0(r)=(u_{\uparrow}(r),u_{\downarrow}(r))^T$ for a single vortex $(n=1)$
in the form of a $2\times 2$ matrix differential equation:
\begin{align}\label{eq:zeroenergy}
\!&\!\left(\!\begin{array}{cc}\!\!-\!\eta (\partial_r^2\!+\!\frac{1}{r}\partial_r)\!+\!V_z\!-\!\mu\!&\! \lambda\Delta(r)\!+\!\alpha (\partial_r\!+\!\frac{1}{r} )\\\\ -\lambda \Delta(r)\!-\!\alpha \partial_r  \!&\! -\!\eta (\partial_r^2\!+\!\frac{1}{r}\partial_r\!-\!\frac{1}{r^2}\!)\!-\!V_z\!-\!\mu\! \end{array}\!\right)\!\!\Psi_0(r)\!=\!0.
\end{align}

To make progress towards an analytic solution, we approximate the radial dependence 
of $\Delta(r)$ by $\Delta(r)=0$ for $r<R$ and $\Delta(r)=\Delta$ for $r\geq R$. 
Let us start by considering the range $(r<R)$, which is the non-superconducting region $(\Delta(r)=0)$. In this range, the wave function is simply that of a spin-orbit 
coupled semiconductor. In the absence of spin-orbit coupling, each spin-component 
has solutions given by Bessel functions $J_0(z r)$ and $J_1(z r)$. 
This suggests that we can include spin-orbit coupling by trying a spinor 
of the form  
\begin{equation}
\Psi(r)=  \left(\begin{array}{c}u_{\uparrow}J_0(z r)\\u_{\downarrow}J_1(z r)\end{array}\right).
\label{eq:Bessel}
\end{equation}
We can determine $(u_\uparrow,u_\downarrow)$ and $z$  by substituting Eq.~(\ref{eq:Bessel}) into Eq.~(\ref{eq:zeroenergy}), which then takes the form: 
\begin{align}
&\left(\begin{array}{cc}\eta  (-\partial_r^2-\frac{1}{r}\partial_r)+V_z-\mu & \alpha  (\partial_r+\frac{1}{r} )\\ -\alpha  (\partial_r)  & \eta (-\partial_r^2-\frac{1}{r}\partial_r+\frac{1}{r^2})-V_z-\mu \end{array}\right)\left(\begin{array}{c} u_{\uparrow} J_0 (z r )\\ u_{\downarrow} J_1(z r)\end{array}\right)\nonumber\\
&=\left(\begin{array}{c}(-\eta z^2+V_z-\mu) u_{\uparrow} J_0 (z r )+  z \alpha u_{\downarrow} J_0(z r) \\ z\alpha u_{\uparrow} J_1(z r)+(-\eta z^2 -V_z-\mu)u_{\downarrow} J_1(z r)\end{array}\right)=0.
\end{align}
This condition simplifies to 
\begin{equation}
\left(\begin{array}{cc}-\eta z^2+V_z-\mu & z\alpha  \\ \alpha z  & -\eta z^2 -V_z-\mu \end{array}\right)\left(\begin{array}{c} u_{\uparrow} \\ u_{\downarrow}\end{array}\right)=0.
\label{eq:matrix}
\end{equation}
The value of $z$ can be determined by setting the determinant of the above matrix to 
zero. This leads to the equation for $z$
\begin{equation}
(\eta z^2-\mu)^2-V_z^2-z^2\alpha^2=0.\label{char:eq0}
\end{equation}
Note that the solutions of the above equation come in pairs $\pm z$. However, 
the Bessel functions $J_0(zr)$ and $J_1(zr)$ are odd and even functions of 
$z$ respectively. Therefore, there are two linearly independent solutions that one 
can obtain for $r<R$.

The $r>R$ region includes non-zero amplitude of superconductivity. This region
is complicated to solve analytically except by a power-series solution in $1/r$. However,
our focus of whether there is a normalizable solution in this region depends on 
the conditions for which there are exponentially decaying solutions as $r\rightarrow\infty$. Inspired by the asymptotic form of Bessel functions $J_n(zr)\propto e^{-zr}/r^{1/2}$, we consider an ansatz 
\begin{align}
&\Psi_0(r)=\frac{e^{-z r}}{r^{1/2}}\left(\begin{array}{c}\rho_{\uparrow}\\\rho_{\downarrow}\end{array}\right).\label{Psi0}
\end{align}
Substituting this ansatz into the BdG equation we get 
\begin{align}
&\left(\begin{array}{cc}\eta  (-\partial_r^2-\frac{1}{4 r^2}+2 z\partial_r-z^2) +V_z-\mu & \lambda\Delta+\alpha  (\partial_r+\frac{1}{2 r} -z)\\ -\lambda\Delta-\alpha  (\partial_r-\frac{1}{2 r}-z)  & \eta  (-\partial_r^2+\frac{3}{4 r^2}+2 z\partial_r-z^2)-V_z-\mu \end{array}\right)\left(\begin{array}{c}\rho_{\uparrow}\\\rho_{\downarrow}\end{array}\right)=0.
\label{eq:64}
\end{align}
We notice that the $r$ dependence disappears in the limit $r\rightarrow \infty$ so that  
\begin{align}
&\left(\begin{array}{cc}-\eta z^2 +V_z-\mu & \lambda\Delta-z\alpha  \\ -\lambda \Delta+z\alpha    & -\eta z^2-V_z-\mu \end{array}\right)\left(\begin{array}{c}\rho_{\uparrow}\\\rho_{\downarrow}\end{array}\right)=0\label{eq:largeRmode_vortex}.
\end{align}
Similar to the case for $r<R$, this equation has a non-trivial solution if $z$ satisfies 
the secular equation:
\begin{align}
&Det\left(\begin{array}{cc}-\eta z^2 +V_z-\mu & \lambda\Delta-z\alpha  \\ -\lambda\Delta+z\alpha    & -\eta z^2-V_z-\mu \end{array}\right)
\nonumber\\
&=(-\eta z^2-\mu)^2-V_z^2+(z\alpha\lambda-\Delta)^2=0\label{eq:char_eq}.
\end{align}
We see from the form of the equation that the solutions of $z$ for the two values of 
$\lambda$ are related by the sign of $\lambda$.

We can determine zero-energy Majorana solutions to the BdG equation 
associated with the vortex by matching boundary conditions for the spinor $\Psi_0(r)$ between solutions 
for $r<R$ and $r>R$ at $r=R$. The boundary conditions for the continuity of the two component 
spinor $\Psi_0(r)$ and its derivative $\Psi_0'(r)$ at $r=R$ constitute four linear 
equations. There would be non-trivial solutions to these equations if there are 
five or more solutions between $r<R$ and $r>R$ out of which to construct 
solutions $\Psi_0(r)$. Following the analysis of Eq.~\ref{char:eq0}, we concluded 
that there were always two linearly independent solutions to use for constructing 
$\Psi_0(r<R)$. Therefore, we can obtain normalizable Majorana modes if there 
are at least three solutions of $z$ with $Re(z)>0$ to use to construct $\Psi_0(r>R)$.

The existence of normalizable zero-energy Majorana solutions depends crucially on the 
number of available roots with $Re(z)>0$ of the above characteristic equation, which has 
four roots in all. Since the
characteristic equation is real, solutions $z$ appear in complex conjugate pairs $(z,z^*)$
or are real.
Furthermore, since the coefficient of $z^3$ in Eq.~\ref{eq:char_eq} vanishes, the sum 
of the four roots must vanish. This implies that at least one of the roots must have a negative real part (i.e. $Re(z)<0$) and another must have $Re(z)>0$. One possibility is that both these roots are complex, in which case we get only two roots with $Re(z)>0$ and 
no Majorana mode. Let us call this case A. If one of these roots is real then it must be accompanied by another real root. Further, if these two real roots have the same sign we are in a similar situation as before and there are no Majorana modes. Let us call this case B. Finally, if these two real roots happen to have opposite signs then three roots 
will have real parts with the same sign. By flipping the sign of $\lambda$ if necessary, we can ensure these three roots satisfy $Re(z)>0$ and there will be a zero-energy Majorana 
mode. We call this case C. 

We can separate the interesting topological case C from the cases A and B by considering 
the sign of the product $C_0=\prod_n z_n=(\mu^2+\Delta^2-V_Z^2)$ of 
the roots $z_n$, which is the polynomial at $z=0$.
  Note that complex conjugate pairs $(z,z^*)$ 
do not change the sign of this product $C_0$ since $z z^*=|z|^2>0$. 
The case B has pairs of real roots with the same sign. These also do not contribute 
to the sign of $C_0$. On the other hand, the topological case C is characterized by 
pairs of real roots with opposite signs, so that the product $C_0$ would be negative.  
This leads to the condition to realize topological zero-energy Majorana modes 
\begin{align}
&C_0=(\mu^2+\Delta^2-V_Z^2)<0.\label{C0}
\end{align}
Therefore Majorana modes are realized only for Zeeman field in excess of a 
critical value $V_Z>\sqrt{\Delta^2+\mu^2}$.
While this gives us the condition for a Majorana mode, we still have not written 
an explicit form of the solution, although there are many constraints that are 
derived. We have almost exact solutions for $r<R$, except for certain coefficients 
that can be written in terms the matrix equations Eq.~\ref{eq:matrix}. However, we 
considered an asymptotic $r\rightarrow\infty$ limit for $r>R$. This has been~\cite{sau2010nonabelian} extended 
to a power-series solution in $1/r$, which is systematic, but does not directly 
impact the qualitative structure of the solution.

\subsection{Domain wall states in the topological phase}
Apart from vortex Majorana states, topological superconductors are characterized 
by interesting states associated with edges of the system that are set by decreasing the chemical potential $\mu(r)$ or changes in the phase of the superconductor $\Delta(r)$.
Similar to the case of vortices, we can understand the spectra of such edges by 
reducing the BdG equation from Eq.~\ref{HBdG} to one dimension by assuming parameters 
to vary only along the $x-$direction. In this case we can assume the BdG spinor 
to be a plane wave with wave vector $k_y$ along the y direction so that the 
BdG Hamiltonian is 
\begin{equation}
H_{BdG}=(-\eta \partial_x^2-\mu(x)+\imath\alpha\sigma_y\partial_x-\alpha k_y\sigma_x)\tau_z + V_Z B\sigma_z+\Delta(x)\tau_x.\label{edgeH}
\end{equation}
Similar to the angular momentum quantum number $m_J$ in Eq.~\ref{eq:theta}, $k_y$ 
transforms to $-k_y$ under particle-hole symmetry $\Xi$ because of the complex 
conjugation operator $K$. As a result $H_{BdG}$ is still particle-hole symmetric and 
the zero-energy Majorana operators appear at $k_y=0$.
The BdG Hamiltonian at $k_x=0$, is particle-hole symmetric and real so it is chiral 
symmetric with the chiral symmetry operator $S$. As before, we can then assume the zero 
energy Majorana mode is an eigenstate of $S$, so that we can replace $\tau_x$ with $\sigma_y\tau_z$.
This reduces the BdG Hamiltonian to a two component form similar to Eq.~\ref{eq:decBdg} as 
 \begin{align}
&\tilde{H}_{BdG}=-\{\eta\partial_x^2+\mu(x)-\imath\frac{\alpha}{2} \sigma_y\partial_x\}\tau_z + V_z\sigma_z+\imath \lambda\sigma_y\tau_z\Delta(x).\label{eqdomain}
\end{align}
The corresponding BdG equation, similar to the case of vortices (i.e. Eq.~\ref{eq:zeroenergy}) is a $2\times 2$ 
coupled differential equation as well: 
\begin{align}
\!&\!\left(\!\begin{array}{cc}\!\!-\!\eta (\partial_x^2\!)\!+\!V_z\!-\!\mu(x)\!&\! \lambda\Delta(x)\!+\!\alpha (\partial_x\!)\\\\ -\lambda \Delta(x)\!-\!\alpha \partial_x  \!&\! -\!\eta (\partial_x^2\!\!)\!-\!V_z\!-\!\mu(x)\! \end{array}\!\right)\!\!\Psi_0(x)\!=\!0.
\end{align}
As in the case of vortices we will consider the parameters $\mu$ and $\Delta$ to be constants at different values across a domain wall at $x=0$. On each side we can expand $\Psi_0(x)$ in terms of plane-waves $\Psi_0(x)\propto e^{zx}\Psi_0$ so that the above equation becomes 
\begin{align}
\!&\!\left(\!\begin{array}{cc}\!\!-\!\eta (z^2\!)\!+\!V_z\!-\!\mu\!&\! \lambda\Delta\!+\!\alpha z\\\\ -\lambda \Delta\!-\!\alpha z  \!&\! -\!\eta (z^2\!\!)\!-\!V_z\!-\!\mu\! \end{array}\!\right)\!\!\Psi_0\!=\!0.\label{eq:Psi0edge}
\end{align}
This equation is identical to Eq.~\ref{eq:largeRmode_vortex} and therefore has three solutions with $Re(z)$ with the same sign only in the case of a topological bulk with $C_0<0$. The properties of domain-walls with various 
boundary conditions depends on the details of the boundary conditions discussed in the two sub-sections.
 
\subsubsection{Edge boundary conditions}
Based on analogy with FQHE and chiral $p$-wave superconductors, one
 expects a chiral gap-less state confined to the edge of the
 semiconductor heterostructure.
We model an edge with a chemical potential $\mu$ towards the edge
 increases from  $\mu(x)=\mu$ for $x<0$
 to $\mu(x)=\infty>|V_z|$ for $x>0$.
We will assume that $\Delta(x)=\Delta$ is independent of $x$.
The wave-function $\Psi(x)$ must then vanish at $x=0$. This leads to the boundary condition 
\begin{align}
&\Psi(x=0)=\sum_n \Psi_n e^{z_n x}=0,
\end{align}
where $\Psi_n$ and $z_n$ satisfy Eq.~\ref{eq:Psi0edge}. The resulting wave-function $\Psi_n(x)$ with be normalizable 
if $Re(z_n)>0$. The boundary condition for the two component spinor $\Psi(x=0)$
 contains two constraints. These can be solved for three values $n$. As we found in the case 
of the vortex, Eq.~\ref{eq:Psi0edge} has three solutions with $Re(z_n)>0$ only if $C_0<0$, which 
was exactly the topological condition that leads to Majorana zero modes in the vortex.

Let us now discuss the solution of the BdG equation associated with the Hamiltonian in Eq.~\ref{edgeH} 
away from $k_y=0$. This Hamiltonian has one zero-energy Majorana state at $k_y=0$. Using perturbation theory, 
we can calculate the correction to the energy at finite $k_y$ as 
\begin{align}
&E(k_y)\sim \alpha k_y \int dx \Psi_0(x)^\dagger \sigma_x\tau_z \Psi_0(x),
\end{align} 
which is linear in $k_y$ at this order. This shows that the dispersion of the Majorana zero mode at $k_y=0$ of 
the topological superconductor for $C_0<0$ is a chiral Majorana with a linear dispersion relation.

\subsubsection{Non-chiral Majorana models in Josephson junction $\pi$ junctions}\label{nonchiralJJ}
Let us now consider another kind of boundary i.e. a Josephson junction with a phase difference of 
$\pi$. In this case, $\mu$ remains the same, while $\Delta(x)$ changes sign at
 $x=0$ from $\Delta(x)=\Delta$ for $x<0$ to $\Delta(x)=-\Delta$ for $x>0$.
Such a $\pi$-phase shift Josephson junction, with $\mu$ being constant is described 
by the  
real BdG Hamiltonian  Eq.~\ref{eqdomain}. In contrast to the case of the system edge, 
$\mu(x)$ is now assumed to be constant with $\Delta(x)$ changing sign at $x=0$.

Similar to the case of the vortex and the edge, we can solve the BdG equation by 
matching plane-wave wave-functions for $x<0$ and $x>0$ at $x=0$. The boundary conditions in this case
consists of matching a two component spinor wave-function $\Psi_0(x=0)$ 
and it's derivative $\Psi_0'(x=0)$ and thus contains four constraints (similar to the 
vortex case). Wave-functions for $x<0$ are identical to those for the edge of the system, 
which consists of three allowed solutions for $C_0<0$ i.e. the topological phase. The 
sign of $\Delta(x)$ flips as we cross the domain wall to $x>0$. 
Using Eq.~\ref{eq:char_eq}, we note that this change of sign of $\Delta$ also changes 
the sign of $z$ in the same $\lambda$ sector. Thus, in the topological regime, $C_0<0$,
we obtain three solutions with $Re(z_n)<0$. These values correspond to three 
normalizable plane-wave solutions for $x>0$. Combining the solutions for $x<0$ and 
$x>0$, in the topological regime $C_0<0$, we have six plane-wave solutions to be used 
to   match four constraints. This leads us to a pair of zero energy Majorana modes 
for the $\pi$ Josephson junction at $k_y=0$. 

We can move away from $k_y=0$ by computing the matrix elements of the $k_y$ 
perturbation proportional to $\sigma_x\tau_z$ in Eq.~\ref{edgeH}, similar to the case of the chiral edge states. However, in this case, conjugating the BdG Hamiltonian with $\sigma_z\tau_z$ together with the application of a mirror symmetry $x\rightarrow -x$ has the effect 
of flipping $k_y$ without any other change of the system. Therefore, the dispersion 
of the system resulting from the perturbation of adding $k_y$, in contrast 
to that of the chiral edge states, must be symmetric $E(k_y)=E(-k_y)$. This dispersion is 
what is called a non-chiral Majorana mode. The two fold degeneracy of the zero energy Majorana modes at $k_y=0$ is broken by going away from the $\pi$ phase difference, similar to those in topological insulators~\cite{fu2008superconducting}.  The resulting dispersion is 
that of a massive Dirac/Majorana mode:
\begin{align}
&E=\pm\sqrt{v^2 k_y^2+b^2 (\phi-\pi)^2},\label{eqEJ}
\end{align}  
where $v,b$ are constants that would be determined from perturbation theory.

\subsection{Relation to bulk phase transition and topological invariant}
The calculation so far showed us that vortices and edges in the semiconductor-superconductor 
heterostructure support Majorana zero modes only when $C_0<0$. Ultimately, this 
condition was obtained from combining the boundary conditions for a vortex together 
with the equation for the evanescent plane wave wave vector $z$ written in Eq.~\ref{eq:char_eq}.
\begin{figure}
\centering
\includegraphics[width=\linewidth]{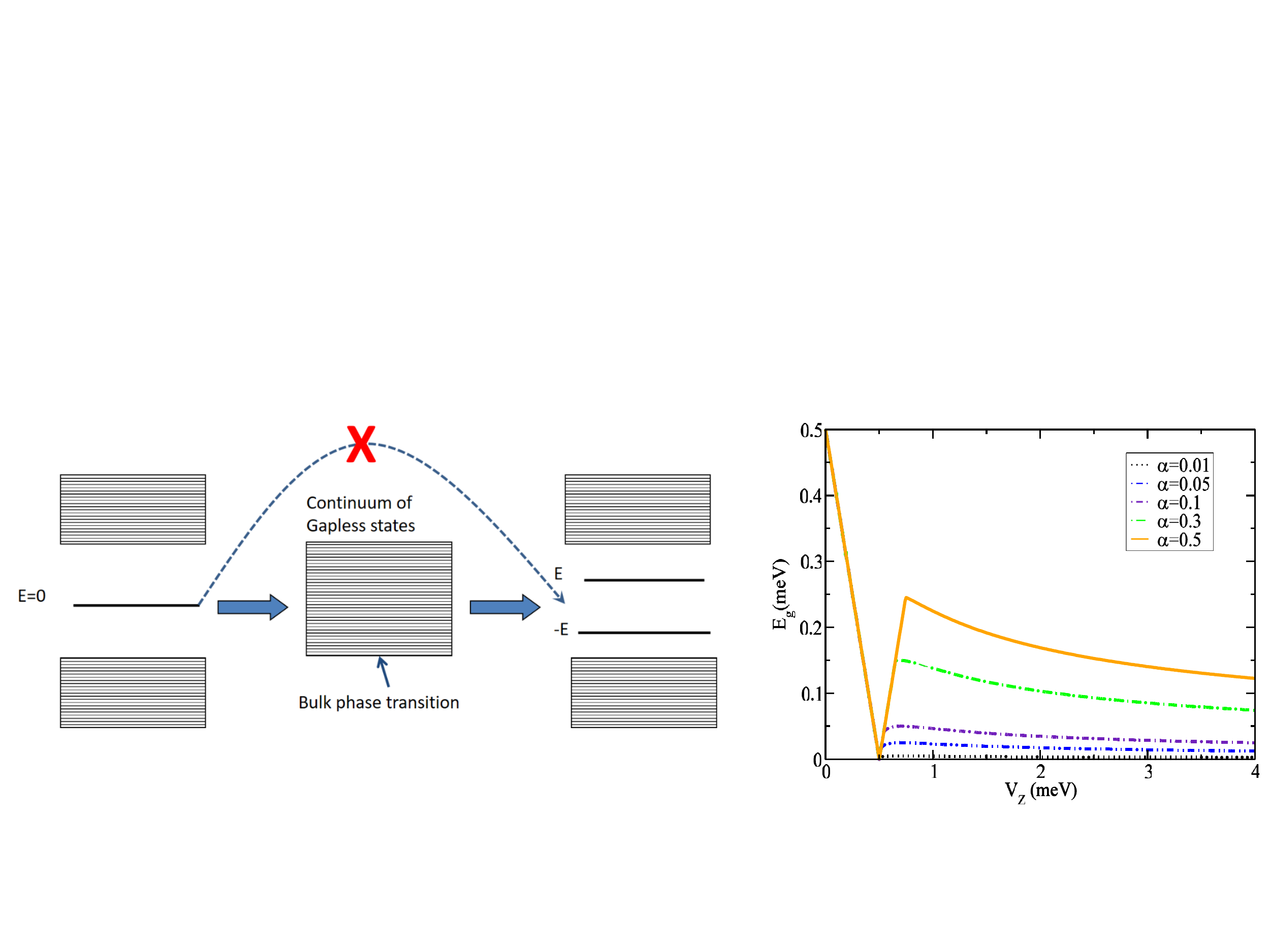}
\caption{Left panel shows the effect of a perturbation on a non-degenerate zero-energy sub-gap Majorana mode. 
Particle-hole symmetry requires energy levels with non-zero energy to come in $(E,-E)$ pairs. Thus, a perturbation 
cannot shift a non-degenerate zero-energy mode since it will not have a partner similar to states in trivial sub-gap states 
shown on the right. These two scenarios must be separated by a bulk phase transition where the gap closure eliminates 
any sub-gap states.  The right panel shows the bulk spectrum of a spin-orbit coupled semiconductor/superconductor (Eq.~\ref{Ek})
as a function of applied Zeeman field $V_Z$. The applied Zeeman field $V_Z$ first closes the bulk 
quasiparticle gap before it reopens as a topological superconducting gap proportional to the strength of spin-orbit coupling $\alpha$.
Right panel reproduced from Sau et. al. arXiv:1006.2829~\cite{sau2010nonabelian}
 }\label{fig:gapclosure}
\end{figure} 
The wavevector $z$ is a complex number representing the $e^{-zr}$ position dependence 
of the wave function in Eq.~\ref{Psi0}. We used this form to conclude that only solutions 
with $Re(z)>0$ of Eq.~\ref{eq:char_eq} could be used to construct normalizable solutions.
We found Majorana solutions only when three such solutions existed of Eq.~\ref{eq:char_eq}. While this feels like a fine-tuned condition, any adiabatic change of the
parameters that changes this condition will necessarily pass through a point in the parameter
space where $Re(z)=0$. Since $z$ is purely imaginary, we can write $z=i k$ for such parameters. Such purely imaginary values of $z$ correspond to propagating plane-wave states. Making this substitution in the characteristic equation one gets,
\begin{align}
&(\eta k^2-\mu)^2-V_z^2+(ik\alpha\lambda-\Delta)^2=0.\label{eqtop}
\end{align}
This equation contains an imaginary part $ik\alpha\lambda\Delta$, which automatically 
forces $k$ to vanish.
The bulk spectrum of the Hamiltonian is 
\begin{equation}
E_k^2=V_z^2+\Delta^2+\tilde{\epsilon}^2+\alpha^2 k^2\pm 2 \sqrt{V_z^2\Delta^2+\tilde{\epsilon}^2(V_z^2+\alpha^2 k^2)}\label{Ek}
\end{equation}
where $\tilde{\epsilon}=\eta k^2-\mu$. 
To associate this relation with bulk properties let us write down the bulk Hamiltonian 
corresponding to Eq.~\ref{HBdG} at the point $k=0$ in the momentum space, 
\begin{align}
&H_{BdG}(k=0)=-\mu\tau_z+V_z \sigma_z\!+\Delta\tau_x.
\end{align}
The spectrum of this Hamiltonian is 
\begin{align}
&E=\pm V_Z\pm\sqrt{\Delta^2+\mu^2}.
\end{align}
These energies vanish precisely when Eq.~\ref{eqtop} is satisfied. Therefore,
the parameters where we go from having zero energy vortex modes in a topological superconductor to the trivial phase of the superconductor 
corresponds to a phase transition where the bulk energy gap of the BdG 
Hamiltonian in Eq.~\ref{HBdG} closes at $k=0$.
In fact, this closure of the gap can be understood in terms of an application of Kitaev's 
topological invariant in Eq.~\ref{topinvKitaev} to the spinful superconductor, which is written as 
\begin{align}
&Q=\textrm{sgn}[Pf[\sigma_y\tau_y H_{BdG}(k=0)]],\label{topinv}
\end{align}
in particle-hole symmetric systems where ``$Pf$'' stands for the Pfaffian. One key difference is 
that because of the different Nambu basis we use for spinful superconductors, the particle-hole matrix 
in the Kitaev Hamiltonian $\tau_x$ is replaced by $\tau_x\rightarrow\sigma_y\tau_y$. In addition, 
for the continuum system used to model systems such as the superconducting nanowire, the $k=\pi$ 
term in Eq.~\ref{topinvKitaev} is trivial and may be dropped.
  By computing this Pfaffian for the $4\times 4$ 
matrix using the 
standard definition~\cite{parameswaran1954skew} we get
\begin{align}
&Q=\mu^2+\Delta^2-V_Z^2,
\end{align}
which is exactly the $C_0$ topological number (see Eq.~\ref{C0}) we obtained from studying vortices 
and edges.

\subsection{Dimensional reduction to one dimensional Nanowires}\label{nanowire}
Majorana modes can also be realized in the one dimensional analog of the 
semiconductor-superconductor structure. This structure is experimentally simpler 
because one does not require a magnetic insulator to generate the Zeeman field. 
In the one dimensional case, the Zeeman field can be generated by a magnetic 
field parallel to the semiconductor superconductor wire. In this case, the 
Zeeman field cannot generate substantial undesired orbital effects that would 
suppress superconductivity, though a finite diameter of the wire can leave 
residual effects~\cite{vaitiekenas2020flux,nijholt2016orbital}. 

\begin{figure}
\centering
\includegraphics[width=0.5\linewidth]{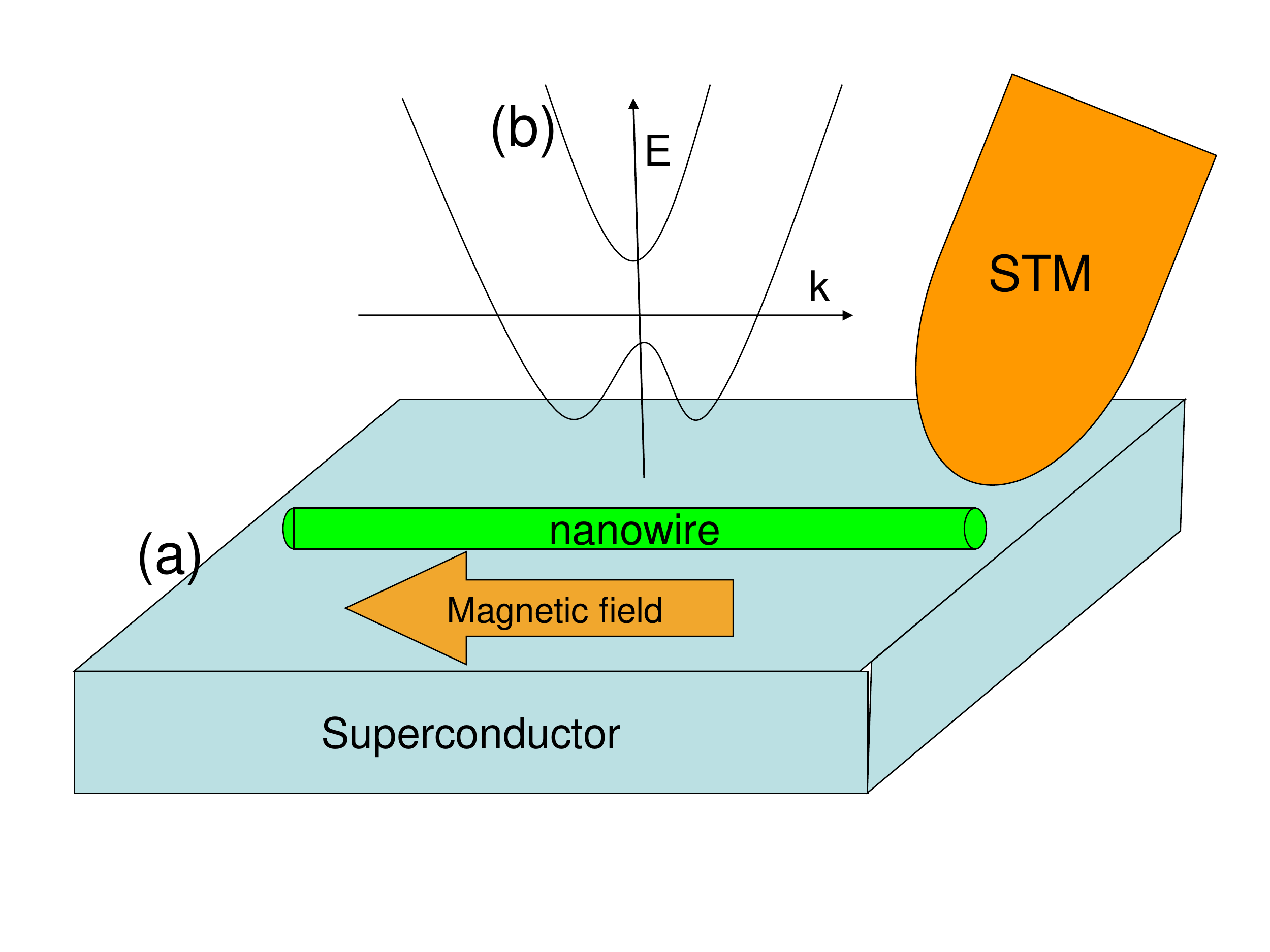}
\caption{ The one dimensional semiconductor 
superconductor heterostructure allows the application of a Zeeman field by a parallel magnetic field.
Such a magnetic field does not suppress superconductivity significantly for a thin-film superconductor. 
Theory (Eq.~\ref{C0}) predicts that the system enters the topological phase for applied Zeeman 
field $V_Z$ in excess of $\sqrt{\mu^2+\Delta^2}$. The zero-energy end Majorana mode that emerges 
at the end of the wire can be measured by tunnel spectroscopy e.g. using scanning tunneling microscopy (STM)~\cite{sau2010nonabelian}. 
Figure reproduced from Sau et. al. arXiv:1006.2829~\cite{sau2010nonabelian}.
 }\label{fig:nanowire}
\end{figure} 

Understanding the one dimensional nanowire does not require any new 
calculations. The BdG Hamiltonian Eq.~\ref{eqdomain} that we used to study the 
spectrum of domain wall states at $k_y=0$ is exactly the Hamiltonian 
for the semiconductor-superconductor nanowire heterostructure. The end of the wire corresponds 
precisely to the edge of the system. The argument from the previous sub-section 
then tells us that a semiconductor nanowire will support topological Majorana modes 
for $C_0<0$. Additionally, the results from the sub-section on Josephson junctions 
implies that a $\pi$-phase Josephson junction will support a pair of such Majorana 
modes. We will discuss signatures of both the end Majorana modes as well as pair 
of Majorana modes in the following section.

One subtlety worth noting is that of the chiral symmetry  (i.e. Eq.~\ref{eqchiralsymmetry}) of the BdG Hamiltonian for the superconducting
semiconductor. This is because the exact form of the topological invariant depends on the symmetry class of the system. 
For example the Kitaev topological invariant Eq.~\ref{topinvKitaev} is only defined for one dimensional superconductors, which have 
particle-hole symmetry by definition and are classified as symmetry class D~\cite{altland1997nonstandard}. This invariant makes no assumption about the 
Hamiltonian being real, which happens to be the case for the specific superconducting semiconductor system being considered. 
In this case the reality of the BdG Hamiltonian leads to a chiral symmetry  Eq.~\ref{eqchiralsymmetry}, which allows 
us to define a more refined topological invariant~\cite{tewari2012topological} in this new symmetry class called BDI. 
To define this operator we split the space into subspaces with projectors $\Sigma_{\pm}$ of eigenvalues $\pm 1$ of 
the chiral symmetry operator $S$. This would imply that $S\Sigma_{\pm}=\pm \Sigma_{\pm}$. The  BdG Hamiltonian is now 
purely off-diagonal in this space so that $\Sigma_+ H_{BdG}(k)\Sigma_+= \Sigma_- H_{BdG}(k)\Sigma_-=0$. We can 
then define the off-diagonal part  
\begin{align}
&A_k=\Sigma_+ H_{BdG}(k)\Sigma_-
\end{align}
of the Hamiltonian $H_{BdG}(k)$, which has the property that $Det[H_{BdG}(k)]=|Det[A_k]|^2$. Thus, $|Det[A_k]|$ can 
only vanish if $H_{BdG}(k)$ has a zero eigenvalue i.e. there is a gap closure.
Therefore, if $H_{BdG}(k)$ is fully gapped, we can define the phase winding of $Det[A_k]$
\begin{align}
&W=\int \frac{dk}{2\pi i}\frac{d Det[A_k]/dk}{Det[A_k]}
\end{align}
as a topological invariant characterizing the topological superconductor in symmetry class BDI~\cite{altland1997nonstandard}. 
This has interesting physical consequences such as it allows multi-channel generalizations of the superconductor semiconductor 
wires to have multiple Majorana modes that are protected as long as the chiral symmetry is preserved~\cite{tewari2012topological}.

\section{Experimental signatures}
\subsection{Transport signature}
As we saw from the schematic set-up in Fig.~\ref{fig:nanowire}, both the zero-energy 
Majorana mode as well as the quasiparticle gap closing associated with the topological quantum 
phase transition may be probed by tunneling transport. Such transport involves transfer of 
electrons from a normal lead into the superconductor which is reflected back as an 
electron or hole according to the scattering matrix:
	\begin{equation}\label{eq:Sij}
		S=\left(\begin{array}{cc}S^{ee} & S^{eh} \\ S^{he} & S^{hh}\end{array}\right).
	\end{equation}
The process of electron reflecting back as a hole from the normal-superconductor interface and transferring a Cooper pair to the superconductor is termed 
Andreev reflection.

\begin{figure}
\centering
\includegraphics[width=\linewidth]{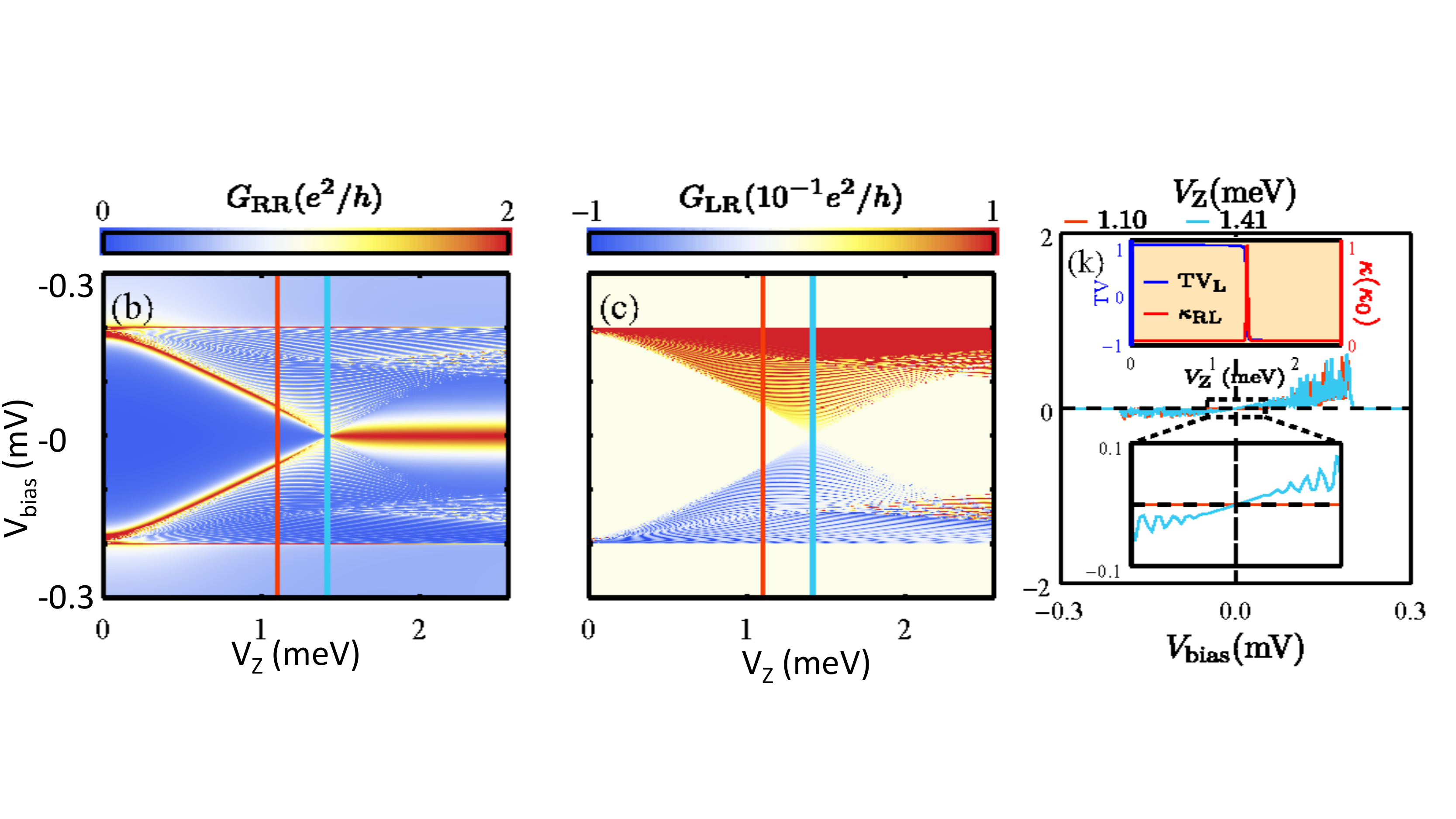}
\caption{(Left panel) End conductance $G=dI/dV$ into a semiconducting/superconducting nanowire 
as a function of applied Zeeman $V_Z$ relative to superconducting gap $\Delta$, shows a 
gap closure at the critical Zeeman field (shown by yellow line) $V_{Z,c}=\sqrt{\Delta^2+\mu^2}$~\cite{pan2020physical}. A zero-bias peak with height quantized 
at $2e^2/h$ emerges about the topological quantum phase transition at $V_Z=V_{Z,c}$. (Middle panel) The non-local conductance between the ends of a semiconducting/superconducting nanowires shows a closure of the superconducting gap where non-zero conductance appears near zero bias before a gap reopening. (Right panel) The non-local conductance at the topological critical point (blue line in the middle panel) shows a conductance which is linearly dependent on bias voltage. The inset shows that the gap 
closure is associated with quantized thermal conductance exactly at the topological phase transition. Figure reproduced from
Pan et. al. arXiv:2009.11809~\cite{pan2020physical}.  }\label{fig:endconductance}
\end{figure} 

Using the Blonder-Tinkham-Klapwijk\cite{blonder1982transition} formalism, the conductance into the superconductor can be written in terms of elements of this scattering matrix as 
\begin{equation}\label{eq:Glocal}
		G=\frac{e^2}{h}\left(N_{ch}-T^{ee}+T^{eh}\right),
	\end{equation}
	where 
\begin{equation}
 T^{\alpha\beta}=\tr\left([S^{\alpha\beta}]^\dagger S^{\alpha\beta}\right).\label{eqT} 
\end{equation} {[$ \tr(..) $ for the trace associated with additional channels such as spin etc.]} and 
	 $ N_{ch}=2 $ is the number of the electron mode in this single channel model, and the transmission. 
In the limit of a long wire, the transmission coefficient $T^{ee}=N-T^{eh}$ 
so that the conductance becomes related to the Andreev reflection probability 
\begin{equation}
		G^{(long)}\simeq\frac{2e^2}{h}T^{eh}.\label{eqGlong}
	\end{equation}
We can understand this conductance as resulting from the transfer of Cooper pairs from the normal lead to the superconductor.

We can compute the conductance into the end of a semiconductor nanowire by first  computing the scattering matrix $S(E)$ numerically using a program 
such as KWANT~\cite{groth2014kwant} and then substituting the answer into Eq.~\ref{eqGlong}.
In addition to a discretized version of the BdG Hamiltonian in Eq.~\ref{eqdomain}, to compute the scattering matrix $S(E)$, we need to specify the  
normal lead and a tunnel barrier. We typically choose the normal lead Hamiltonian to be similar to Eq.~\ref{eqdomain} except that $\Delta=0$ and 
the chemical potential $\mu_{lead}$ is much higher than that in the semiconductor nanowire.
The result shown in Fig.~\ref{fig:endconductance} shows a gap at small Zeeman potential, which closes 
as the Zeeman potential is increased and merges into a zero-energy peak which 
persists beyond the topological phase transition. The closure of the gap seen in the spectrum is consistent with the spectrum that we saw in Fig.~\ref{fig:gapclosure}
in the discussion of the topological quantum phase transition. Note that we do not 
typically see the reopening of the gap in these plots similar to experiments. Actually the measurement of a gap closure followed by 
the emergence of a zero-bias conductance peak qualitatively similar to that seen in Fig.~\ref{fig:endconductance}, seen 
in experiments~\cite{mourik2012signatures} a few years after the prediction has been one of the main motivating drivers of the field. 
However, several quantitative features are yet to be observed. As we will discuss later these together with certain alternative scenarios that 
may arise in these systems have made the search for Majorana modes somewhat controversial.

One of the quantitative features in the conductance plot shown in Fig.~\ref{fig:endconductance} is the quantized value of the height 
of the zero-bias conductance peak associated with the end Majorana mode. To understand this quantization of conductance into a Majorana 
mode, we eliminate the lead by using the Mahaux-Wiedenmuller transformation~\cite{beenakker1997randommatrix} to write the scattering matrix $S$ 
in terms of the Hamiltonian $H$ of the nanowire as: 
\begin{align}
&S=1-2\pi i W^\dagger(E-H+i\pi W W^\dagger)^{-1}W,\label{SMW}
\end{align}
where $W$ is an $N_{ch}\times N$ matrix and $N_{ch}$ is the number of channels and $N$ is the size of the BdG Hamiltonian of the nanowire.
Since the conductance in the tunneling limit is not expected to depend on the details of the lead (which can be checked numerically), 
\begin{equation}\label{eq:W}
W_{mn}=\delta_{m,n}, \quad 1\le m\le M, \quad 1\le n \le N.
\end{equation}
We can then write this scattering matrix in terms of the nanowire Green function $g_0(E)=(E-H)^{-1}$ by expanding as a formal power-series 
in $W$ as 
\begin{align}
&S=1-2\pi i W^\dagger [g_0(E) -i\pi W W^\dagger g_0(E)+\dots]W.
\end{align}
By writing $W=\Gamma w$, we can then formally resum the power-series to write $S$ in terms of the local part of the Green function $g_l(E)=w^\dagger g_0(E) w$
so that 
\begin{align}
&S=1-2i\pi \Gamma[g_l(E)^{-1}-i\Gamma^{-1}\pi]^{-1}.\label{eqSf}
\end{align}

The local Green function $g_l(E)$ can be thought of the Green function at the end of the wire and is only  an $N_{ch}\times N_{ch}$ matrix, 
where $N_{ch}=2$ in the simplest case of a spin-polarized lead. The two components of the channel are the particle and hole of the Nambu space.  
The zero-energy Majorana mode, which has a particle-hole symmetric wave-function, appears as a zero-energy pole of both $g_0(E)$ as well as 
$g_l(E)$. We can include this pole structure of $g_l(E)$, by approximating $g_l(E)$ near $E\sim 0$ as 
\begin{align}
&g_l(E)\simeq (\bm 1+\tau_x)u^2/E+a\tau_z.\label{eqgl}
\end{align}
Here we use a Nambu basis where the particle-hole symmetry of the Green function takes the 
form $\tau_x g_l(E)\tau_x=-g_l(-E)$. 
Substituting $g_l(E)$ into Eq.~\ref{eqSf} we get the amplitude of the electron-hole 
reflection amplitude to be 
\begin{align}
&r_{eh}=S_{eh}(E)=\frac{2i\pi\Gamma[u^2-i\pi \Gamma a^2 E]}{E(\pi^2\Gamma^2 a^2-1)-i\pi \Gamma u^2}\approx -\frac{i\pi\Gamma[u^2]}{E+i\pi \Gamma u^2},
\end{align}
where for the last step we took the limit of small $a$ i.e. contribution from other states.
Note that $r_{eh}(E=0)=-1$, which is referred to the phenomenon of perfect Andreev reflection.
This is a hall-mark signature of topological superconductors and indeed can be derived from the topological invariant~\cite{Wimmer2011Quantum}. 
In contrast, the non-topological superconducting case, which does not have any such zero energy pole can be understood from Eq.~\ref{eqgl} by assuming $u(E)\propto E$, is characterized by an 
 Andreev reflection amplitude $r_{eh}(E\sim 0)\rightarrow 0$ that vanishes.
The reflection amplitude $r_{eh}$ determines the reflection probabilty $T_{ii}^{eh}=|r_{eh}|^2$, so that the conductance resonance for the Majorana mode is given by 
\begin{equation}
		G_{Maj}(E)\simeq\frac{2e^2}{h}\frac{(\pi\Gamma u^2)^2}{E^2+(\pi\Gamma u^2)^2}.
	\end{equation}
This is the standard form for the conductance resonance associated with tunneling into a Majorana 
zero mode. What is remarkable is that the height of the peak is quantized i.e. $G_{Maj}(E=0)=2e^2/h$ independent of the value of the tunneling to the normal lead $\Gamma$. However, the total weight under the conductance peak, which is related to the current 
$I_{Maj}$ at bias voltages larger than the peak width $\Gamma u^2$ is proportional to 
$\Gamma$ i.e. 
\begin{align}
I_{Maj}=\int_0^{\infty}dE G_{Maj}(E)\propto \Gamma u^2.
\end{align}
Since the current ultimately vanishes with $\Gamma$, this resolves the apparent paradox of the zero-bias conductance being independent of tunneling 
even as the tunneling rate $\Gamma$ vanishes.

\begin{figure}
\centering
\includegraphics[width=\linewidth]{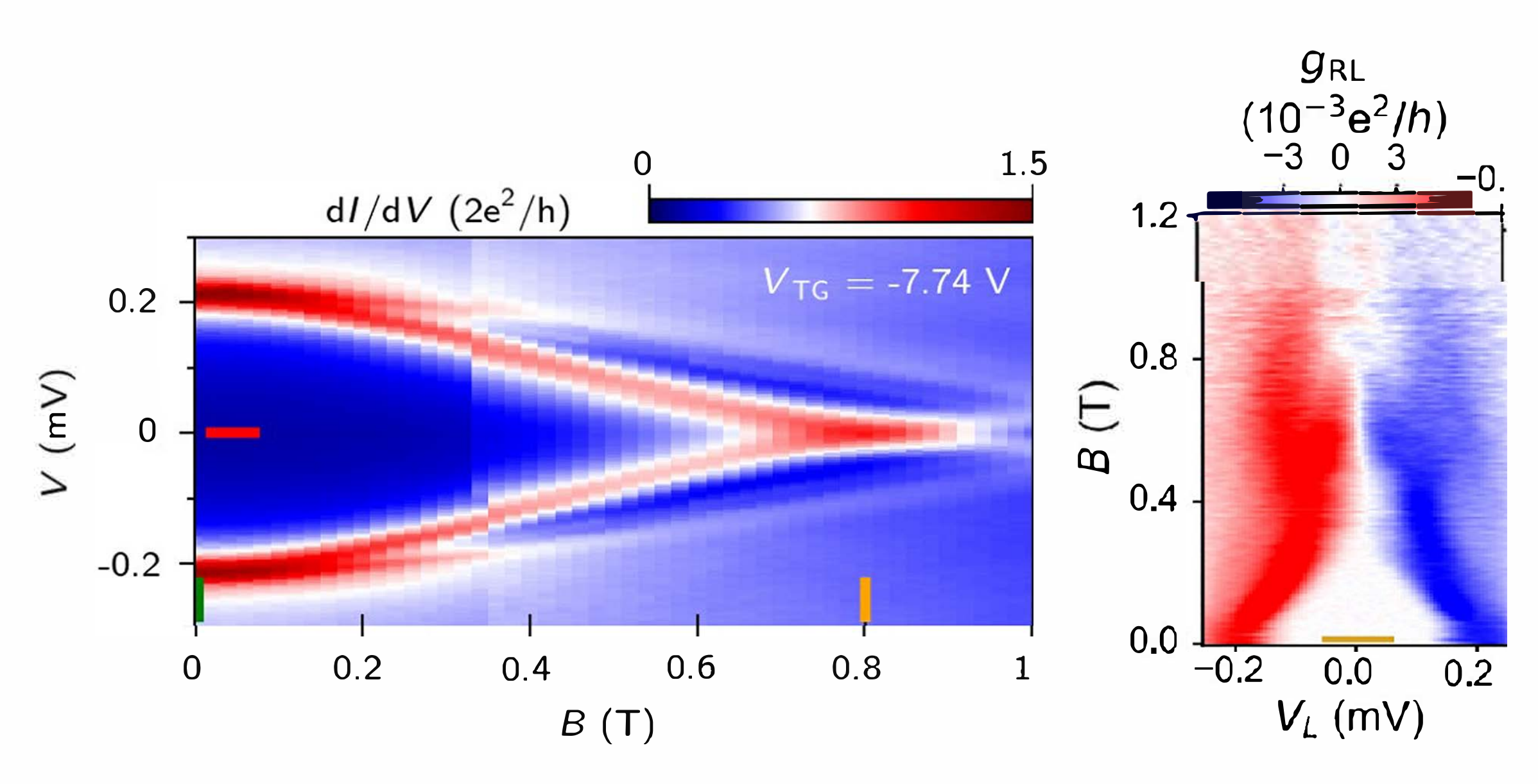}
\caption{(Color online) (Left panel) Conductance into the end of a semiconductor/superconductor nanowire~\cite{zhang2021large} similar to that measured in Mourik et al Science (2012)~\cite{mourik2012signatures}, 
shows a ZBCP signature theoretically expected from Majorana modes above a critical Zeeman field though gap closing and reopening signatures are not seen. Experimental figure reproduced from Zhang et al arxiv:2101.11456~\cite{zhang2021large}. (Right panel) Conductance across a similar superconducting nanowire in a three-terminal configuration 
shows evidence for a gap in transmission at smaller Zeeman field that is closed by increasing the field~\cite{puglia2020closing}. Hints of a reopening are seen at higher magnetic field, but the gap appears to be weak. Experimental figure reproduced from Puglia et al 	arXiv:2006.01275~\cite{puglia2020closing}}\label{fig:exptconductance}
\end{figure}

As already mentioned, the field received a large boost from early, though preliminary verification of the transport and Josephson predictions~\cite{mourik2012signatures,das2012zerobias,deng2012anomalous,churchill2013superconductornanowire,finck2013anomalous}. 
Specifically, as we see in the recent conductance data seen in the left panel of Fig.~\ref{fig:exptconductance}, the conductance as a function of bias voltage $V$ 
and applied Zeeman potential $V_Z$, shows a gap at small Zeeman field, which closes at about a magnetic field of $B\sim 0.7 T$ followed by the emergence of a 
zero-bias conductance peak with conductance near the quantized value predicted for Majorana zero modes. This is qualitatively similar to the first measurements 
of the conductance into the device~\cite{mourik2012signatures,das2012zerobias,deng2012anomalous,churchill2013superconductornanowire,finck2013anomalous}, though improvements in device fabrication since then have significantly enhanced the quality of the features.
Qualitatively similar results have been obtained by several groups since the first observation confirming that these results are quite reproducible ~\cite{deng2016majorana,zhang2017ballistic,nichele2017scaling,puglia2020closing,vaitiekenas2018effective,vaitiekenas2020flux,zhang2021large,yu2020nonmajorana}.
While these features are in qualitative agreement with the theory predictions seen in Fig.~\ref{fig:endconductance}, there are discrepancies. The first discrepancy 
worth noting is that neither the theoretical conductance (left panel of Fig.~\ref{fig:endconductance}) or the experimental measurement (left panel of Fig.~\ref{fig:exptconductance}) shows evidence of the bulk gap reopening expected from the right panel of Fig.~\ref{fig:gapclosure}. While this is not technically a 
discrepancy in the sense that the theoretical conductance in most models also do not show this feature because of competition from the zero-bias peak
it is an important feature to confirm. Secondly, the height of the conductance peak, although near the predicted value~\cite{nichele2017scaling,zhang2021large}, shows significant deviations both above 
and below the predicted value with changing parameters and therefore does not appear to be as robust as predicted by theory. We will elaborate on the 
implications of this discrepancy for the field in the last section of the chapter.

\subsection{Bulk gap closure}
The conductance from the end shown in Fig.~\ref{fig:endconductance} only shows us an apparent gap closure but not the gap reopening expected 
from the right panel of Fig.~\ref{fig:gapclosure}. Additionally, even the gap closure feature is typically expected to be obscured by the presence 
of Andreev bound states associated with complicated end potentials that we will discuss later. 
Instead, we can consider a more direct measure of the gapless states at the phase transition by studying transport through such states. 
Such an experiment can be performed by adding another lead on the right end $R$ of the semiconductor wire, in addition to the lead $L$ at 
the left end. Since the superconductor has to be grounded this is referred to as a three-terminal configuration. 
The scattering matrix $S$ must now be doubled to include scattering from both leads 
\begin{equation}\label{eq:S}
		S=\left(\begin{array}{cc}S_{\text{LL}} & S_{\text{LR}}\\S_{\text{RL}} & S_{\text{RR}}\end{array}\right), 
	\end{equation}
where each block $S_{ij}$ has the particle-hole structure of the scattering matrix in Eq.~\ref{eq:Sij} from the previous sub-section.
We can characterize the transport properties of such a three-terminal device by a conductance matrix 
\begin{equation}\label{eq:condmat}
		\hat{G}=\left(\begin{array}{cc}G_{\text{LL}} & G_{\text{LR}}\\ G_{\text{RL}} & G_{\text{RR}}\end{array}\right)=\left(\begin{array}{cc}dI_L/dV_L & -dI_L/dV_R \\ -dI_R/V_L & dI_R/dV_R\end{array}\right),
	\end{equation}
	where $ I_{\text{L}} $ ($ I_{R} $) is the current entering the left (right) normal lead from the scattering region, and $ V_{\text{L}} $ ($ V_{\text{R}} $).
In the limit of a long wire, $ G_{\text{LL}} $ and $ G_{\text{RR}} $ are the local conductances at each end that we discussed in the 
last sub-section and are plotted in the left panel of Fig.~\ref{fig:endconductance}. The transport properties 
across the wire would be measured from the nonlocal conductances ($ G_{\text{LR}} $ and $ G_{\text{RL}} $).
 These non-nonlocal conductances can be written in terms of the transmission probabilities defined in Eq.~\ref{eqT}
	\begin{equation}\label{eq:Gnonlocal}
		G_{ij}=\frac{e^2}{h}(T_{ij}^{ee}-T_{ij}^{eh}), \qquad i\neq j,
	\end{equation}
and are therefore expecteed to vanish for a gapped wire (away from the topological phase transition). 

The appearance 
of a finite non-local conductance near zero bias would signal such a bulk gap closure. This expectation is verified~\cite{rosdahl2018andreev}
is verified by the numerical results for the non-local conductance shown in the middle panels Fig.~\ref{fig:gapclosure}. These results were 
obtained for a scattering matrix $S$ computed using KWANT~\cite{groth2014kwant} using the BdG Hamiltonian Eq.~\ref{HBdG} for a nanowire. 
An unfortunate complication of these results is that the non-local conductance actually, even at the topological phase transition, vanishes at exactly zero bias
as a consequence of particle-hole symmetry~\cite{akhmerov2011quantized}. While the vanishing at the critical point is linear in voltage 
as shown in the right panel of Fig.~\ref{fig:gapclosure} as opposed to a gapped signal away from the phase transition, introduction of disorder etc 
might make the bulk gap closure difficult to identify in a definitive way. It is worth mentioning that an advantge of the scattering matrix formalism 
is that we do not need the bulk Hamiltonian or the topological condition Eq.~\ref{C0}, which are limited to clean single-band systems, to identify 
the topology of the nanowire. We can compute the topological invariant $TV_L$ directly from the scattering matrix $S_{LL}(E=0)$ from the scattering 
matrix topological invariant 
\begin{align}
&TV_L=det(S_{LL}(E=0)).
\end{align}
The numerical result for this invariant is shown in the inset of the right panel of Fig.~\ref{fig:gapclosure} and confirms that the topological invariant $TV_L$ 
vanishes at the quantum critical point.

The vanishing of $TV_L$ at the topological quantum critical point, suggests a zero-mode in the scattering matrix $S_{LL}(E=0)$. This means that 
there is a mode that suffers no reflection when incident from the left lead $L$ and must thus be completely transmitted. This seems 
to contradict the linear vanishing of the non-local conductance $G_{RL}$ that we discussed in the last paragraph.
As will become clear when we discuss the teleportation process through a Majorana wire, the vanishing of the non-local conductance through a Majorana 
wire occurs because of the transformation of an electron into an equal superposition of electron and hole during the transmission process. While this 
destroys the charge associated with the transfer of the electron, it does not reduce the entropy transfer associated with this transmission.
This entropy transfer at the bulk gap closure contributes to the heat conductance 
	\begin{equation}\label{eq:Gth}
		\kappa=\kappa_0(T_{ij}^{ee}+T_{ij}^{eh}), \qquad i\neq j,
	\end{equation}
which can also be computed from the transmission probabilities $T_{ij}^{ab}$ computed from the scattering matrix $S$.
The perfect transmission of quasiparticles at the bulk gap closure associated with the topological quantum phase transition appears as a quantized peak 
in thermal conductance with height $ \kappa_0=\pi^2k_B^2\tau/6h $ at temperature $ \tau $ happens at TQPT~\cite{senthil1999spin,senthil2000quasiparticle,evers2008anderson} {($ h $ for the Planck constant and $ k_B $ for the Boltzmann constant)}.

The non-local conductance, for which theoretical results were shown in the middle panel of Fig.~\ref{fig:endconductance}, has also been measured 
in recent experiments~\cite{puglia2020closing}. The results, though not as extensively as the end conductance, shown in the right panel of Fig.~\ref{fig:exptconductance}
 indicate the existence of a gap at low magnetic fields consistent with the local conductance. In this plot, we infer a gap from the range of bias voltage over which 
the conductance vanishes. Increasing a magnetic field appears to suppress this gap near the critical value where a zero bias conductance peak appears. 
However, the data is not definitive about a re-entrant gap beyond the critical value, where only a slight suppression of non-local conductance is seen.

\subsection{Fractional Josephson effect}
\begin{figure}
\centering
\includegraphics[width=\linewidth]{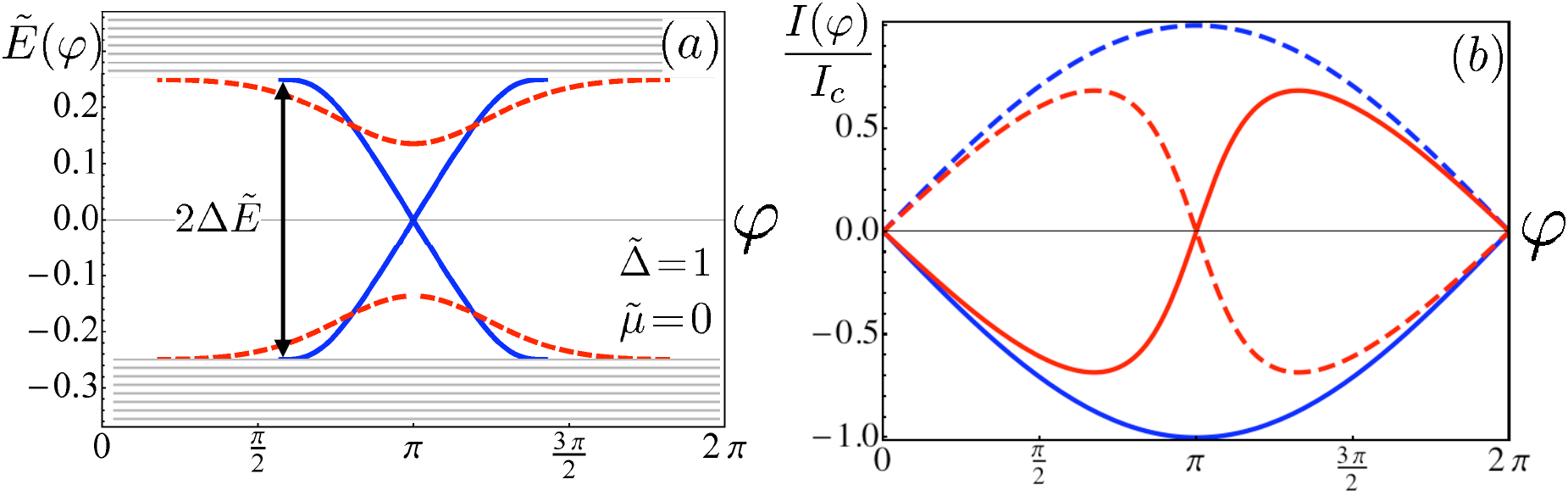}
\caption{(a) Andreev bound state spectrum of a JJ in a superconducting  in the  topologically trivial (dashed line: $\tilde V_x\!=\!0.75$) and  topologically nontrivial (solid line: $\tilde V_x\!=\!1.25$) superconducting phases. 
The spectrum as a function of superconducting phase $\phi$, in the trivial case, shows a gap while that in the nontrivial case shows a crossing of zero-energy with a pair 
of zero-energy Majorana modes at phase $\phi=\pi$~\cite{lutchyn2010majorana}.
(b) The zero-energy crossing of the Andreev state spectrum of the topologically non-trivial superconducting phase  leads to a supercurrent (shown in blue) that changes 
sign i.e. going from the solid to the dashed line as the phase $\phi$ advances by $2\pi$. The topological superconducting supercurrent is thus $4\pi$ periodic, leading to the 
fractional Josephson effect, as opposed to the topologically trivial supercurrent (shown in red) which is $2\pi$ periodic as expected~\cite{lutchyn2010majorana}. Figure courtesy of Lutchyn et al arXiv:1002.4033~\cite{lutchyn2010majorana}.  }\label{fig:spectrumtnph}
\end{figure}

As we saw earlier from the spectrum of Josephson junction in topological superconductors (i.e. Eq.~\ref{eqEJ}), a $\pi$ phase difference leads 
to a degenerate pair of states, which splits in energy as the phase difference $\phi$ deviates from $\pi$. Formally Eq.~\ref{eqEJ} contains a wave-vector $k_y$. 
However, we saw in the dimensional reduction argument to one dimensional nanowire (see subsection \ref{nanowire}) that the results for the nanowire can be obtained by setting 
$k_y=0$ in the domain wall results for the two dimensional superconductor. A detailed numerical calculation of the spectrum for a Josephson junction 
using the BdG Hamiltonian of a nanowire (i.e. Eq.~\ref{HBdG}), which is shown in panel (a) of Fig.~\ref{fig:spectrumtnph}, confirms the existence of the 
pair of zero-energy bound states at phase $\phi=\pi$ on in the topological superconducting (i.e. topologically nontrivial)  phase. The topologically trivial phase shows a 
gapped spectrum as a function of phase $\phi$ as expected.

While the value of phase $\phi$ where the zero-energy crossing occurs is not protected against changes in detail of the
Hamiltonian of the junction, 
the crossing itself is protected by particle-hole symmetry. Let us start by writing the dispersion of the Andreev bound states (i.e. Eq.~\ref{eqEJ}) near phase $\phi\sim\pi$ 
as 
\begin{align}
&E_{\pm}\simeq \pm \zeta (\phi-\pi),
\end{align}
where the index $\pm$ refer to the two branches of the spectrum for the topologically nontrivial phase seen in Fig.~\ref{fig:spectrumtnph}(a). 
The level crossing of Andreev states as the phase $\phi$ changes by $2\pi$ can be understood in terms of the topological invariant in Eq.~\ref{topinvKitaev} 
by considering the topological superconductor Josephson junction in a ring. The $2\pi$ superconducting phase difference is generated across 
the Josephson junction by introducing a superconducting flux quantum $\Phi=\Phi_0$, where $\Phi_0$ is actually half an electron flux quantum $2\Phi_0=hc/e$ 
( by virtue of the electron charge being half a Cooper pair charge ). Thus, changing the superconducting phase $\phi$ by $2\pi$ changes the boundary conditions 
around the ring from periodic i.e. $k=0$ to anti-periodic i.e. $k=\pi$, where $k$ is the wave-vector for Bogoliubov quasiparticles.
For topological superconductors with a non-trivial value for the topological invariant in Eq.~\ref{topinvKitaev}, such a change 
in boundary condition leads to a change in the Pfaffian of the BdG Hamiltonian with Josephson junction phase $\phi=0$ and $\phi=2\pi$.
This change in the sign of the Pfaffian  according to Eq.~\ref{PfHBdG} from $\phi=0$ to $\phi=2\pi$ guarantees a zero-energy level crossing 
of the Andreev bound states in the junction between $\phi=0$ and $2\pi$ for the topologically nontrivial phase as seen in Fig.~\ref{fig:spectrumtnph}(a). An odd number of such crossings is uniquely associated with 
topological superconducting phase 
based on the invariant in Eq.~\ref{topinv}. 

\begin{figure}
\centering
\includegraphics[width=\linewidth]{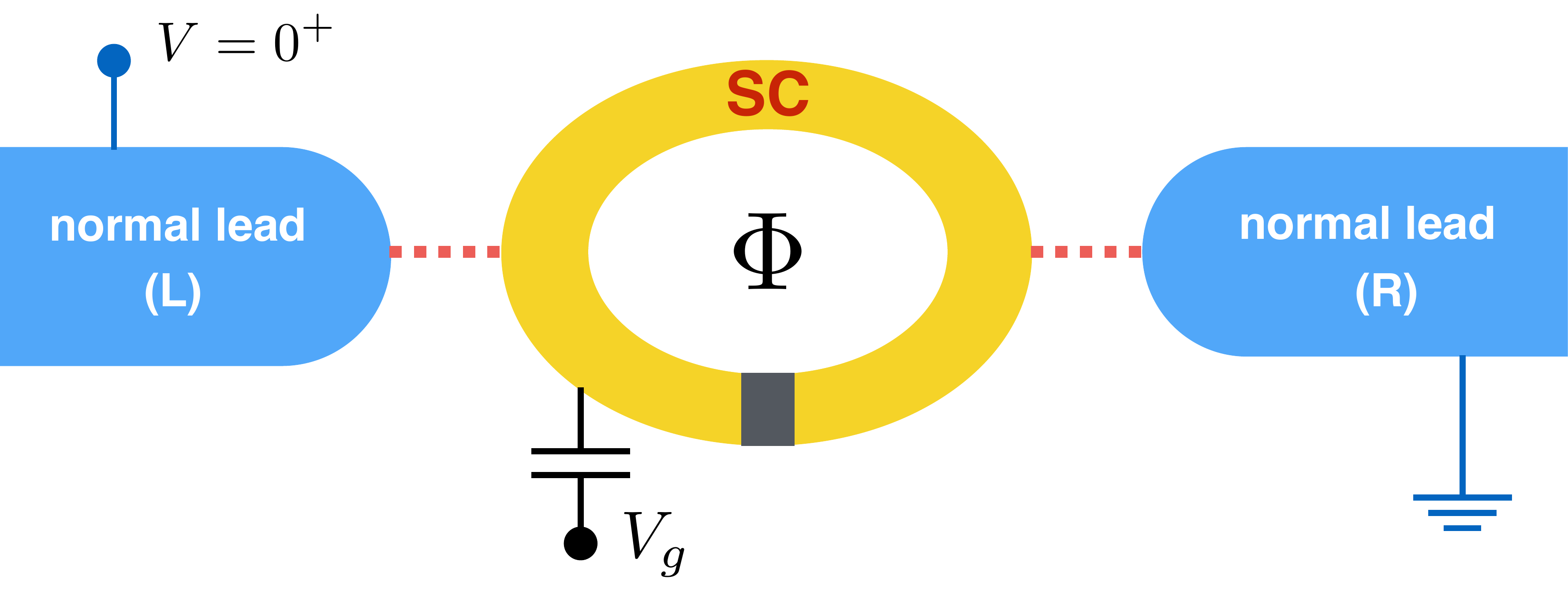}
\caption{(Color online) Andreev level crossing in a topological Josephson junction is associated with fermion parity 
change in topological superconducting ring. fermion parity change can in principle be detected by Coulomb blockade transport~\cite{liu2019proposal}. Figure reproduced from Liu et. al. arXiv:1803.01872~\cite{liu2019proposal}}\label{fig:spectrumtnph2}
\end{figure}

A physical consequence of the change of the Pfaffian from a change in flux is the change in ground state fermion 
parity~\cite{Read2000,Kitaev2001,Stone_2006} associated with such a flux in the set-up shown in Fig.~\ref{fig:spectrumtnph2}.
To understand the connection between the Pfaffian and ground state fermion parity, note that 
the two branches of the spectrum that are seen near the zero-energy crossing in Fig.~\ref{fig:spectrumtnph}(a) are related 
by particle-hole symmetry that transforms $E\rightarrow -E$ and also transforms the creation operator $\psi^\dagger$ for the state to $\psi^\dagger\rightarrow \psi$. 
Thus, we can combine these two branches in terms of the creation operator to
\begin{align}
&H_{JJ}(\phi)\approx \zeta(\phi-\pi)\psi_\phi^\dagger\psi_\phi.
\end{align}
Noting that the fermion operator is continuous across $\phi\sim\pi$ i.e. $\psi^\dagger_{\phi\sim\pi}\sim\psi^\dagger_\phi$, 
the crossing of the pair of Andreev bound 
states seen in Fig.~\ref{fig:spectrumtnph}(a) is really a zero crossing of the energy $\zeta(\phi-\pi)$ of the fermion state with creation 
operator $\psi^\dagger$. 
Once the energy of such a state goes 
from being negative to positive, we can lower the energy of the system by emptying the fermion state. 
Thus, zero-energy level crossings, which are associated 
with changes in the sign of the Pfaffian according to Eq.~\ref{PfHBdG}, are also associated with a change in fermion parity.
This suggests that the sign of the Pfaffian of the BdG Hamiltonian is related to it's ground state fermion parity, as can 
also be established by direct computation~\cite{Stone_2006}.
Thus the change in Pfaffian in going from phase $\phi=0$ to $\phi=2\pi$ implied by Eq.~\ref{topinvKitaev}, also implies a change in
the ground state fermion parity of the Josephson junction as the superconducting phase $\phi$is changed by $2\pi$. 
 The change in the ground state fermion parity of the ring represents a change 
in the number of electrons in the ring from even to odd. Such a change in fermion parity can be 
detected by measuring Coulomb blockade transport in the superconducting 
ring~\cite{liu2019proposal}.

The superconducting ring used in the setup in Fig.~\ref{fig:spectrumtnph2} to detect the fermion parity change is practically challenging to construct. Alternatively, 
we can consider a case where the JJ is isolated from external leads so that the fermion parity of the system remains fixed as one changes the phase $\phi$ by $2\pi$.
This necessarily forces a topological superconducting system with a changing ground state fermion parity to enter an excited state when the phase changes by $2\pi$.
Introducing a second change of phase $\phi$ by $2\pi$ restores the system to the ground state. This leads to a $4\pi$ periodicity of the supercurrent of the form
\begin{align}
&I_{JJ}(\phi)=I_{2\pi}\sin{\phi}+I_{4\pi}\sin{\phi/2},\label{Ifrac}
\end{align}
where $I_{2\pi}$ is a conventional contribution to the supercurrent from topologically trivial channels and $I_{4\pi}$ is the amplitude of the $4\pi$ periodic component from the topologically superconducting channel.
Such a $4\pi$ periodic supercurrent that arises from the change in quasiparticle occupation in the junction is termed the fractional Josephson effect~\cite{Kwon2004Fractional} and 
is a hall-mark of topological superconductivity.
The supercurrent for the topologically nontrivial phase in Fig.~\ref{fig:spectrumtnph}(b) is consistent with the form of $I_{JJ}$ in the limit where $I_{4\pi}\gg I_{2\pi}$.

\begin{figure}
\centering
\includegraphics[width=\linewidth]{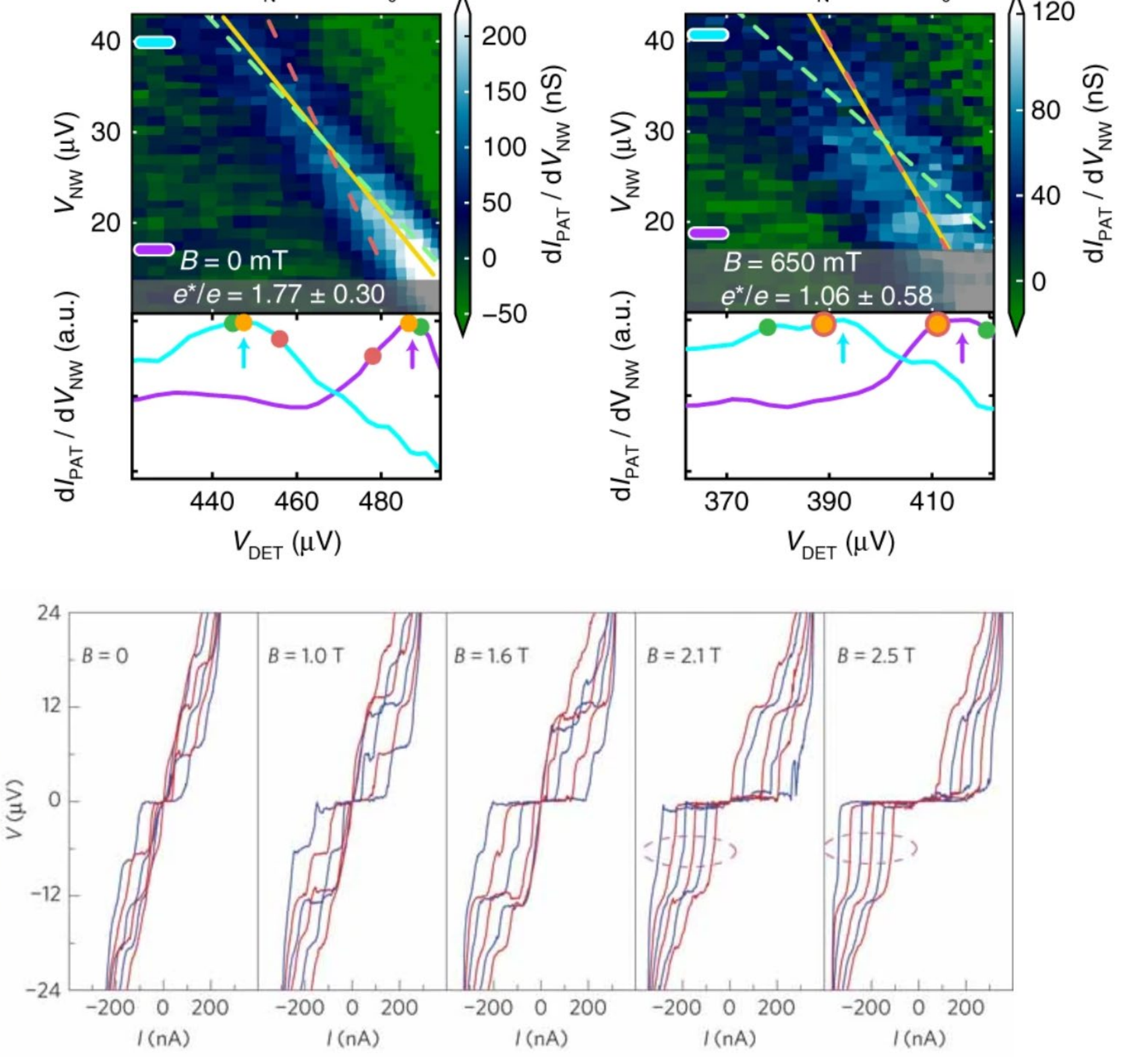}
\caption{(Color online) (Top panel) The spectrum of radiation emitted by a Josephson junction in a a superconducting nanowire in the absence of a magnetic field (i.e. $B=0$), shows a conventional Josephson relation between frequency of radiation (related to $V_{Det}$ and the voltage $V=V_{NW}$ applied across the Josephson junction. The right panel, which is in the presence of a large magnetic field, shows the frequency of the emitted radiation (i.e. $V_{DET}$) drop by half relative to $V$~\cite{laroche2019observation}. (Bottom panel) Applying an external ac radiation of frequency $\omega$ generates Shapiro steps at applied voltage $V$ that are multiples of the frequency $\omega$. In the topologically trivial phase (i.e. at $B=0$) one sees a conventional sequence of steps in voltage $V$ at integer multiples of $\omega$. Applying a magnetic field leads to the disappearance of the first 
shapiro step suggesting voltage steps at $2\omega$~\cite{Rokhinson2012Fractional}. Figures reproduced from Laroche et. al. arXiv:1712.08459 ~\cite{laroche2019observation} and Rokhinson et al. arXiv:1204.4212~\cite{Rokhinson2012Fractional}}\label{fig:exptJosephson}
\end{figure}

The $4\pi$ periodic supercurrent $I_{JJ}(\phi)$ associated with the fractional Josephson effect may be measured from the radiation of a voltage biased 
Josephson junction~\cite{laroche2019observation}. Applying a voltage $V$, the phase across the Josephson junction varies as $\phi=V t$, which leads to an 
ac Josephson supercurrent $I_{JJ}(\phi=V t)=I_{2\pi}\sin{Vt}+I_{4\pi}\sin{Vt/2}$. The radiation from the ac current   
for the topologically trivial phase at zero magnetic field, which is shown in the top left panel of Fig.~\ref{fig:exptJosephson}, is peaked at a frequency 
that has a conventional slope with respect to the applied voltage across the junction. In contrast the frequency of the radiation in the 
topologically non-trivial phase at large Zeeman field has slope with voltage, which is half of the conventional case. 
Being able to detect the radiation from the Josephson junction in a single nanowire requires sensitive on-chip detector technology. Because of this, the first measurements of the 
fractional Josephson effect was based on the 
  Shapiro voltage steps where a microwave irradiated Josephson junction is used to generate finite voltage steps~\cite{Tinkham1996introduction}.
The origin of the voltage steps can be understood by considering a phase of the form $\phi=V t+\phi_{ac}\sin{(\omega t)}$, 
where one can check that a dc current is supported 
in the limit of $\omega$ and 
$V$ becoming commensurate. Similar to the half frequency radiation from the fractional Josephson effect,
 the voltage steps $V$ (relative to $\omega$) in a potentially topologically non-trivial superconducting phase was observed to be larger 
 by a factor of 2~\cite{Rokhinson2012Fractional} as seen in the right panel of 
Fig.~\ref{fig:exptJosephson}. Since these measurements involve measurement of DC voltages, 
they preceded the radiation measurements and were observed 
in 2012~\cite{Rokhinson2012Fractional} about the same time as the zero-bias conductance peak measurements.
 In fact both these effects have also been claimed to be observed in other systems as well~\cite{wiedenmann20164,deacon2017josephson}. 
 However, the constraint of a fixed fermion 
parity, which is required for the validity of the fractional Josephson effect (i.e. Eq.~\ref{Ifrac}), requires a dynamical measurement 
on a time-scale shorter than the quasiparticle poisoning time. Realistic experimental systems have a finite density of sub-gap states that would allow the excited state from the zero-energy crossing of the Andreev state  
to relax by a process that is referred to as quasiparticle poisoning unless the experiment is done on a sufficiently short time-scale ~\cite{lutchyn2010majorana,houzet2013dynamics}. On the other hand, rapidly changing the phase $\phi$ 
drives the system out of equilibrium through Landau-Zener transitions~\cite{sau2017detecting} that can also produce unconventional current 
phase relations in topologically trivial superconductors. In fact, such non-topological fractional Josephson effects from processes have already been claimed 
to be observed both in conventional JJ~\cite{billangeon2007ac} as well as InAs nanowires~\cite{dartiailh2021missing}.

\subsection{Teleportation}\label{teleportation}

\begin{figure}
\centering
\includegraphics[width=\linewidth]{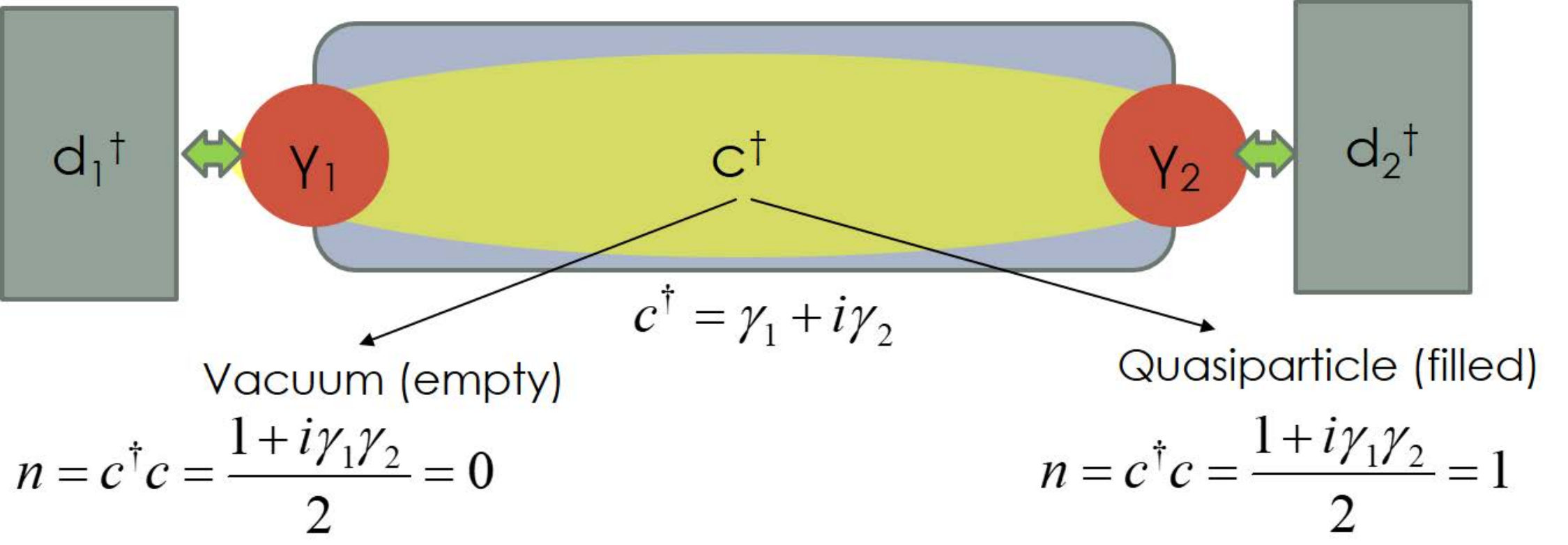}
\caption{(Color online) Pair of Majorana modes $\gamma_{1,2}$ form conventional fermion $c^\dagger=(\gamma_1+i\gamma_2)/2$. This form of fractionalization of a fermion into Majorana modes leads to non-local transport or teleportation~\cite{Fu2010Teleportation} of electrons from a left lead to a right lead in the scenario that the topological superconductor is Coulomb blockaded.}\label{fig:teleportation}
\end{figure}

The fact that a pair of Majorana modes $\gamma_{1,2}$ can be used to construct a single fermion mode $c^\dagger=\gamma_1+i\gamma_2$ 
leads to a unique transport property of Coulomb blockaded Majorana wires~\cite{Fu2010Teleportation} when in the set-up shown in Fig.~\ref{fig:teleportation}. 
Before considering the effect of Coulomb blockade on the central wire, let us first assume that the pair of end Majorana modes $\gamma_{1,2}$
 have a small splitting $\epsilon$.
The Hamiltonian for such a system is 
\begin{align}
&H=\sum_{\alpha}t_\alpha[d_\alpha^\dagger \gamma_\alpha+h.c]+\epsilon_c c^\dagger c,
\end{align}
where $t_\alpha$ is the coupling to the lead fermions $d^\dagger_{\alpha=1,2}$.
The above Hamiltonian can be written compactly in a Majorana basis consisting of $\tilde{\gamma}_\alpha=d_\alpha^\dagger+d_\alpha$
\begin{align}
&H=i\sum_{\alpha}t_\alpha[\tilde{\gamma}_\alpha \gamma_\alpha]+i\epsilon_c \gamma_1\gamma_2.
\end{align}
We can eliminate the Majorana modes in the wire $\gamma_{1,2}$ from the low-energy states in the tunneling limit $E, t_\alpha \ll \epsilon_c$, 
so that the Hamiltonian can be approximated by an effective tunneling of the lead Majorana modes 
\begin{align}
&H_{eff}=i\tilde{\epsilon}\tilde{\gamma}_1\tilde{\gamma}_2=i\tilde{\epsilon} (d_1+d_1^\dagger)(d_2+d_2^\dagger),
\end{align}
where $\tilde{\epsilon}=t_1 t_2/\epsilon_c$. 
The above Hamiltonian implies an effective direct transmission between the two leads similar to that discussed in the subsection 
on bulk gap closure. In fact, an electron $d_1^\dagger$ in lead 1 transmits to an equal mixture of electron and hole
consistent with the observation of a vanishing non-local conductance at zero bias in Fig.~\ref{fig:endconductance} (middle and right panels)~\cite{Bolech2007}.

Let us now consider the limit of a long wire, where $\epsilon_c$ is generated by Coulomb blockade as 
the energy difference between different electron number states. In this case the operator $c^\dagger$ has the same charge as an electron 
and is indeed exactly  an electron creation operator.
To understand the effect of the charging energy, we eliminate the Majorana modes $\gamma_{1,2}$ from $H$ in favor of the electron operator $c$ so that 
\begin{align}
&H=\sum_{\alpha}t_\alpha[(d_\alpha^\dagger-d_\alpha) (s_{\alpha}^*c+s_{\alpha}c^\dagger)]+\epsilon_c c^\dagger c,
\end{align}
where $s_\alpha=1,i$.
In the case of strong Coulomb blockade charge conservation violating terms such as $d_\alpha^\dagger c^\dagger$ are projected out of the intermediate state so that
the system is described by  
\begin{align}
&H_C=\sum_{\alpha}t_\alpha[(s_\alpha d_\alpha^\dagger c-s_\alpha^* d_\alpha c^\dagger) ]+\epsilon_c c^\dagger c.
\end{align}
The Hamiltonian for the system is now exactly equivalent to transmission of electrons between leads through a  non-interacting quantum dot i.e. a 
Fano resonance with an amplitude $t_1t_2/\epsilon_C$ at $E=0$. Even wires that are significantly longer than the coherence length can have a large charging energy with a large energy $\epsilon_c$.
In this case, the above Hamiltonian $H_C$ describes a process of teleportation of electrons between the leads $1$ and $2$ through the Majorana wire.
This process is actually a coherent transfer of an electron and may be used to propose an interferometric signature of Majorana modes. Preliminary 
evidence of such interferometric signatures have recently been observed by the Copenhagen group~\cite{whiticar2020coherent}.

\begin{figure}
\centering
\includegraphics[width=\linewidth]{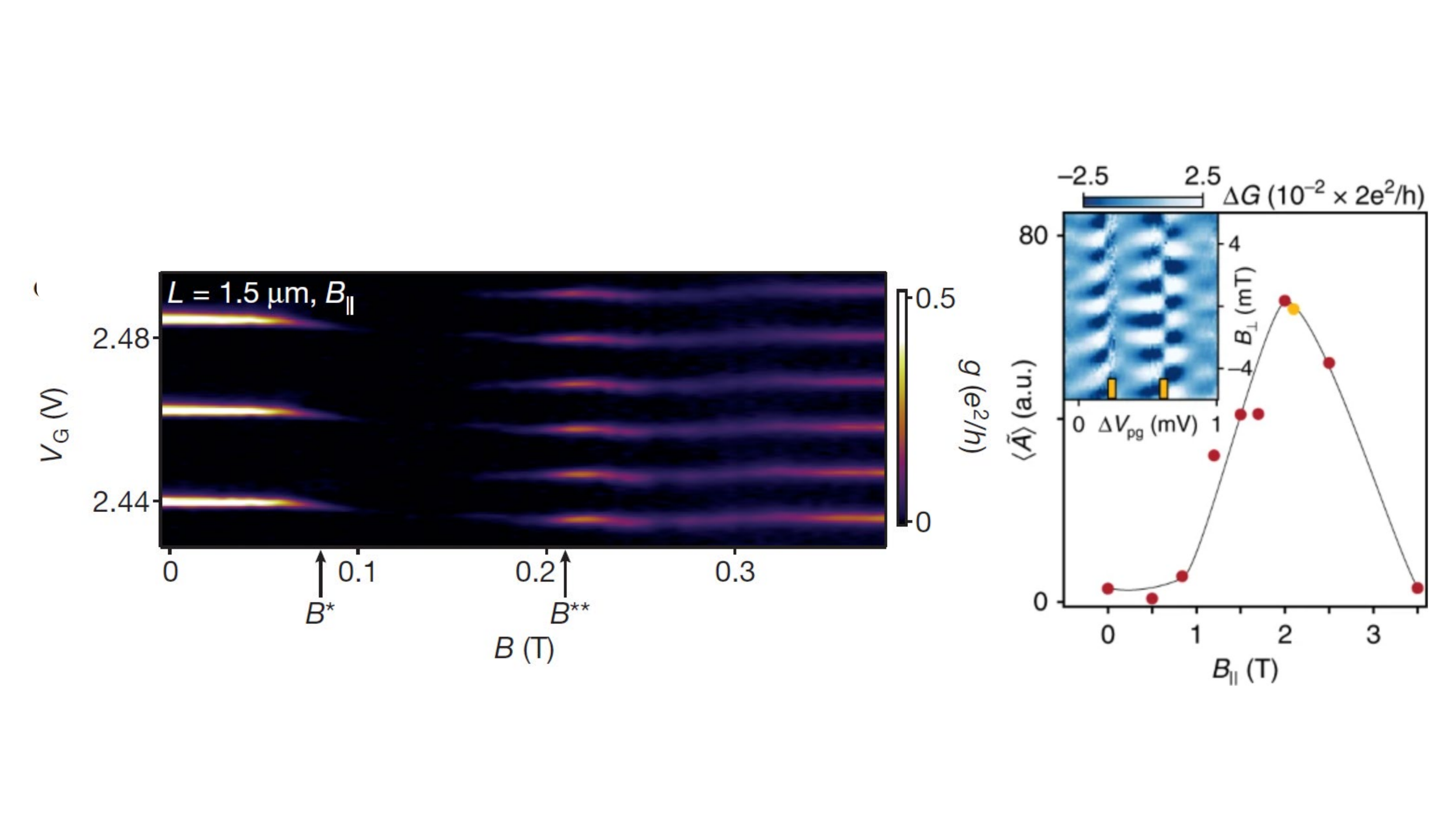}
\caption{(Left panel) Shows the zero-bias conductance as a function gate voltage shows $2e$ periodic Coulomb blockade oscillations that evolve into 
$e$ periodic oscillations with increasing magnetic field where a topological superconducting phase is possible~\cite{albrecht2016exponential}. In the case of a long gapped wire, such a conductance 
could represent teleportation through Majorana modes. (Right panel) Shows interference of electrons transmitted through a superconducting semiconductor wire 
in a putative topological regime as a flux dependent conductance oscillation~\cite{whiticar2020coherent}. The conductance oscillation shifts phase as the gate tunes the energy of the bound 
state $c^\dagger$ (see Fig.~\ref{fig:teleportation}) through the Fermi energy. Figures courtesy of Albrecht et al arXiv:1603.03217
~\cite{albrecht2016exponential} and Whiticar et al. arXiv:1902.07085~\cite{whiticar2020coherent}.
}\label{fig:exptteleportation}
\end{figure}

The teleportation process  can be used to measure a Majorana qubit, which is one of the key ingredients of the braiding protocols discussed earlier
in subsection~\ref{braiding}. 
To understand how we would use $H_C$ to measure a Majorana qubit, imagine that the superconducting island in Fig.~\ref{fig:teleportation} 
had an additional semiconductor wire with its own set of Majorana modes $\gamma_{3,4}$. The parity of the extra pair of Majorana modes would 
affect the fermion parity of the island, which ultimately affects the sign of $\epsilon_C$. Since the Coulomb blockade can be chosen to favor a particular 
total fermion parity $i\gamma_1\gamma_2\gamma_3\gamma_4$, we can view the sign of $\epsilon_C$ as being set by the fermion parity of the 
Majorana wire $i\gamma_1\gamma_2$. Thus, the effective tunneling between 
the leads is modified to
\begin{align}
&t_{eff}=\frac{t_1t_2}{\epsilon_C}(i\gamma_1\gamma_2).
\end{align}
The sign of this effective tunneling amplitude reflects the phase of the transmitted electron, which can be measured by interference
~\cite{plugge2017majorana}. 
Measurement of the interference phase would be equivalent to the measurement of the fermion parity of the Majorana qubit.

Interestingly, preliminary experimental evidence for teleportation has been seen in recent experiments as seen in Fig.~\ref{fig:exptteleportation}.
The conductance shown in the left panel of  Fig.~\ref{fig:exptteleportation}, shows that conductance through such a wire in the Coulomb blockade regime
which is $e$ periodic in gate voltage. This is characteristic of transport through an electronic bound state in the wire. A relatively trivial possibility for the 
origin of such a state, which is difficult to rule out experimentally~\cite{liu2019proposal}, is a conventional bound state in the middle of a short wire.
The signal from such a state in the case of a long wire would be negligible and can only arise if there are a pair of bound states similar to the Majorana states seen in 
  Fig.~\ref{fig:teleportation}. However, the same signal can arise from Andreev bound states (ABSs) at the ends of the wire that we will discuss later~\cite{Sau2015Proposal}. 
The non-local conductance from  ABSs is expected to be different from those arising from Majorana modes i.e. the conductance from ABSs
 would not show interference as seen in 
the right panel of Fig.~\ref{fig:exptteleportation}. However, as mentioned, it is difficult to rule out the short wire scenario without more careful experiments such as 
those that have been proposed~\cite{liu2019proposal}.

\section{Discussion and Conclusion}
After a decade of intense search for MZMs in SM-SC heterostructures, the field is right now at a crossroads. Tremendous efforts have been invested in the past few years to improve the quality of the various interfaces and reduce non-magnetic and magnetic disorders in the heterostructures. This has resulted in the conversion of a soft gap characterized by a substantial non-zero-bias background sub-gap conductance induced by disorder scattering in the early Majorana experiments \cite{mourik2012signatures} to a near perfect hard gap in the recently realized full epitaxy InAs/Al hybrid nanowires \cite{Chang2015Hard}. Experimental progress in reducing dissipation has also resulted in substantial enhancement in the height of the zero bias conductance peak from about $\sim 0.1 \frac{2e^2}{h}$ in the first generation Majorana devices \cite{mourik2012signatures} to zero bias conductance peak height approaching and exceeding $\frac{2e^2}{h}$ in the recent experiments \cite{nichele2017scaling,zhang2021large,yu2020nonmajorana}. However, despite the claims of several breakthroughs, the field has perennially seemed to remain on the cusp of a confirmatory evidence of MZM that has not yet been materialized in experiments.          

The main obstacle to a confirmatory evidence of MZM is that ends of a nanowire are often locations where robust zero energy states are induced by various non-topological effects that have little or no connection to MZMs. To distinguish such robust zero energy states of non-topological origin from topological MZMs, experiments investigating the quantization of the zero bias peak heights at value $\frac{2e^2}{h}$, and the persistence of this quantized peak height with variation in the experimental parameters (the so-called quantized conductance plateau), have recently attracted vigorous attention. While ballistic Andreev reflection \cite{blonder1982transition} from an ordinary zero mode leads to a conductance quantization with zero bias peak height $\frac{4e^2}{h}$, MZMs should lead to a quantized peak height of $\frac{2e^2}{h}$ as they  effectively behave as ``half a fermion''. Moreover, topological MZMs being insensitive to weak  perturbations, the persistence of the ZBP height with variations in the tunnel gate potential and applied magnetic field are taken as spectacular transport evidence unique to MZMs. 

In recent experiments \cite{nichele2017scaling,zhang2021large,yu2020nonmajorana} the height of the zero bias conductance peak has indeed been observed to approach or exceed $\sim \frac{2e^2}{h}$. However, a quantized conductance plateau for a convincing range of tunnel barrier or magnetic field is yet to be realized. Unfortunately, quantized conductance peak alone at isolated points in the parameter space cannot be taken as confirmatory evidence for MZMs. Furthermore, the absence of a convincing quantization plateau around the points in parameter space with ZBP height $\sim \frac{2e^2}{h}$  may be an indication that the robust zero energy states in these systems may in fact have originated from either,  (a) disorder-induced weak anti-localization that leads to robust class-D conductance peaks with peak height between $0$ and $\frac{2e^2}{h}$,~\cite{Pikulin2012Zero,bagrets2012class,mi2014xshaped,pan2020generic} or (b) partially separated Andreev bound states (ps-ABS) \cite{Moore2018}, also known as quasi-Majorana modes \cite{vuik2018reproducing,Stanescu_Robust}, whose low bias conductance signature depends on the overlap of the wave functions of component Majorana bound states of a conventional fermionic state. The observation  of zero bias conductance peaks with peak height $\sim \frac{2e^2}{h}$ only at isolated points in the parameter space and absence of convincing quantized plateaus around them may be an indication of significant residual disorder in the hybrid nanowires. With sustained improvements in material parameters by reducing disorder and interface inhomogeneity, we hope that quantized conductance plateau with peak height $\frac{2e^2}{h}$ will eventually be found in experiments, not just along a single tuning parameter but in islands in a higer-dimensional parameter space, confirming the existence of topological MZMs in SM-SC heterostructures.

\bibliography{3terminal.bib}

\end{document}